\documentclass[twocolumn]{aastex631}
\usepackage{float}
\usepackage{hyperref}

\begin{document}

\title{Photometric Mapping of Carbonaceous/Siliceous Dust and Water Ice in the ISM with JWST:\\ 
Applications to the Dense Sightlines
\footnote{Released on xxx, 1st, 2025}}

\author[0000-0002-2449-0214]{Burcu G\"{u}nay} 
\correspondingauthor{Burcu G\"{u}nay}
\email{burcu.gunay@armagh.ac.uk} 
\affiliation{Space Telescope Science Institute, 3700 San Martin
  Drive, Baltimore, MD 21218, USA}
\affiliation{Department of Physics $\&$ Astronomy, Johns Hopkins University, 3400 N. Charles Street, Baltimore, MD 21218}
\affiliation{Armagh Observatory and Planetarium, College Hill, Armagh, BT61 9DB, NI, UK}
\author[0000-0001-5340-6774]{Karl D. Gordon}
\affiliation{Space Telescope Science Institute, 3700 San Martin
  Drive, Baltimore, MD 21218, USA}
\author[0000-0003-4797-7030]{Joshua E. G. Peek} 
\affiliation{Space Telescope Science Institute, 3700 San Martin
  Drive, Baltimore, MD 21218, USA}
\affiliation{Department of Physics $\&$ Astronomy, Johns Hopkins University, 3400 N. Charles Street, Baltimore, MD 21218}
\author[0000-0001-9462-5543]{Marjorie Decleir} 
\altaffiliation{ESA Research Fellow}
\affiliation{European Space Agency (ESA), ESA Office, Space Telescope Science Institute (STScI), 3700 San Martin Drive, Baltimore, MD 21218, USA}
\author[0000-0002-5895-8268]{Dries Van De Putte} 
\affiliation{Department of Physics \& Astronomy, The University of Western
Ontario, London ON N6A 3K7, Canada}
\author[0000-0003-0789-9939]{Kirill Tchernyshyov} 
\affiliation{University of Washington, Seattle, WA, USA}
\author[0000-0001-7289-1998]{Michael G. Burton} 
\affiliation{Armagh Observatory and Planetarium, College Hill, Armagh, BT61 9DB, NI, UK}

\begin{abstract} 

We introduce a new photometric mapping method for the James Webb Space Telescope (JWST) to measure the spatial distribution of carbonaceous dust, siliceous dust and water ice by using absorption features arising from the grains in the dense interstellar medium (ISM). Employing NIRCam and MIRI imaging filters, low-resolution spectroscopic data can be obtained to measure the optical depths of the 3.0--$\mu$m water ice --OH feature, the 3.4--$\mu$m aliphatic hydrocarbon --CH feature, and the 10--$\mu$m silicate --SiO feature for large fields of view. This method provides extensive statistical data of the grains across wide fields in the ISM at minimal observing cost. In this study, we present its application on observational data from the literature to validate the measured optical depths and simulations to assess the accuracy of the method under various conditions. We showed that the photometric method can be employed to obtain reasonably accurate measurements of optical depth. We demonstrate that JWST optical depth maps enable the independent exploration of abundance distributions of major grain components across a wide spatial coverage in the ISM.

\end{abstract}



\section{Introduction}\label{sec:introduction}

The evolution of a galaxy is driven by the cycle of material between stars and the ISM. During this cycle, the refractory compounds form interstellar dust particles and volatile compounds form interstellar ices. The ISM grains are characterized by their size, morphology and chemical composition. As a result of combined laboratory and observational studies, it has long been known that the grains of the ISM predominantly consist of three major components: carbonaceous dust, siliceous dust, and water ice \citep[e.g.,][]{Jones2017,HensleyDraine2023, Ysard2024}, alongside other minor dust and ice compositions. However, the physical structure and chemical composition of ISM grains are variable and remain only partially understood within the framework of current grain models \citep[e.g.,][]{Jones2017,HensleyDraine2023, Ysard2024} due to observational challenges and resulting lack of statistical data.

Analyzing the components of bulk solids poses additional challenges compared to gases due to their complex chemistry and diverse structures, often requiring a combination of multiple complementary techniques in laboratory studies (e.g., \citealt{Henning1998, Dorschner1986, Boogert2015, Gunay2018, Gerber2025}). In observational studies, the remote analysis of ISM particles is inherently constrained by available observational techniques, the presence of other intervening material in the line of sight, and instrumental sensitivity limitations. Consequently, the physical structure and chemical composition of interstellar grains remain incompletely characterized, owing to their intrinsic heterogeneity, environmental variability, and evolutionary processing within the interstellar medium. This leads to significant uncertainties regarding the elemental abundances and limits our understanding of the chemical evolution of the Universe. 

The total chemical abundances of elements observed in both the gas and solid phases in the ISM should align with cosmic abundances (see \citealt{Wang2015, DraineHensley2021, HensleyDrain2021, Zuo2021a, Zuo2021b}). Differences in elemental abundances in the gas phase relative to cosmic levels are explained by the condensation of elements into the solid phase. 

A direct way to predict elemental abundances locked in grains is to model their wavelength-dependent interactions with electromagnetic radiation through absorption, scattering, emission, and polarization, using their proposed compositions and particle size distributions (e.g., \citealt{Mathis1977, Draine2003b, Zuo2021a, Hirashita2023, Zubko2004, HensleyDraine2023, Ysard2024}). 

However, significant discrepancies exist between elemental depletions from the gas phase and the abundances predicted by grain models \citep[e.g.][]{Kim1996, Cardelli1996, Jenkins2009, Wang2015, Poteet2015, HensleyDrain2021, Zuo2021a, Zuo2021b, Psaradaki2024, Decleir2025}. It is also important to note that cosmic abundance estimations are still debated, as different studies are not fully consistent (see \citealt{Zuo2021a, HensleyDrain2021, DeCiaNature2021, RamburuthHurt2024}).

In the ISM, carbon and oxygen dominate the solid phase in terms of atomic abundance \citep{Savage1996, Zubko2004, HensleyDrain2021} (along with other major dust forming elements, such as silicon, magnesium, and iron). They are the most abundant elements in the ISM after hydrogen and helium because they form efficiently during stellar synthesis processes. Therefore, their elemental abundances are powerful clues to understanding stellar and galactic evolution.

However, the abundances of carbon and oxygen remain problematic, despite improvements in cosmic abundance estimations based on solar and stellar abundance studies and spectroscopic measurements of gas-phase abundances \citep{Poteet2015, Zuo2021a, Zuo2021b, Psaradaki2024}. These abundance discrepancies are known as the carbon crisis and the missing oxygen problem: the carbon abundance estimates for the solid phase exceed the amount of carbon depleted from the gas phase, while the proposed oxygen abundances for the solid phase are insufficient to explain the depleted oxygen from the gas phase  \citep[e.g.][]{Kim1996, Cardelli1996, Wang2015, Poteet2015, Jones2017, HensleyDrain2021, Zuo2021a, Zuo2021b, Psaradaki2024}.

The degree of carbon depletion into solid grains varies across different sightlines \citep{Costantini2019, Gunay2022}. Notably, recent dust models that incorporate mixed components, such as composite grains or core/mantle grains, together with nano-sized carbon grains can reproduce observed dust properties while remaining consistent with interstellar carbon abundance constraints \citep{HensleyDraine2023, Ysard2024}. Oxygen depletion into solid grains varies with physical conditions, increasing from diffuse clouds to dense regions \citep{Jenkins2009, Whittet2010, Jones2017, Draine2020, Dishoeck2021, Potapov2024}. 

To estimate the carbon and oxygen budget of solid grains in the ISM, prominent infrared (IR) spectral absorption features are employed to trace chemical groups. The 3.0--$\mu$m water ice O$-$H stretching feature, the 3.4--$\mu$m aliphatic hydrocarbon C$-$H stretching feature and the 10.0--$\mu$m silicate Si$-$O stretching feature (see Figure~\ref{fig:model-filters} and Figure~\ref{fig:AllSpectraG04}) exist along multiple sight-lines and are suitable for quantitative studies (e.g. \citealt{Hendecourt1986, Pendleton2002, DraineISD2003, Dartois2004, Boogert2015, Bouilloud2015, Gunay2018, Gerber2025}). 

The optical depth of an absorption feature allows for the determination of the number of absorbing chemical groups along the line of sight, in other words, column densities ($N$, cm$^{-2}$), as described by the Beer-Lambert law \citep{Swinehart1962, Hendecourt1986}. The total absorption of the carrier species can be estimated by measuring the area of the feature (\(\tau \Delta \bar{\nu}\)). Accordingly, the column density (\(N\)) can be derived from the optical depth (\(\tau\)) using the equivalent width (\(\Delta \bar{\nu}\), cm\(^{-1}\)) of the absorption feature and the integrated absorption coefficient (\(A\), cm\,molecule\(^{-1}\)), which is determined through laboratory measurements, using the following equation \citep{Swinehart1962, Hendecourt1986}.

\begin{equation}\label{eq:columndensity}
N = \frac{\tau \Delta \bar{\nu}}{A} 
\end{equation}

In the ISM, the column densities of dust and ice can vary significantly across sightlines over small spatial scales \citep{Murakawa2000, Chiar2002, Moultaka2004, Gibb2004, Noble2017, Gordon2021, Gunay2022, Ginsburg2023, Shao2024, Decleir2025, Smith2025}. This spatial variability can be studied using single-point, long-slit, or integral field spectroscopy techniques for limited spatial scales. 

However, exploring large areas of the ISM using spectroscopic measurements is time-consuming, requiring long observation periods and high costs. Instead, a broader spatial coverage can provide insights into how ISM components vary across different environments. 

Wide-field photometry offers a solution for tracing ISM components by capturing both absorption and emission features (e.g. \citealt{Murakawa2000, Gunay2020, Gunay2022, Ginsburg2023, Sandstrom2023, Sandstrom2023A, Rieke2024, Alberts2024, Bolatto2024, Ballering2025, Grant2025}). Although integral field spectroscopy is an important alternative, the choice between integral field spectroscopy and imaging depends on the balance between spatial coverage and spectral detail. 

Integral field unit (IFU) spectroscopy, as provided by JWST (NIRSpec and MIRI/MRS) offers detailed spatially resolved spectral data but with a more limited FoV compared to that provided by the NIRCam and MIRI imagers. NIRSpec provides a FoV of \(3'' \times 3''\), and MIRI/MRS provides a FoV up to \(6.6'' \times 7.7''\), which are much smaller compared to the imaging fields provided by NIRCam and MIRI. NIRCam provides a FoV of \(129'' \times 129''\) for each module (A and B), and MIRI provides a FoV of \(73'' \times 113''\). 

Photometric method can be applied to many sources within the field of view (FoV) simultaneously, providing spatially resolved low-resolution spectral information for large areas, enabling the exploration of the interplay between ISM components. This makes photometric measurements with JWST more observing cost-effective and suitable for large-scale surveys (such as \citealt{Ginsburg2023}), in particular when high spectral resolution is not a critical requirement for the science goal. 

We present an efficient method using JWST photometry to measure the \textit{optical depth} of the 3.0--$\mu$m water ice $-$OH feature, the 3.4--$\mu$m aliphatic hydrocarbon $-$CH feature and the 10.0--$\mu$m silicate $-$SiO feature. This method has the potential to provide extensive statistical data for grain models and trace abundance variations across large fields in the ISM, especially for the transition regions from diffuse to dense sightlines.

We present the methodology along with the details of optical depth calculations in Section~\ref{sec:Methodology}. In Section~\ref{sec:TestingwithModelSpectra}, we use model spectra to constrain methodological biases and evaluate potential uncertainties in the photometric optical depths arising from spectral variations. We examine its practical application by simulating photometric and spectroscopic optical depth measurements using observational spectra from the literature in Section~\ref{sec:TestingwithObservationalData}. We explore potential calibration equations to correct for methodological biases using model spectra in Section~\ref{sec:Calibrations}. Finally, we apply the method to a synthetic FoV to create optical depth maps, as described in Section~\ref{sec:MappingApplications}, and evaluate the feasibility of the independent mapping of the grain components. We summarize and discuss our results in Section~\ref{sec:SummaryandDiscussion}.

\section{Methodology} \label{sec:Methodology}

The methodology for measuring optical depth is based on \textit{three-band photometry} for each feature: one filter to sample the absorption flux and two filters to estimate the continuum flux levels at the absorption wavelengths. This approach is used because the study aims to keep the number of filters minimal (thereby reducing the required observing time), while maximizing observational efficiency.

We analyzed whether the optical depths of IR absorption features due to water ice, aliphatic hydrocarbons, and silicates can be accurately obtained using the selected JWST filters (Table~\ref{tab:features-filters}). We used a set of observed and model spectra with known absorption feature strengths and derived two distinct optical depth measurements: one using photometric fluxes and the other using spectroscopic fluxes (see Section~\ref{sec:Measurements}). 

We compared the known optical depths values with the optical depths measured from photometric and from spectroscopic measurement simulations (hereafter referred to as \emph{photometric} and \emph{spectroscopic} optical depths, or $\tau_{\text{p}}$ and $\tau_{\text{s}}$ for short) to assess biases due to throughputs of filters that do not perfectly isolate the desired fluxes compared to precise spectroscopic measurements.

First, we use model spectra with known applied optical depths (hereafter referred to as \emph{reference} optical depths; $\tau_{\text{0}}$) to investigate \emph{methodological biases} due to use of filters and the \emph{uncertainties} introduced by limited spectral resolution depending on the employed filter set, which affect the ability to capture the full shapes of continuum and absorption feature profiles (see Section~\ref{sec:TestingwithModelSpectra}).

Then we use observational spectra from the literature. The $\tau_{\text{p}}$ and $\tau_{\text{s}}$ values were compared with the reported values from the literature (hereafter referred to as \emph{reported} optical depths; $\tau_{\text{r}}$, \citealt{Gibb2004}) to compare methodologies and evaluate the differences (see Section~\ref{sec:TestingwithObservationalData}). 

\subsection{Synthetic Photometry} \label{sec:SyntheticPhotometry}

\begin{figure*}[ht]
  \begin{center}
    \begin{tabular}{cc}
      {\includegraphics[angle=0,scale=0.44]{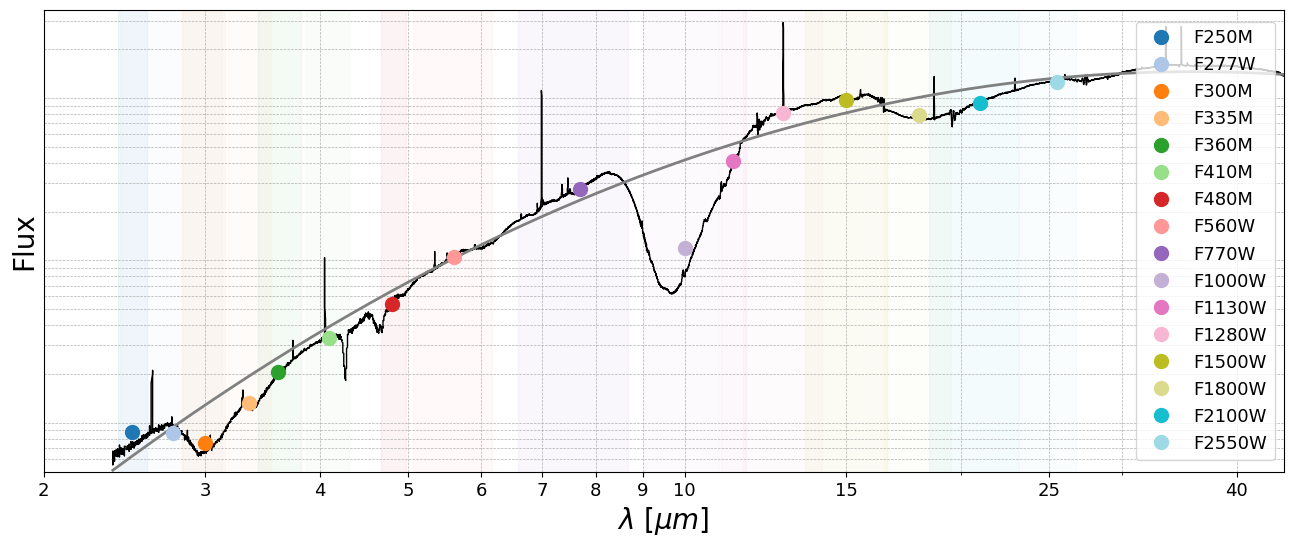}} \\
        \end{tabular}
    \caption{The photometric fluxes are illustrated, as an example, on the spectrum of the Galactic Center source Sgr A$^{*}$ \citep{Gibb2004}, with colored dots indicating the JWST filters used. The half power pass-band wavelengths of the filters are also illustrated in the background using filled areas with matching colors. The gray line represents the general continuum fit using photometric flux measurements from the selected continuum filters.}   
    \label{fig:SgrASpectra}
      \end{center}
\end{figure*}

We used synthetic photometry \citep{Gordon2022Calibration} to simulate flux measurements (see Section~\ref{sec:FluxCalculations} in Appendix) through the NIRCam and MIRI filters. To simulate photometric flux measurements we developed a code that applies the throughput profiles \citep{Koornneef1986} of the NIRCam and MIRI imaging filters, which give the electrons out of the detector per photon into the instrument, as a function of wavelength (the Photon Conversion Efficiency - PCE), to the spectra. The filter throughputs are interpolated (see the profiles in the bottom panel in Figure~\ref{fig:AllSpectraG04} and Figure~\ref{fig:BB} in Appendix) to match the wavelengths elements of the model or observed spectra. Then the photon integrations and normalizations are performed using integrated filter throughputs to obtain photometric fluxes at the central wavelengths of the filters (see Section~\ref{sec:Measurements}). Photometric flux measurements through selected NIRCam and MIRI filters are illustrated on the spectrum of a representative source, Sgr A$^{*}$, in Figure~\ref{fig:SgrASpectra} and the model spectrum in Figure~\ref{fig:model-filters}. Measurements of the observational spectra set are also shown in Figure~\ref{fig:AllSpectraG04} in Appendix.

\begin{deluxetable*}{cccc}
\tabletypesize{\small}  
\setlength{\tabcolsep}{10pt}
\tablecaption{NIRCam filters for NIR measurements and MIRI filters for MIR measurements used in this work.}
\label{tab:features-filters}
\tablehead{
\colhead{} & \colhead{Absorption Features} & \colhead{Absorption Filters} & \colhead{Continuum Filter Options}
}
\startdata
NIR & 3.0--$\mu$m / 3.4--$\mu$m &F300M / F335M & F250M, F277W, F360M, F410M, F480M\\
MIR & 10.0--$\mu$m & F1000W & F560W, F770W, F1130W, F1280W, F1500W \\
\enddata
\end{deluxetable*}

\subsection{Filter Sets} \label{sec:FilterSets}

We aimed to identify the optimal combination of filters for sampling the NIR and MIR spectrum. We defined three filters to measure each absorption feature: one primary filter to measure the absorption and two filters to estimate the continuum. We combined selected continuum filters to make six different filter sets in the NIR and MIR regions (Table~\ref{tab:filter-sets}). We illustrated the use of filters in three-band photometry through example filter sets (Filter Set 1 for the NIR and Filter Set 1 for the MIR) in Figure~\ref{fig:continuumfitting}.

\begin{deluxetable}{ccc}
\tabletypesize{\small}  
\setlength{\tabcolsep}{8pt}
\tablecaption{\textit{Filter Sets} for continuum estimation.}
\label{tab:filter-sets}
\tablehead{
 & \colhead{NIR} & \colhead{MIR}
}
\startdata
Set 1 & F250M$-$F410M & F770W$-$F1280W \\
Set 2 & F250M$-$F360M & F770W$-$F1500W \\
Set 3 & F277W$-$F410M & F560W$-$F1280W \\
Set 4 & F277W$-$F360M & F560W$-$F1500W \\
Set 5 & F250M$-$F480M & F770W$-$F1130W \\
Set 6 & F277W$-$F480M & F560W$-$F1130W \\
\enddata
\end{deluxetable}

We selected the filters based on their bandwidths and their effectiveness in sampling both the features of interest and the continuum. We prefer medium-band filters, since narrow-band filters require more exposure time to obtain the required SNR. We avoid using the wide-band filters in the NIR region where possible, as they can cover other spectral features. Since all narrow-band filters and some medium-band/wide-band filters are specifically tailored to measure certain spectral features, we avoid using them where possible. We also excluded filters far from the absorption feature, as they are ineffective at capturing the spectral profiles.

However, it may also be feasible to use some of the excluded filters to estimate optical depths. We further investigated the effect of filter set selection in the following sections by simulating measurements on the model spectra (Section~\ref{sec:TestingwithModelSpectra}) and the observed spectra (Section~\ref{sec:TestingwithObservationalData}).

\subsection{Measurements} \label{sec:Measurements}

The fluxes, optical depths, and total absorptions were derived using analysis codes (developed in Python), following the computational procedures based on the calculations detailed below.

\subsubsection{Fluxes} \label{sec:FluxCalculations}

\textit{Photometric Fluxes:} Photometric fluxes (\(F(\Delta\lambda)\)) derived from a flux density distribution (\(F(\lambda)\)) over a wavelength range \((\Delta\lambda)\) are determined by the photometric filter's bandwidth (BW), which depends on the throughput function (\(T(\lambda)\)) of the filter \citep{Rieke2008}, as presented in Equation~\ref{eq:BW}. 

\begin{equation}
\label{eq:BW}
BW = \frac{\int d\lambda \, T(\lambda)}{T_{\text{max}}}
\end{equation}

A photometric measurement in a filter can be described as the photon-weighted average flux density ($\langle F(\lambda) \rangle$) \citep{Koornneef1986, Gordon2022Calibration}. We calculated photometric fluxes by applying the throughputs of filters (\(T(\lambda)\)) to spectra (\(F(\lambda)\)) using Equation~\ref{eq:photometricflux} \citep{Gordon2022Calibration}. 

\begin{equation}
\label{eq:photometricflux}
F(\Delta\lambda) = \langle F(\lambda) \rangle = \frac{\int F(\lambda) T(\lambda) \lambda \, d\lambda}{\int T(\lambda) \, d\lambda}
\end{equation}

\textit{Spectroscopic Fluxes:} We used the approximate spectral flux levels corresponding to the central wavelength (\(\lambda_{0}\)) of each filter as spectroscopic fluxes (\(F(\lambda_{0})\)) \citep{Koornneef1986, Tokunaga2005}. For this, we linearly interpolated the nearest flux data to the central ($\sim$pivot) wavelengths (\(\lambda_{0-}\): the nearest short wavelength and \(\lambda_{0+}\): the nearest long wavelength ) in the data array to calculate spectroscopic fluxes using Equation~\ref{eq:spectroscopicflux}.

\begin{equation}
\label{eq:spectroscopicflux}
F(\lambda_{0}) = \frac{F(\lambda_{0+}) - F(\lambda_{0-})}{\lambda_{0+} - \lambda_{0-}} (\lambda_{0} - \lambda_{0-}) + F(\lambda_{0-})
\end{equation}

\textit{Continuum Fluxes:} The continuum fluxes at absorptions are estimated as a function of wavelength (\(F_0(\lambda\))). we applied analytical function fitting using the flux measurements obtained with continuum filters, which are referred to as short-wavelength continuum fluxes (\(F_{0}(\lambda_{s})\)) and long-wavelength continuum fluxes (\(F_{0}(\lambda_{l})\)). After testing alternative analytical functions, we adopted a local linear continuum fitting approach, as demonstrated in Figure~\ref{fig:continuumfitting}, and subsequently calculated the continuum fluxes using Equation~\ref{eq:continuumflux}.

\begin{equation}
\label{eq:continuumflux}
F_{0}(\lambda) = \frac{F_{0}(\lambda_{l}) - F_{0}(\lambda_{s})}{\lambda_{l} - \lambda_{s}} (\lambda - \lambda_{s}) + F_{0}(\lambda_{s})
\end{equation}

\subsubsection{Optical Depths} \label{sec:OpticalDepthCalculations}

\textit{Photometric Optical Depths:} We obtained photometric optical depths (\(\tau_p\)) using the photometric fluxes (\(F(\Delta\lambda)\)) measured through each filter and the corresponding continuum fluxes (\(F_0(\lambda_{0})\)) using Equation~\ref{eq:OD-photometric}.

\begin{equation} \label{eq:OD-photometric}
\tau_p = -\ln{\frac{F(\Delta\lambda)}{F_0(\lambda_{0})}}
\end{equation}	

\textit{Spectroscopic Optical Depths:} Then spectroscopic optical depths ($\tau_s$) are obtained using the spectroscopic fluxes (\(F(\lambda_{0})\)) and the corresponding continuum fluxes (\(F_0(\lambda_{0})\)) using Equation~\ref{eq:OD-spectroscopic}.

\begin{equation} \label{eq:OD-spectroscopic}
\tau_s = -\ln{\frac{F(\lambda_{0})}{F_0(\lambda_{0}) }}
\end{equation}	

\subsubsection{Total Absorption} \label{sec:IntegratedAbsorption}
The total absorption of the carrier species, also known as the integrated absorption ($\mathcal{A}$, cm$^{-1}$), is obtained by integrating the optical depth across the absorption feature's frequency range ($\Delta \bar{\nu}$), using Equation~\ref{eq:IntegratedAbsorption}.

\begin{equation} \label{eq:IntegratedAbsorption}
\mathcal{A} = \int_{} \tau(\bar{\nu}) \, d\bar{\nu}
\end{equation}

\begin{figure*}[htbp] 
  \begin{center}
    \begin{tabular}{cc}
         {\includegraphics[angle=0,scale=0.26]{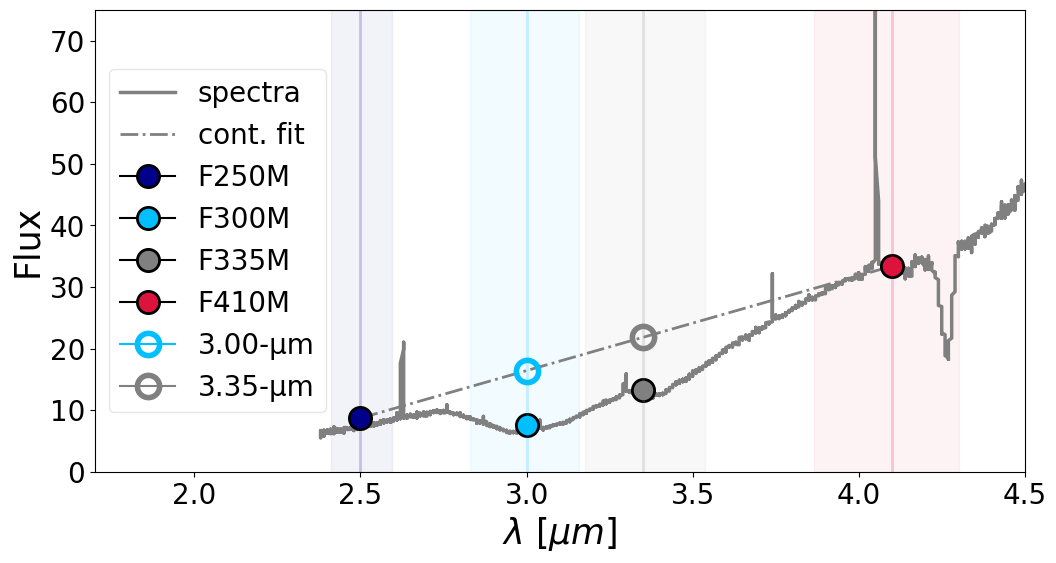}} &
      {\includegraphics[angle=0,scale=0.26]{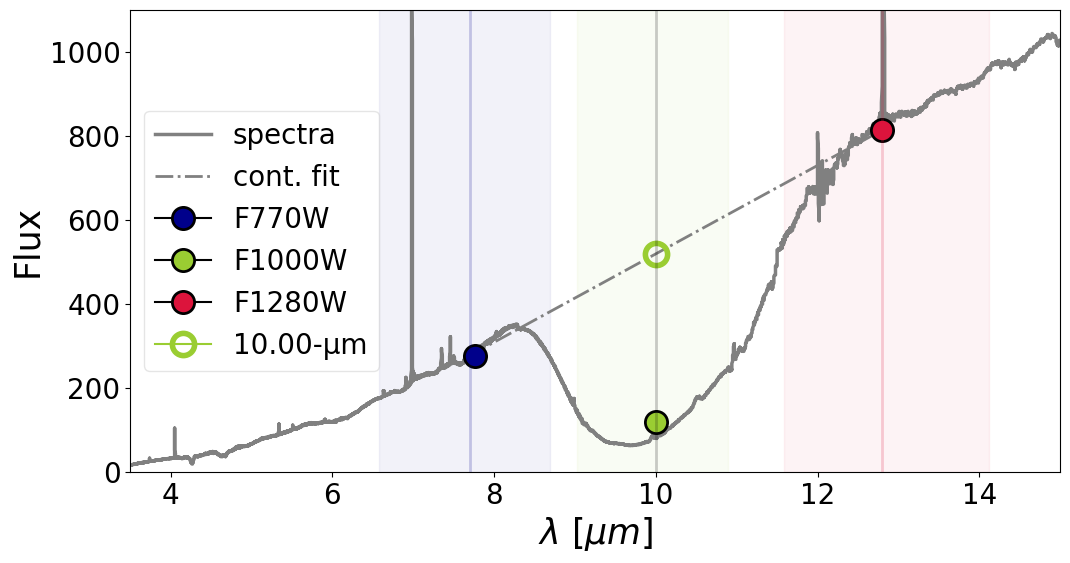}} \\
       \end{tabular}
    \caption{Local linear continuum fitting (dashed lines) examples with Set 1 for the NIR region (left panel) and Set 1 for the MIR region (right panel) of the spectrum of Sgr A$^{*}$ \citep{Gibb2004}. The photometric fluxes measured at continuum and absorptions are indicated with filled circles, and the corresponding continuum flux estimations are indicated with unfilled circles. The half power pass-band wavelengths of the filters are also illustrated in the background using filled areas.}   
    \label{fig:continuumfitting}
      \end{center}
\end{figure*}

\section{Testing with Model Spectra} \label{sec:TestingwithModelSpectra}

Models allow us to isolate individual factors that influence the accuracy of photometric optical depths. We used model spectra to constrain methodological biases due to use of filters and estimate uncertainties arising from spectral variations that cannot be captured due to limited resolution. 

There could be large variations in spectra through the ISM sightlines, arising from the intrinsic properties of background sources and the nature of the intervening material. The profile of absorption features arising from solid grains can vary depending on variations in chemical composition or the presence of impurities, and several physical factors, such as morphology (crystalline, amorphous), structure (porous, fluffy, etc.), shape, size, and temperature of the grains, can also contribute to changes in the peak absorption (central wavelength: CW) and width (full width at half maximum: FWHM) of absorption features (such as \citealt{Ysard2024, Rocha2024, Dartois2024, Bergner2024, Shao2024}). The continuum profiles of spectra in the IR region can primarily vary depending on the thermal radiance of background stellar sources \citep{Rieke2008, Gordon2022Calibration}, which is governed by the spectral type (temperature) and also influenced by the evolutionary stage (such as from YSOs to AGBs) of the source, particularly in the presence of shells, envelopes, and disks, which can also contribute to changes in the slope (spectral index: $\alpha$) of a spectrum (such as \citealt{Crapsi2008}, \citealt{Evans2003}, \citealt{Robitaille2017}, \citealt{Richardson2024}).

As we are not able to capture the detailed shapes of the absorption and continuum profiles, the variations in the profiles of absorption features and continuum cannot be fully accounted for, resulting in differences between the photometric optical depth measurements and the true values. Using spectral models, we separately analyzed the impact of continuum and absorption variations to avoid degeneracies caused by their combinations.

\subsection{Spectral Models} \label{sec:SpectralModels}

 We used analytical equations to model the spectra, treating the continuum and absorption features as distinct spectral elements. We described the details of modeling spectra in Section~\ref{sec:ModelingSpectra} in the Appendix. We generated absorption feature templates (Section ~\ref{sec:AbsorptionFeatures}) using \textit{Gaussian Functions} with a range of CWs and FWHMs. We created a range of continuum profiles (Section~\ref{sec:Continuum}) using \textit{normalized blackbody curves} (BBs; see Figure~\ref{fig:BB}), polynomial functions, and a flat continuum to represent different spectral index ranges. We presented CW$\&$FWHM ranges used for absorption profile templates and continuum types in Table~\ref{tab:model-parameter-ranges} in Appendix. 

 Then using these spectral elements and applying a range of reference optical depths $\tau_{\text{0}}$ ($A$), we generated model spectra sets representing different scenes. For this we continuum fluxes (\(F_0(\lambda)\)) and transmitted fluxes (\(F(\lambda)\)) by accounting for the absorption features with a range of optical depths ($\tau$), using the equation presented below.

\begin{equation} 
\label{eq:tau}
F(\lambda) = F_0(\lambda) \exp(-\tau)
\end{equation}

First, we generated models to represent idealized cases. For the absorption features, the CW values were set to the $\lambda_{0}$ values of the filters, and approximate average FWHM values were adopted based on literature values for the absorption templates (see Section~\ref{sec:AbsorptionFeatures} for details). These absorption feature templates (see, for example, Figure~\ref{fig:calibration-models}) were then combined with a flat continuum. These models named as \textit{standard spectral models} and used isolate influence of variations in absorption and continuum profiles. 

\begin{figure*}[htbp]
  \begin{center}
    \begin{tabular}{c}
      {\includegraphics[angle=0,scale=0.44]{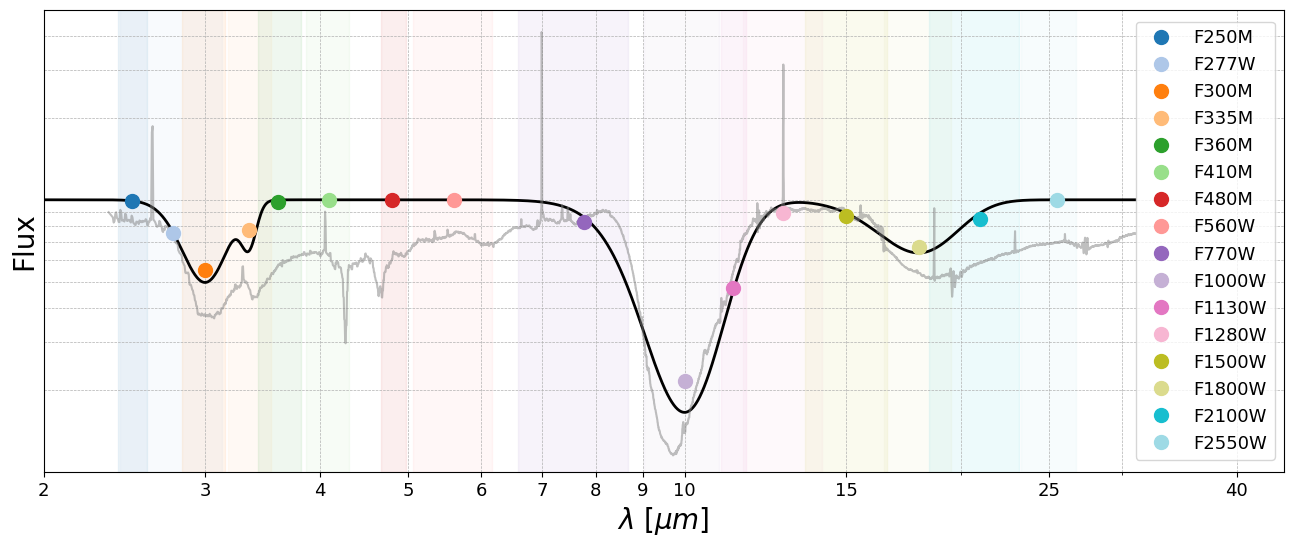}} \\
        \end{tabular}
    \caption{An example spectral model for 1--30--$\mu$m wavelength range, including the water ice (3.0--$\mu$m), carbonaceous dust (3.4--$\mu$m), and silicate dust (10.0--$\mu$m, 18.0--$\mu$m) absorption features, is presented on a flat continuum. The normalized spectrum of a Galactic Center source (Sgr A$^{*}$) is also shown for comparison. The photometric fluxes are shown on the model spectrum with colored dots indicating the employed JWST filters. The half power pass-band wavelengths of the filters are also illustrated in the background using filled areas with matching colors. }
    \label{fig:model-filters}
      \end{center}
\end{figure*}

An example standard spectral model is shown in Figure~\ref{fig:model-filters}, in comparison with the normalized spectrum (2.5--25.5 $\mu$m) of the Galactic Center source (Sgr A$^{*}$). We illustrated all major absorption features, including the $-$OH stretching vibration of water ice around 3.0--$\mu$m, the $-$CH stretching vibration of carbonaceous dust around 3.4--$\mu$m, and the $-$SiO stretching vibration of siliceous dust around 10.0--$\mu$m. The $-$SiO bending vibration of siliceous dust near 18.0--$\mu$m \citep{vanBreemen2011} is also shown solely to provide a more realistic visualization and to highlight its potential impact on the continuum, but it is not considered in the analysis in this work. However, we note that, as expected, it can significantly influence the optical depth results, depending on the selected Filter Set, particularly if it includes long wavelength filter that overlap with this feature.

We note that, in our analysis, each absorption feature (3.0--$\mu$m, 3.4--$\mu$m and 10.0--$\mu$m) was treated independently to overcome potential degeneracies, except the aliphatic hydrocarbon feature, as this approach enabled more realistic spectra modeling, representative of dense sightlines. For the dense sightlines, modeling the NIR region presents additional challenges due to the overlap of the aliphatic hydrocarbon feature with the water ice feature, which impacts continuum fluxes. To achieve more accurate values, we modeled the aliphatic hydrocarbon feature together with a proportional ice feature component, using the average optical depth ratio ($\langle\tau_{3.0}\rangle / \langle\tau_{3.4}\rangle \approx 2$) derived from the observational spectral dataset (see Table~\ref{tab:comparison-all}).

We also generated spectral models with variations in the continuum and absorption features to estimate uncertainties associated with these differences. We aim to cover the likely ISM scenarios based on the studies reported in the literature (see Section~\ref{sec:ModelingSpectra}), though we do not necessarily include the exceptional continuum, CW, FWHM, and $\tau$ values at the extremes of the reported ranges. These models are named as \textit{exploratory spectral models}.

The parameters used for both the standard and exploratory spectral models (see Section~\ref{sec:ModelingSpectra}) are presented in Table~\ref{tab:model-parameter-ranges}. We evaluated the accuracy of the method under various conditions by comparing results obtained using exploratory spectral models with those from standard spectral models.

\subsection{Testing Accuracy of the Method} \label{sec:TestingtheAccuracyoftheMethod}

Limited resolution in absorption and continuum profiles introduces unaccounted variations, leading to uncertainties in the resulting photometric optical depth measurements. We discussed the accuracy of the photometric method, considering the influence of variations in the continuum profile, which depends on the type of background sources and the influence of the absorption profiles, which depends on the nature of the medium. 

We separately analyzed the impact of continuum and absorption variations on model spectra to avoid degeneracies caused by their combinations. To analyze continuum-related uncertainties, we used constant absorptions intensities with various continuum shapes to represent a variety of background sources with different spectral types, luminosities, and evolutionary stages located within or behind the ISM cloud. To analyze absorption-related uncertainties, we used a flat continuum combined with a range of absorption strengths across various CWs and FWHMs, representing different densities and characteristics of the intervening material. We emphasize that the results obtained using spectral models also depend on the continuum and the absorption model parameters: the selected CWs and FWHMs, which are derived from values reported in the literature (see Section~\ref{sec:ModelingSpectra} in the Appendix).

First, we compared Filter Sets based on their accuracy in capturing continuum and absorption profiles as the reliability of the photometric method is depended to the choice of filter set, which determines the extent to which spectral details are captured or missed. Then, for the selected Filter Set, we examined the uncertainties arising from continuum estimation and absorption measurements.

\subsubsection{Filter Set Influence} \label{sec:FilterSetSelection}

The efficiency of Filter Sets with linear continuum fit approximation was evaluated using different background radiances modeled with BBs. As shown in Figure~\ref{fig:BB-OpticalDepths-AllFilterSets}, we compared the photometric optical depths ($\tau_{p}$) with the reference optical depths ($\tau_{0}$). To isolate continuum related effects, the continuum capturing efficiency was tested using $\tau_{0} = 0$, while the absorption capturing efficiency was evaluated using $\tau_{0} = 1$.

Filter Sets sampling broader continuum ranges are more sensitive to continuum profiles and may provide a fair estimate of the continuum. In contrast, filter sets sampling narrower continuum ranges are more influenced by overlap with absorption profiles and may offer a fair estimate of the absorption.

We also compared the efficiency of the filter sets using the standard model spectra, assuming a flat continuum, as shown in Figure~\ref{fig:Method-Biases}. Based on the comparison with the reference optical depths (see Figure~\ref{fig:BB-OpticalDepths-AllFilterSets} and Figure~\ref{fig:Method-Biases}), the most suitable filter sets are identified according to their ability to capture both absorption features and the continuum. 

In theory, alternative Filter Set options can also be used for optical depth estimations by applying correction factors derived from correlations with the reference optical depths (see Section~\ref{sec:Calibrations}). In practice, however, additional uncertainties may arise due to the presence of other spectral features (see Figure~\ref{fig:AllSpectraG04} and Figure~\ref{fig:OD-Data-FilterSets}), as further discussed in Section~\ref{sec:TestingwithObservationalData}.

In this study, we selected Filter Set 1 for the NIR and Filter Set 1 for the MIR region, as they provide one of the most suitable combinations for estimating the continuum and capturing absorption profiles, being largely free from contaminating wings, shoulders, or other absorption/emission features (see Section~\ref{sec:TestingwithObservationalData}).

\subsubsection{Methodological Biases} \label{sec:MethodologicalBiases}

We analyzed the methodological biases in the photometric optical depths using the selected filter sets and standard spectral models, including those arising from filter use ($\Delta\tau_{ps} = |\tau_{p} - \tau_{s}|$), limited resolution ($\Delta\tau_{s0} = |\tau_{s} - \tau_{0}|$), in addition to the total bias ($\Delta\tau_{c0} = |\tau_{p} - \tau_{0}|$). No significant differences were observed between $\Delta\tau_{s0}$ and $\Delta\tau_{c0}$, suggesting that, under ideal conditions, biases primarily due to the use of filters and those stemming from limited resolution are comparatively small for the selected Filter Sets (e.g. for standard model, for $\tau_{0}$ = 1, $\tau_{s}$ values are 0.99, 0.99 and 0.92 for water ice, aliphatic hydrocarbon and silicate, respectively). 

Under ideal conditions, methodological biases can be largely mitigated by applying calibration equations derived from correlations between photometric and true values (see Section~\ref{sec:Calibrations}). However, in practice, there will be uncertainties in the derived optical depths, arising from the intrinsic shapes of the absorption features and the continuum cannot be fully captured by the photometric method.

\subsubsection{Uncertainties Due to Absorption Profiles}
\label{sec:Variations-absorptions}

Variations in absorption profiles introduce inherent uncertainties in the resulting photometric optical depths due to limited number of measurements at the fixed-wavelengths. We explored how the variations in CWs $\&$ FWHMs (Table~\ref{tab:model-parameter-ranges}) influence optical depth measurements by testing most likely scenarios on model spectra based on the reported values in the literature (see references in Section~\ref{sec:AbsorptionFeatures}). To isolate the effects of variations in absorptions on the resulting optical depths, we used a flat continuum. 

We applied a range of optical depths to each feature to examine how photometric optical depths deviate from the reference (accurate) optical depths under different variations.  The applied optical depth ranges are consistent with the average spectroscopic optical depth levels ($\tau_{\text{0[3.0]}} \approx 2$, $\tau_{\text{0[3.4]}} \approx 1$, and $\tau_{\text{0[10.0]}} \approx 2.5$) obtained in this study (see Section~\ref{sec:TestingwithObservationalData} and Table~\ref{tab:comparison-all}).

\textit{Variations in CWs}: We analyzed the deviations in photometric optical depths (${\Delta \tau_{\text{CW}}}$) as function of CW shifts presented in Table~\ref{tab:model-parameter-ranges}. To simulate the CW shifts reported in the literature (see Section~\ref{sec:AbsorptionFeatures}), we modified the model parameter ($\mu$) in the Gaussian Function (Equation~\ref{eq:gaussian}). The resulting photometric optical depths were compared with the reference optical depth values, as shown in the upper panels of Figure~\ref{fig:Variations-Abs}. As we can see in the figure, shifts in CWs of absorption features from the $\lambda_{0}$ values can cause uncertainties in the photometric optical depth values ($\Delta$$\tau_{CW}$). First, we used frequently reported CWs in the literature (3.05--$\mu$m for water ice, 3.40--$\mu$m for aliphatic hydrocarbon and 9.80--$\mu$m for the silicate feature). We found that $\Delta \tau_{CW}$ is $\sim$0.1 for water ice, $\sim$0.05 for aliphatic hydrocarbons, and $\sim$0.17 for silicate in the model spectra, for the maximum optical depth range applied in this study. This corresponds to ${\Delta \tau_{\text{CW}}}/{\tau_{\text{0}}}$ reaching approximately \( 4.5\% \), \( 3.4\% \), and \( 6.7\% \) for the water ice, aliphatic hydrocarbons, and silicate features, respectively. Then we extend this analysis using more extreme CWs reported in the literature (3.10--$\mu$m for water ice, 3.45--$\mu$m for aliphatic hydrocarbon and 9.70--$\mu$m for the silicate feature). We found that $\Delta \tau_{CW}$ is $\sim$0.3 for water ice, $\sim$0.05 for aliphatic hydrocarbons, and $\sim$0.7 for silicate within the optical depth ranges applied in this modeling study. This corresponds to ${\Delta \tau_{\text{CW}}}/{\tau_{\text{0}}}$ reaching approximately \( 15.1\% \), \( 2.7\% \), and \( 3.0\% \) respectively.

\textit{Variations in FWHMs}: We analyzed the deviations in photometrical optical depths (${\Delta \tau_{\text{FWHM}}}$) as a function of variations in FWHM values presented in Table~\ref{tab:model-parameter-ranges}. To simulate FWHM ranges reported in the literature (see Section~\ref{sec:AbsorptionFeatures}), we modified the corresponding model parameter ($\sigma$) in the Gaussian function (Equation~\ref{eq:gaussian}). The resulting photometric optical depths were compared with the reference optical depth values, as shown in the bottom panels of Figure~\ref{fig:Variations-Abs}. As we can see in the figure, variations in FWHMs can cause uncertainties in the photometric optical depth values ($\Delta$$\tau_{FWHM}$). We found that $\Delta$$\tau_{FWHM}$ is $\sim$0.11 for water ice, $\sim$0.11 for aliphatic hydrocarbon, $\sim$0.29 for silicate, for the model spectra at the maximum optical depth ranges applied in this study. This corresponds to ${\Delta \tau_{\text{FWHM}}}/{\tau_{\text{0}}}$ reaching approximately \( 5.5\% \), \( 8.9\% \), and \( 12.4\% \)  for the water ice, aliphatic hydrocarbon, and silicate features, respectively, within the optical depth range applied in this modeling study.

\subsubsection{Uncertainties Due to Continuum Profiles}

\label{sec:Variations-continuum}

The linear continuum fit approximation can result in underestimation or overestimation of optical depths, depending on the shape of the true continuum. We previously demonstrated how continuum related differences in optical depth values vary with the spectral type (temperature) of the background sources (see Figure~\ref{fig:BB-OpticalDepths-AllFilterSets}). 

For dense sightlines, observable background sources are relatively cool IR-bright stars at various evolutionary stages (see Section~\ref{sec:Continuum}). Since their continua peak at longer wavelengths, the IR absorption features fall within the slope of the spectrum (see Figure~\ref{fig:BB}), in contrast to diffuse sightlines where the IR absorption features typically fall within the flatter Rayleigh-Jeans tail. Therefore, the curvature of the spectral continuum introduces uncertainties in the photometric optical depths when approximated with a linear continuum fit. 

By assuming the background radiation can be roughly represented as a blackbody radiation, the differences in optical depths introduced by the linear continuum approximation can be estimated, using the slope values at NIR and MIR. The slope of the continuum can be calculated using photometric fluxes (e.g., for Filter Set 1, the fluxes at 2.5--$\mu$m and 4.1--$\mu$m in NIR and the fluxes at 7.7--$\mu$m and 12.8--$\mu$m in  MIR). 

We used BBs to derive correlations between these slopes values and differences arising from the linear continuum fit approximation as demonstrated in Figure~\ref{fig:SpectralIndex-Error}. This approach allows us to identify and exclude cases where significant continuum related deviations occur, particularly for spectra with steeper slopes, where the continuum profile diverges substantially from flat-like cases (e.g., the NIR spectrum of W3 IRS5, which is in fact a massive protocluster rather than a single source).

Alternatively, these large differences can be partially mitigated by applying corrections based on the differences ($\Delta \tau_{\text{cont.}}$) derived from spectral models with BBs, as shown in Figure~\ref{fig:SpectralIndex-Error}. However, we note that this approximation lacks the sensitivity required to account for small deviations, as the intrinsic shapes of individual spectra cannot be accurately represented by blackbody models (see Section~\ref{sec:Continuum}). Consequently, residual discrepancies persist, contributing to uncertainties in the derived optical depths.

We analyzed the deviations in photometric optical depths depending on continuum variations. We used model spectra created with different continuum profiles based on observational data ([G04]), BBs for selected temperatures (T = 500\,\text{K}, 3000\,\text{K} and T = 10000\,\text{K}) and a flat continuum to explore the continuum dependent deviations in photometric optical depths (${\Delta \tau_{\text{cont.}}}$). The resultant photometric optical depths were compared with the reference optical depth values, as shown in Figure~\ref{fig:Variations-Cont}. As we can see in the figure, the continuum related variations can cause some degree of differences in the photometric optical depths ($\Delta$$\tau_{cont.}$ = $\tau_{p}$ - $\tau_{0}$). The average $|\Delta\tau_{\text{cont.}}|$ values remained almost constant (within $\sim \pm 0.01$) across the optical depth range applied in this modeling study ($\tau_{0}$ = 0--1), as the continuum-related difference is additive rather than multiplicative (see Figure~\ref{fig:Variations-Cont}). We found that the average $|\Delta \tau_{\text{cont.}}|$ is 0.43 for the water ice feature, 0.22 for the aliphatic hydrocarbon feature, and 0.04 for the silicate feature. If we exclude the source with a large variation, $|\Delta\tau_{\text{cont.}}|$ reduces to 0.34 for water ice and 0.16 for aliphatic hydrocarbon. If we keep the source with a large variation but apply the correction, $|\Delta\tau_{\text{cont.}}|$ decreases to 0.21 for water ice and 0.11 for aliphatic hydrocarbons. Based on the corrected values and average optical depths obtained from the observational data set ($\tau_{\text{0[3.0]}} \approx 2$, $\tau_{\text{0[3.4]}} \approx 1$, and $\tau_{\text{0[10.0]}} \approx 2.5$), the resulting ${\Delta \tau_{\text{cont.}}}/{\tau_{\text{0}}}$ ratios are approximately \( 10\% \) for water ice, \( 11\% \) for aliphatic hydrocarbons, and \( 0.05\% \) for silicates. We emphasize that, since the continuum of the sources in the MIR region generally exhibits a flat profile (see Figure~\ref{fig:SpectralIndex-Error}), the continuum estimates for the silicate feature are already in close agreement with those obtained using linear continuum fitting approximations. Therefore, the application of further corrections does not improve the results. It should be noted that the results presented here are based on a model that does not incorporate the Si--O bending vibration feature near 18.0~$\mu$m. While filter overlap with this feature can cause differences in practice, the Filter Set used here was chosen to minimize such overlap, so no significant impact is expected (see also Section~\ref{sec:TestingwithObservationalData}).

\subsubsection{Limitations} \label{sec:Limitations}

While the brightness of background sources and the density of the medium can be constrain to optimize data quality in observations, spectral variations remain unavoidable across different ISM sightlines. We explored the impact of each variation separately using models, and we do not observe large deviations from the $\tau_{\text{0}}$ values. The ${\Delta \tau}/{\tau_{\text{0}}}$ values remain below \( \sim 15\% \) within the range of expected variations we applied to the standard model spectra. However, in practice, continuum and absorption profiles will be different from those we used here and the magnitude of differences in the measured $\tau_{p}$ values can be larger due to rare intrinsic properties of background sources or the combined influence of variations in the continuum and absorption profiles. Nonetheless, by applying statistical tests (such as the resampling methods applied in \citealt{Gunay2020} and \citealt{Gunay2022}), significant source-to-source variations arising from intrinsic properties of the sightlines can be identified and excluded from ISM optical depth maps.

\section{Testing with Observational Data} \label{sec:TestingwithObservationalData}

We then tested the accuracy of the method using available spectra (or spectral energy distributions, SEDs) of sources with previously reported optical depth measurements for the water ice $-$OH feature, the aliphatic hydrocarbon $-$CH feature, and the silicate $-$SiO feature. For this, we employed the Infrared Space Observatory spectra of stars presented in \cite{Gibb2004}, where detailed information on observations and a brief description of the nature of the sources can also be found.

We compared the photometric optical depth measurements with spectroscopic optical depth measurements to quantify differences arising from use of filters. We also examined them in relation to their corresponding reported values in the literature \citep{Gibb2004} to analyze differences arising from methodological limitations in capturing spectral profiles.

\subsection{Spectra Set}\label{sec:SpectraSet}

We employed spectra of 17 background sources from \cite{Gibb2004}: Sgr A$^{*}$, Elias 16, Elias 29, AFGL 989, R CrA IRS 1, R CrA IRS 2, W3 IRS 5, AFGL 490, Orion BN, Orion IrC2, Mon R2 IRS 2, Mon R2 IRS 3, AFGL 2136, AFGL 2591, S140, NGC 7538 IRS 1, NGC 7538 IRS 9, aW33A, AFGL 7009S. Among them, Sgr A$^{*}$, a supermassive black hole located at the Galactic Center, provides a sightline that probes dense clouds through the Galactic arms \citep{BryantKrabbe2021}. The rest of the sources are classified into five groups in \cite{Gibb2004} according to source type: a field star located in a quiescent cloud (Elias 16), low-mass YSO (Elias 29), intermediate-mass YSO (AFGL 989, R CrA IRS 1, R CrA IRS 2), high-mass YSO with weak processing (W3 IRS 5, AFGL 490, Orion BN, Orion IrC2, Mon R2 IRS 2, Mon R2 IRS 3, AFGL 2136, AFGL 2591, S140, NGC 7538 IRS 1, NGC 7538 IRS 9), and high-mass YSO with strong processing (W33A, AFGL 7009S). However, more recently, W3 IRS5 has been reported as a massive protostellar cluster \citep{Wang2013}. 

The optical depths reported in the literature \citep{Gibb2004} for all sources are listed in Table~\ref{tab:comparison-all}. Among these sources, Elias 16's spectrum covers only the NIR region, while R CrA IRS 2's spectrum is noisy in the MIR region. In the spectrum of Orion IRc2, aliphatic hydrocarbons were not detected. The spectrum of W33A is saturated throughout the ice feature in the NIR region. The spectrum of AFGL 7009S is highly saturated, and because of this, the water ice feature and aliphatic hydrocarbons were not reported. However, we use the entire spectral dataset for optical depth measurement simulations to test the efficiency of the photometric method.

\subsection{Optical Depth Measurements}\label{sec:PhotometricOpticalDepthMeasurements}

The photometric flux measurements through the suitable NIRCam and MIRI filters are shown on the spectra set from \cite{Gibb2004} in Figure~\ref{fig:AllSpectraG04} in Appendix (also see Figure~\ref{fig:SgrASpectra}). The photometric and spectroscopic fluxes are calculated using approximations presented in Equation~\ref{eq:photometricflux} and Equation~\ref{eq:spectroscopicflux}, respectively. 

The continuum fluxes at corresponding wavelengths are calculated based on different continuum estimates. First, we applied a polynomial fit for \textit{general continuum} detection using multiple continuum filters (Table~\ref{tab:features-filters}). Then we applied power-law and line fitting for the \textit{local continuum detection} by employing different Filter Sets listed in Table~\ref{tab:filter-sets}. By comparing photometric optical depths and spectroscopic optical depths, we found that local continuum fitting is preferable as the results are more consistent for each source than those obtained with general continuum estimation approximations. Local linear continuum fitting generally yielded higher optical depth values, with average $\Delta \tau / \tau$ values of approximately 0.14, 0.23, and 0.01 for the water ice, aliphatic hydrocarbon, and silicate features, respectively, compared to the power-law continuum. However, local linear continuum fitting provides more consistent (less scattered) results compared to those obtained using local power-law continuum estimation approximations (we also discussed local linear continuum fitting in detail in Section~\ref{sec:TestingwithModelSpectra}). For the rest of the analysis we applied \textit{local linear continuum} using the Equation~\ref{eq:continuumflux} to obtain optical depths. 

We tested the efficiency of different Filter Sets (listed in Table~\ref{tab:filter-sets}) by comparing the photometric and spectroscopic optical depths (see Figure~\ref{fig:OD-Data-FilterSets}). The comparison of resultant optical depths indicates that almost all Filter Sets exhibit varying degrees of success in estimating optical depths. This suggests that, when necessary, different combinations of suitable filters can be utilized for such estimations, albeit with differing levels of accuracy and precision. We further discussed filter set selection in Section~\ref{sec:TestingwithModelSpectra}. Our findings indicate that Filter Set 1 for the NIR and Filter Set 1 for the MIR region of the spectra represent the one of the optimal choices for constraining absorption features and fitting the continuum. We illustrated the photometric fluxes and local linear continuum fit obtained by Filter Set 1 for NIR and Filter Set 1 for MIR  on the spectrum of Sgr A$^{*}$ in Figure~\ref{fig:continuumfitting}. 

Then we computed photometric and spectroscopic optical depth using the selected Filter Sets. We present the optical depth results obtained by photometric and spectroscopic measurements, with a comparison to reported values in the literature \citep{Gibb2004} in Table~\ref{tab:comparison-all}. 

\subsection{Comparison with Spectroscopic Method}\label{sec:ComparisonwithSpectroscopicMethod}

\begin{figure*}
  \begin{center}
    \begin{tabular}{ccc}
    {\includegraphics[angle=0,scale=0.20]{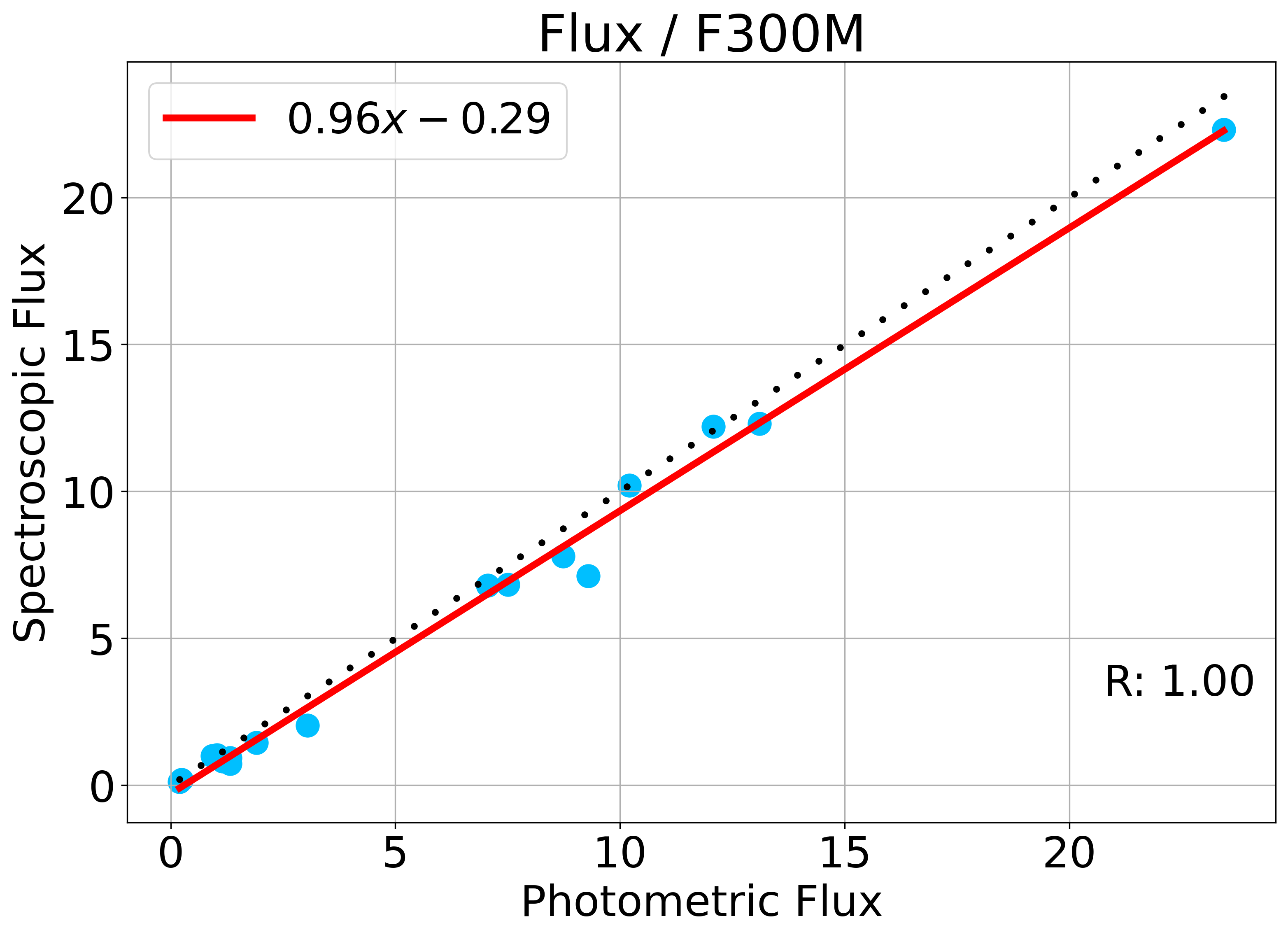}} &
    {\includegraphics[angle=0,scale=0.20]{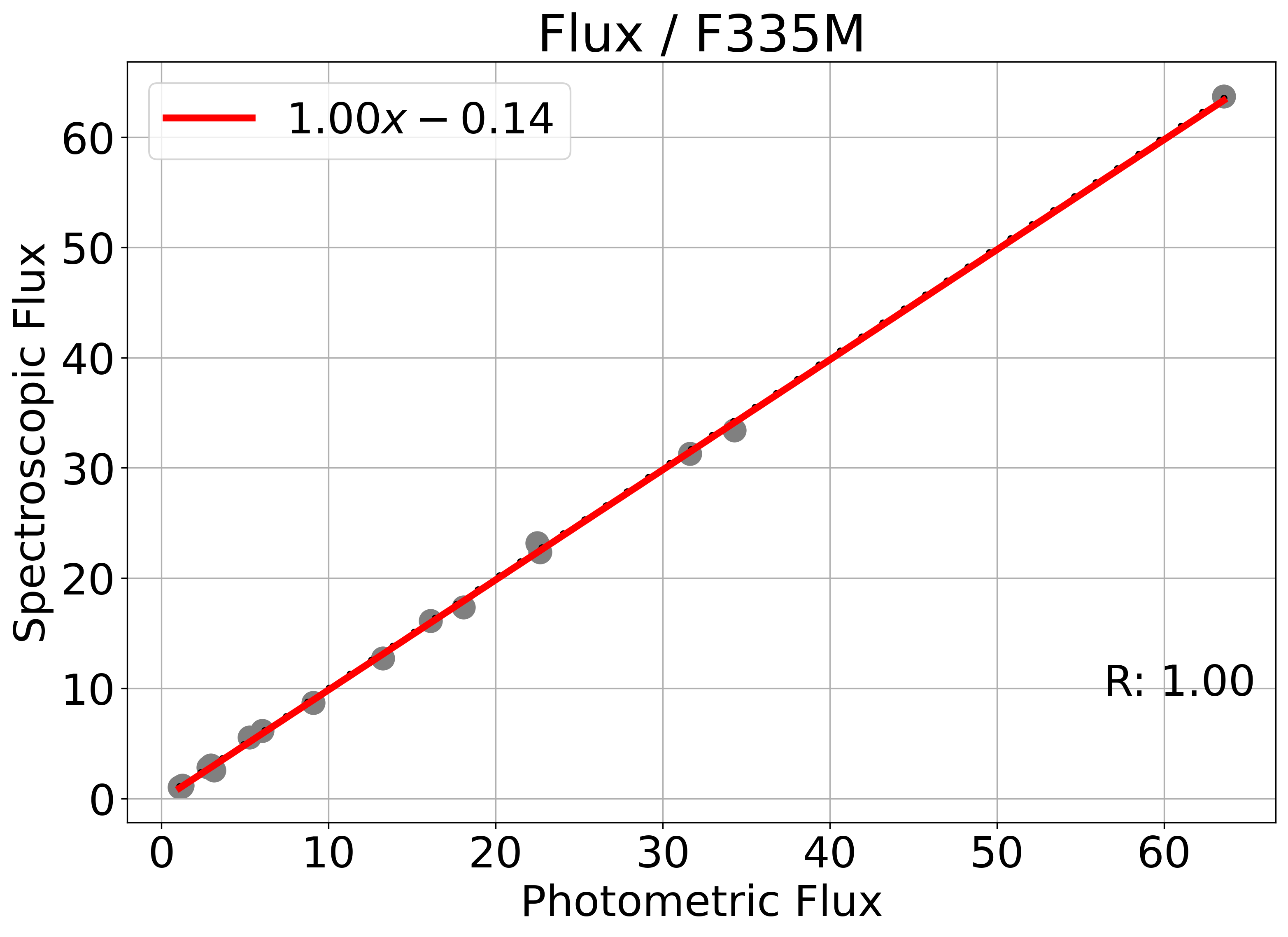}} &
    {\includegraphics[angle=0,scale=0.20]{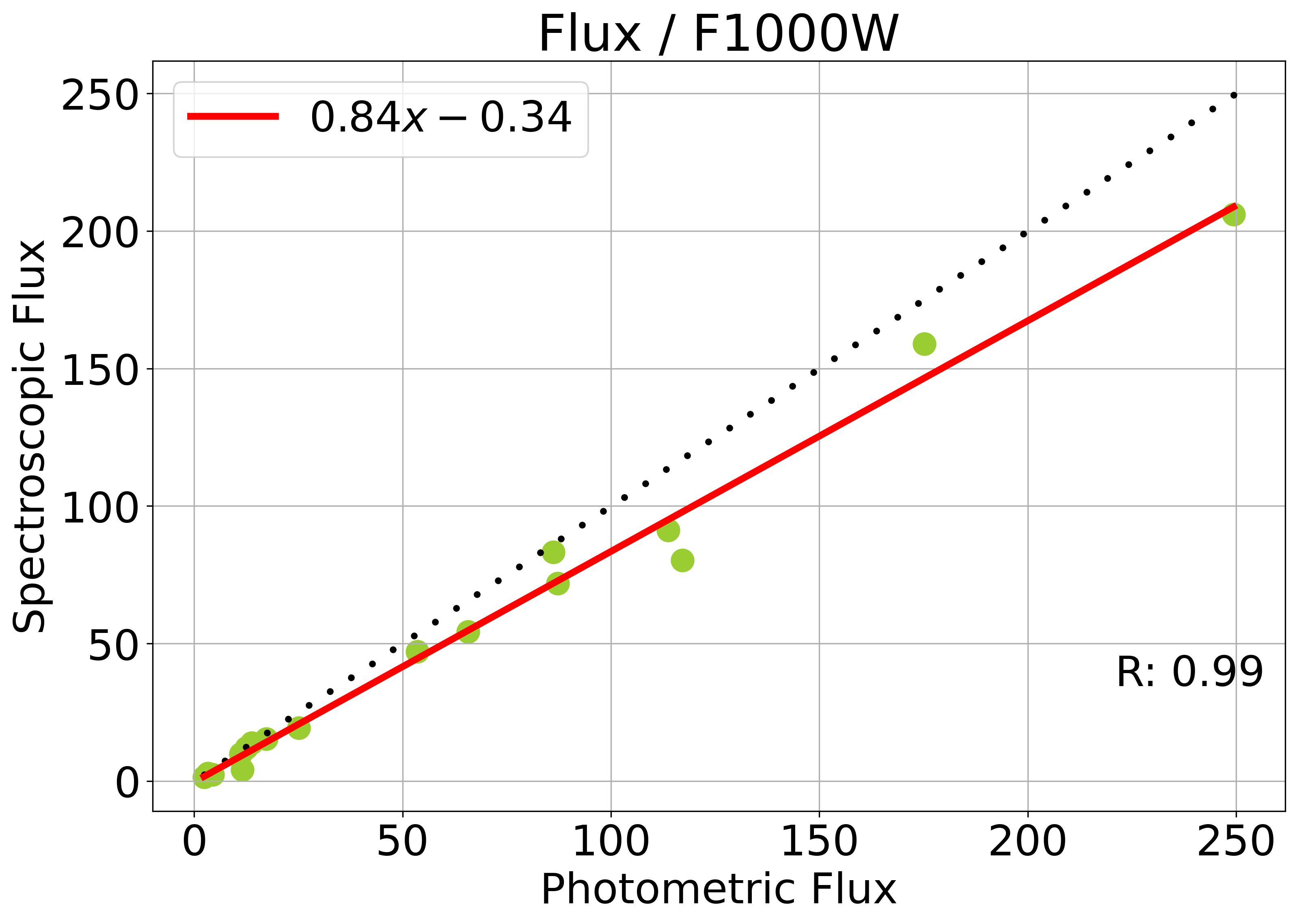}}\\
    {\includegraphics[angle=0,scale=0.20]{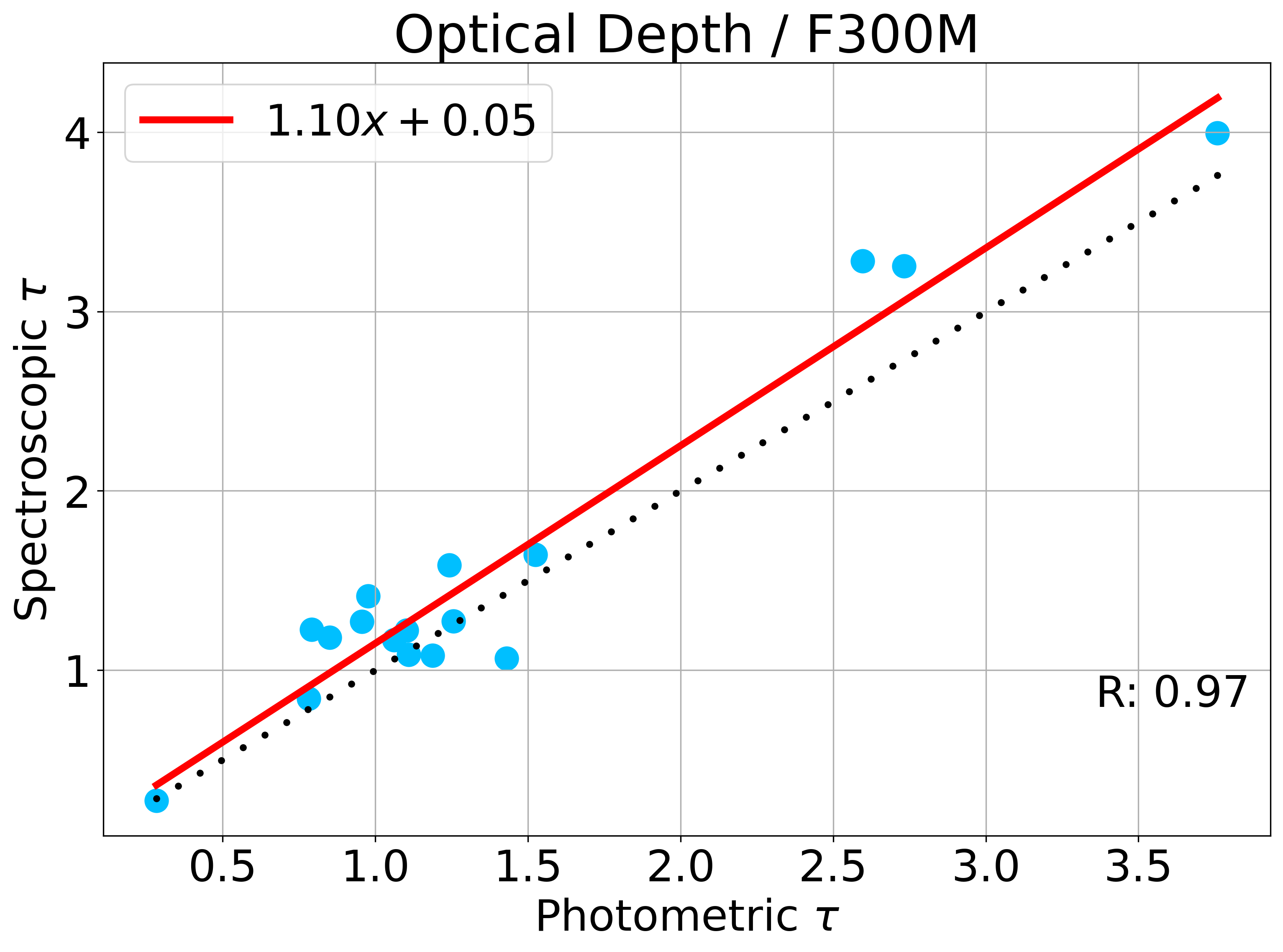}} &
    {\includegraphics[angle=0,scale=0.20]{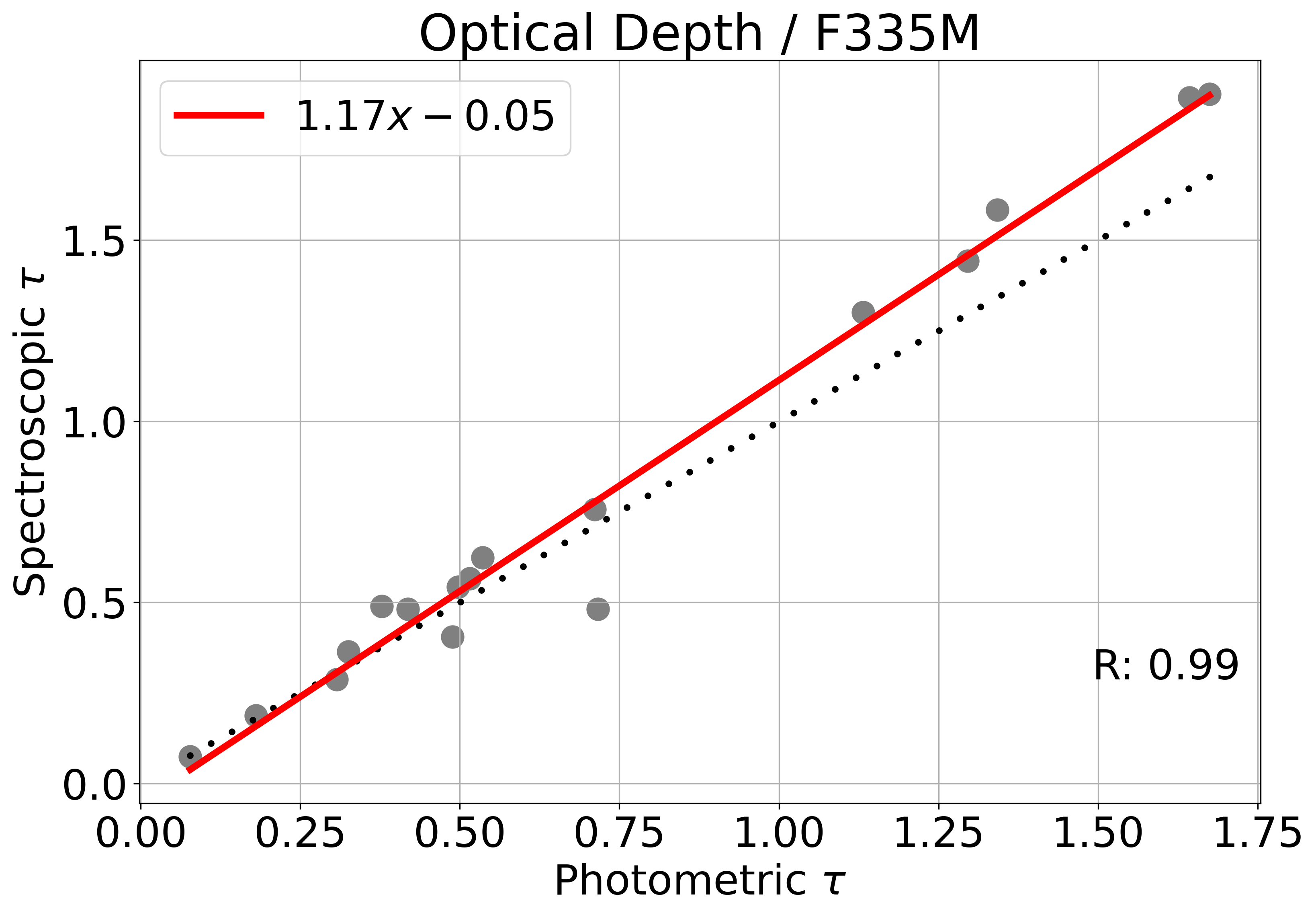}} &
    {\includegraphics[angle=0,scale=0.20]{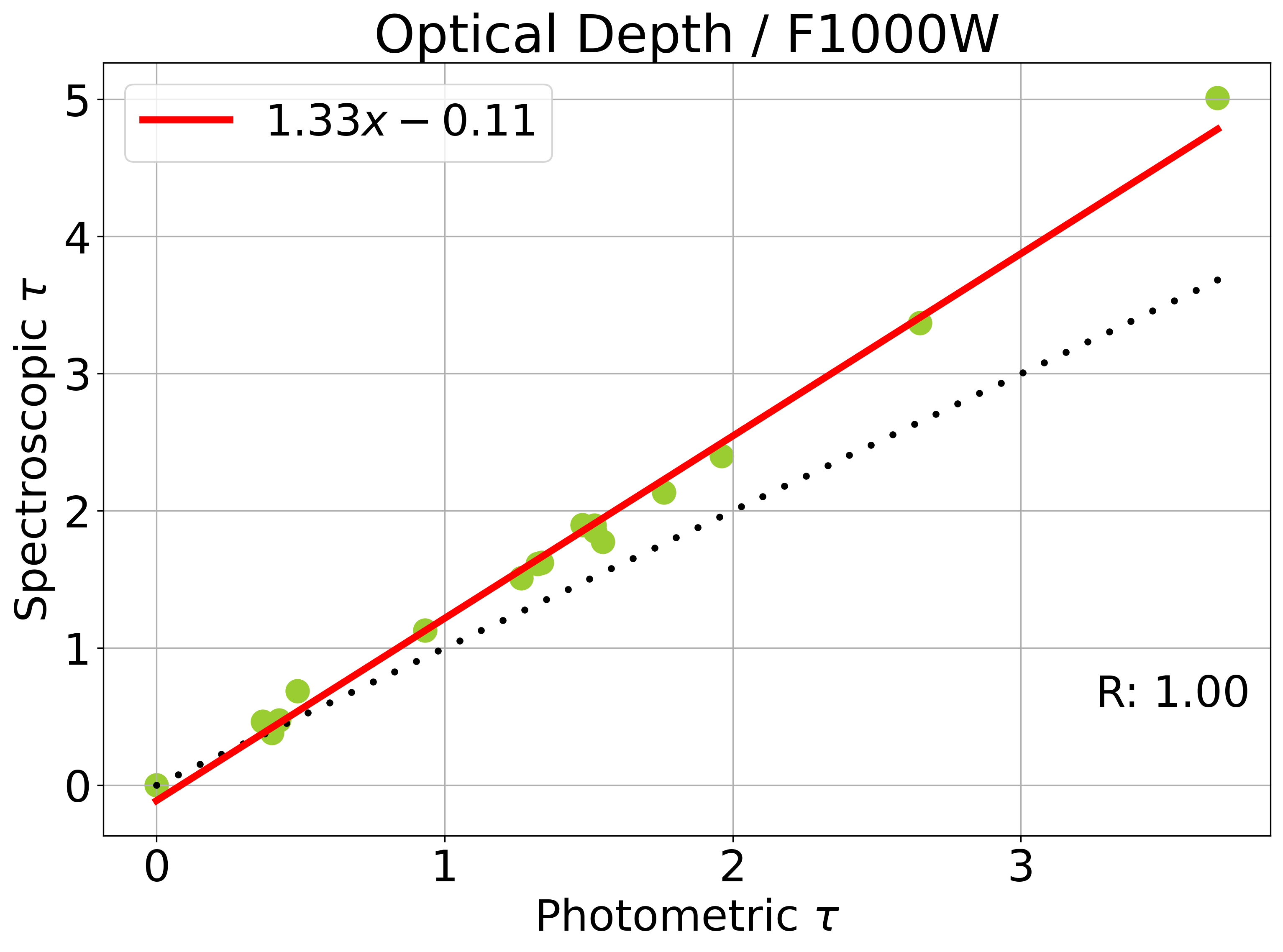}}\\ 
    {\includegraphics[angle=0,scale=0.20]{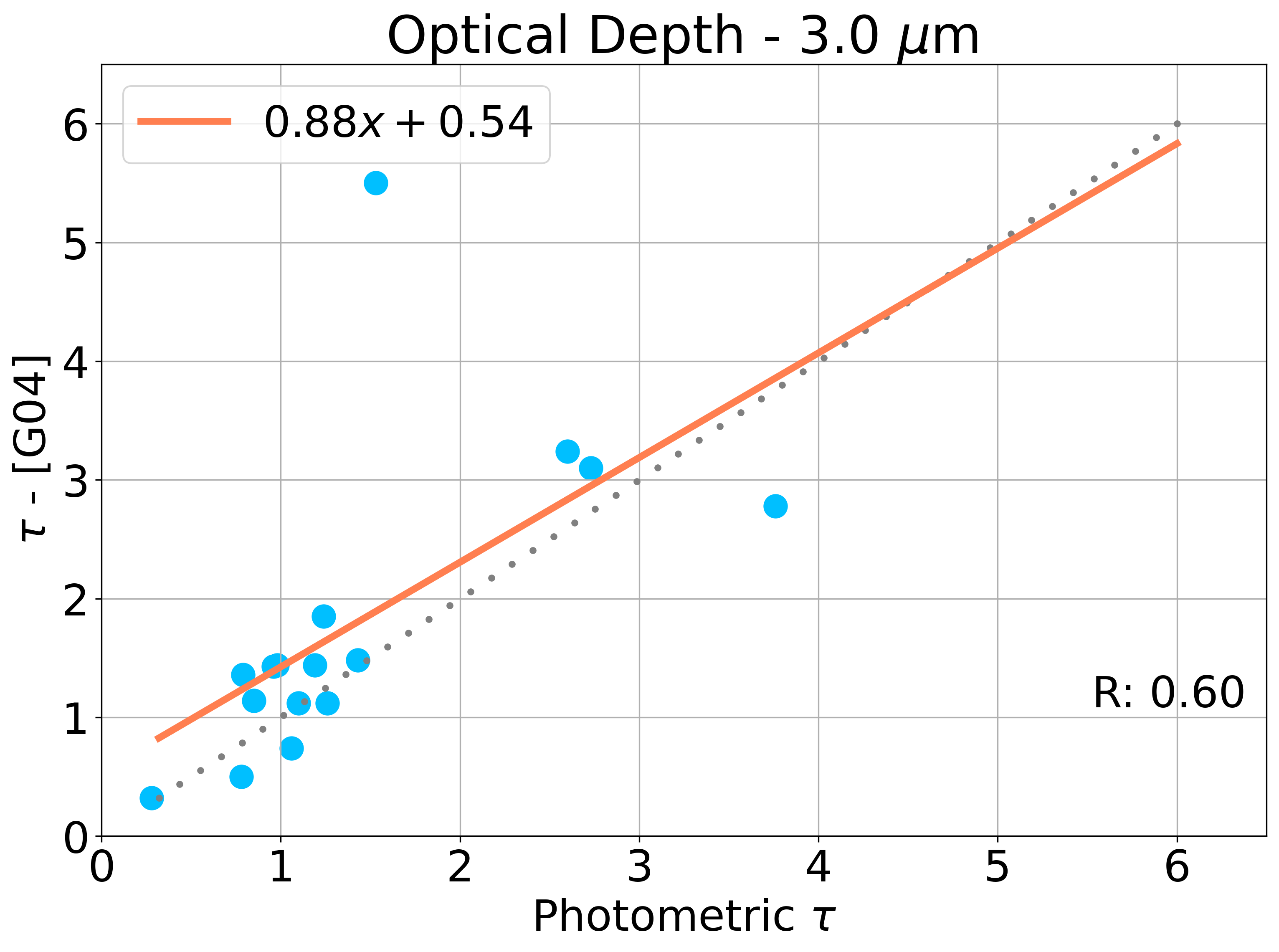}} &
    {\includegraphics[angle=0,scale=0.20]{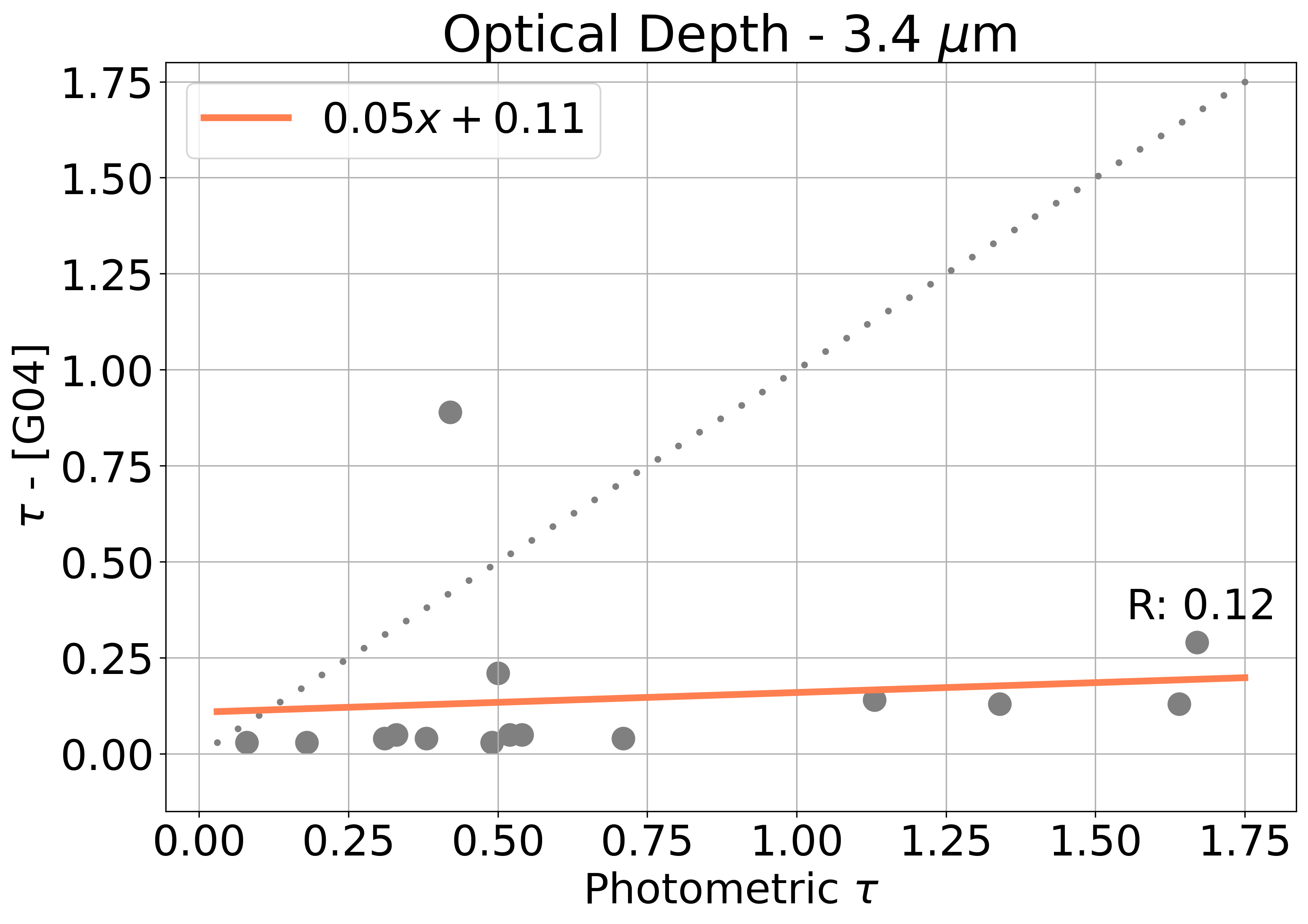}} &
    {\includegraphics[angle=0,scale=0.20]{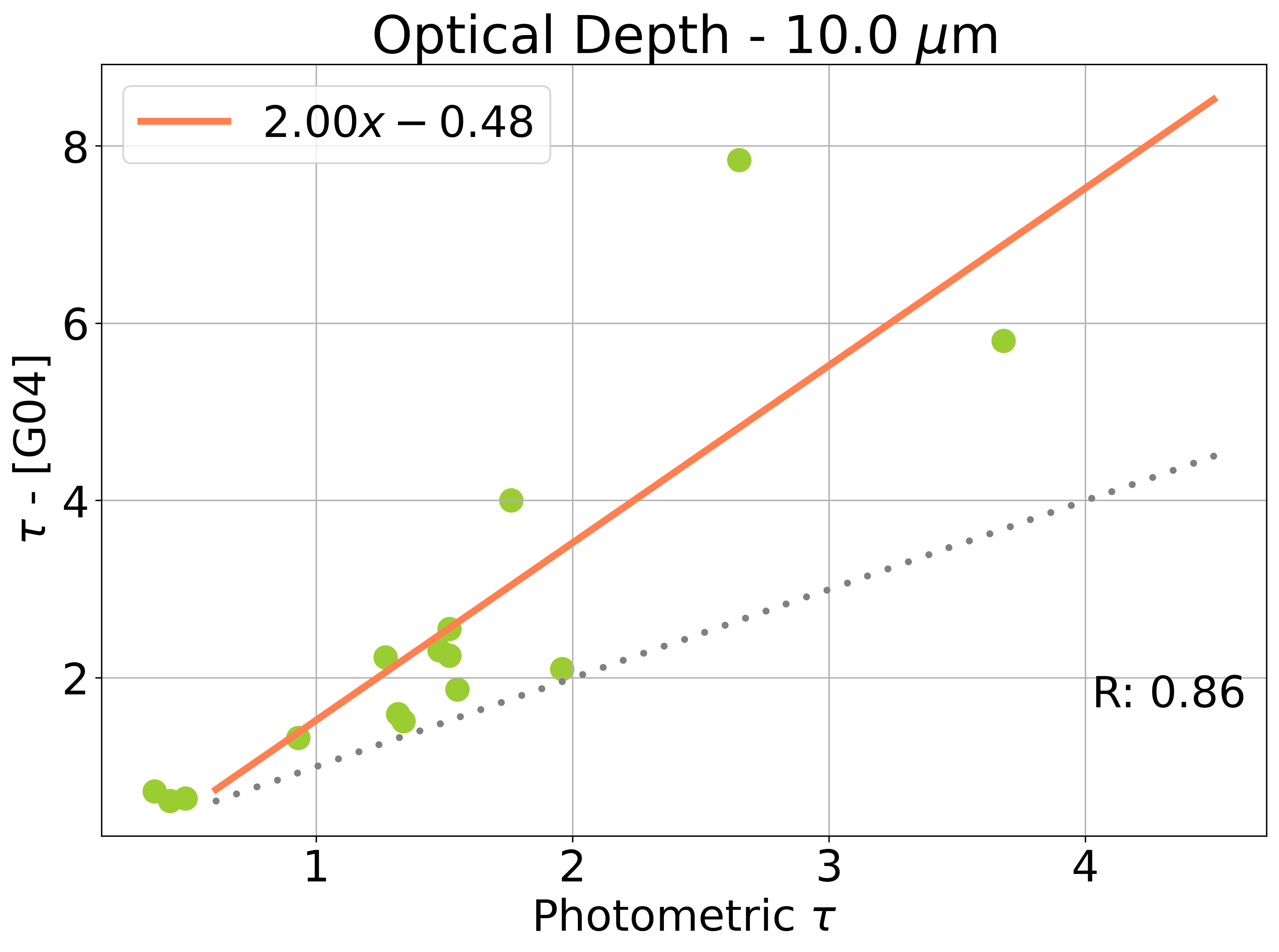}}\\
    {\includegraphics[angle=0,scale=0.20]{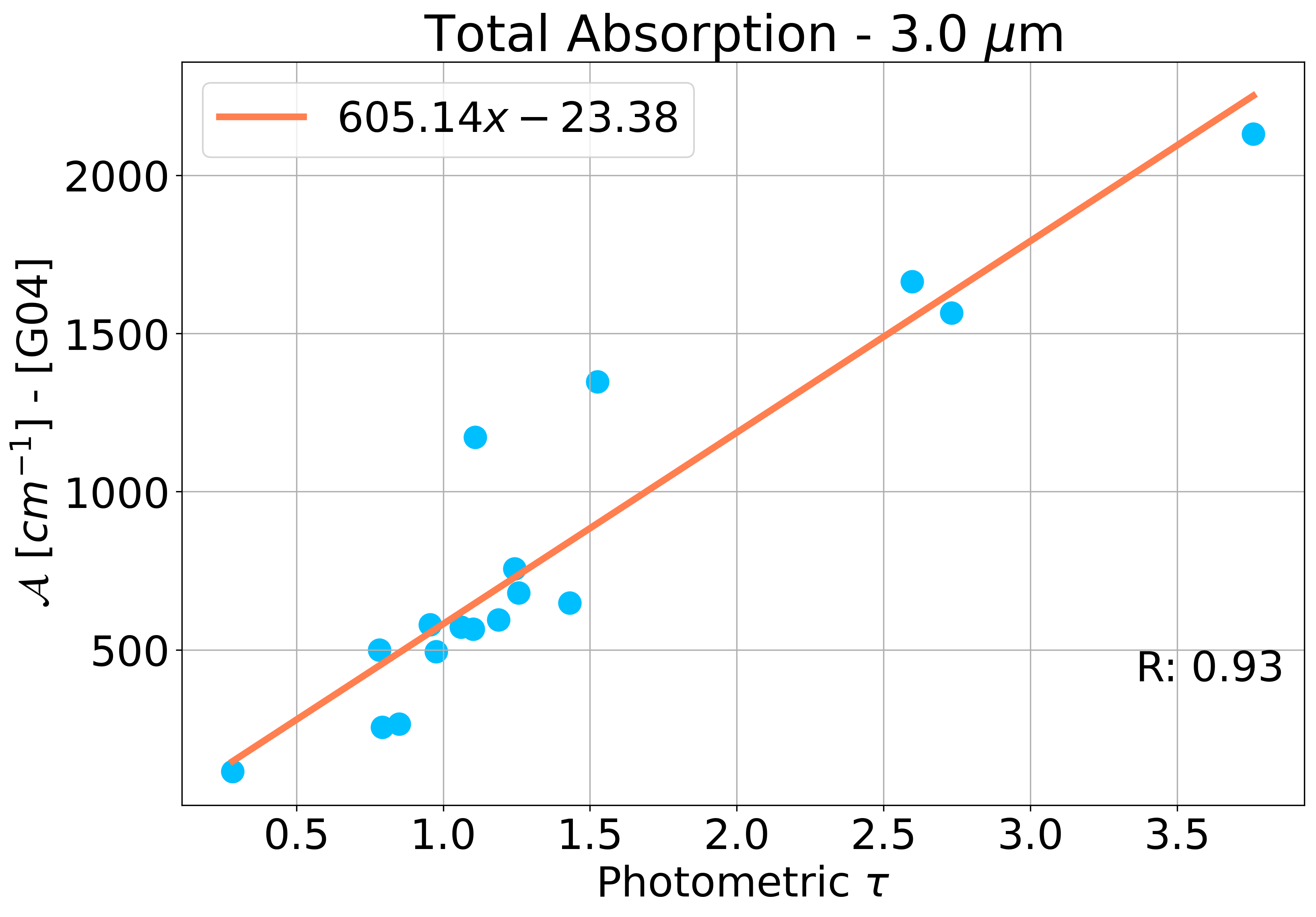}} &
    {\includegraphics[angle=0,scale=0.20]{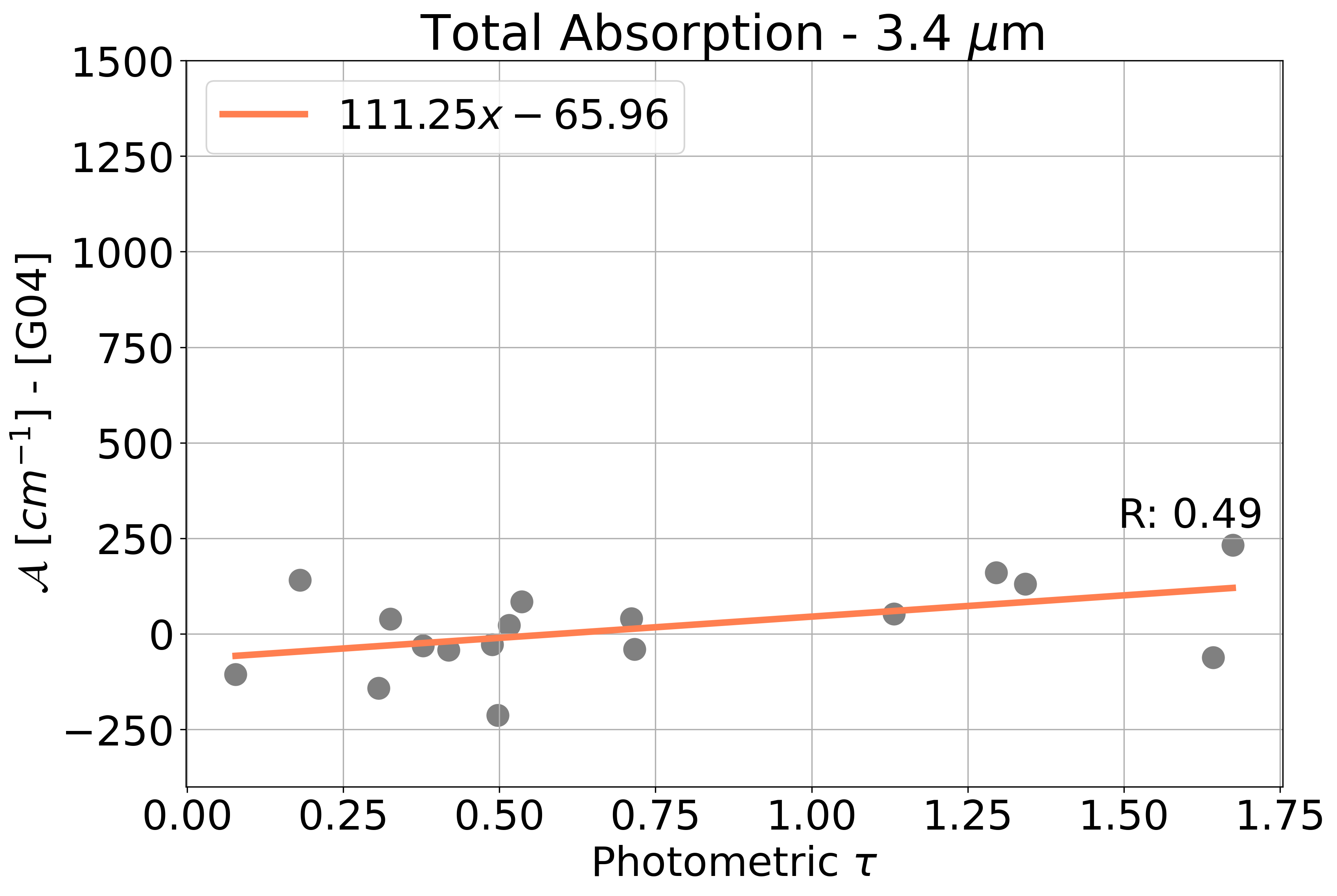}} &
    {\includegraphics[angle=0,scale=0.20]{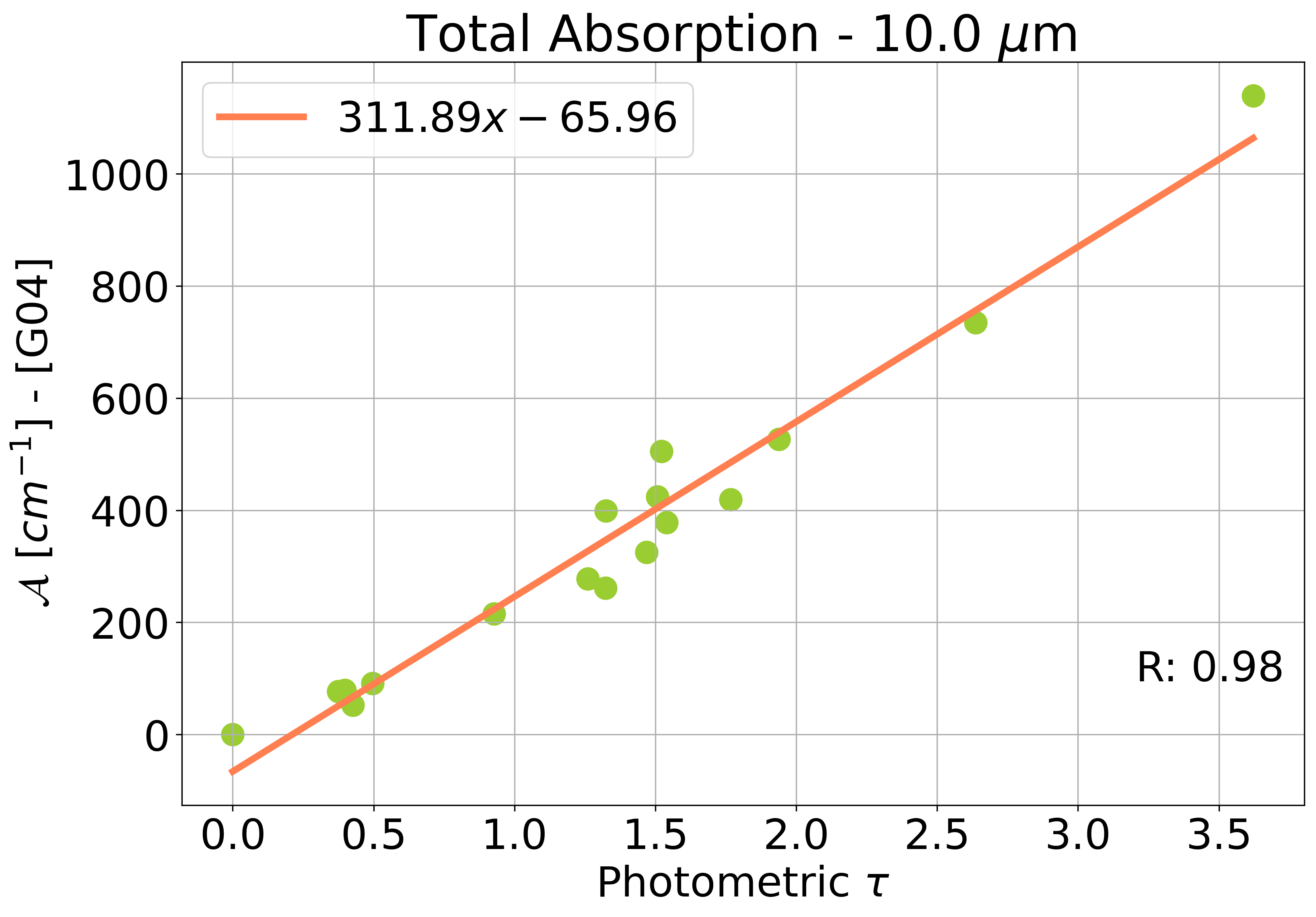}}\\
        \end{tabular}
    \caption{The photometric measurements obtained using the spectral dataset from the literature \citep{Gibb2004} are compared with their corresponding reference values to examine correlations (R is the correlation coefficient). The dotted lines indicate the equal $\tau_{p}$ and reference values. \textit{First row}: Linear fits describing the correlation between photometric and spectroscopic fluxes at absorptions derived from simulations. \textit{Second row}: Linear fits describing the correlation between $\tau_{p}$ and $\tau_{s}$ values. \textit{Third row}: Linear fits describing the correlation between $\tau_{p}$ and $\tau_{r}$ ([G04]) values. \textit{Fourth row}: Linear fits describing the correlation between $\tau_{p}$ and total absorption ($\mathcal{A}$) values.}  
    \label{fig:comparison-SPR}
      \end{center}
\end{figure*}

In this section, we analyzed the differences between fluxes, photometric optical depths, spectroscopic optical depths, and reported optical depths from the literature \citep{Gibb2004} that were obtained with a \textit{spectroscopic method}. We presented correlations in Figure~\ref{fig:comparison-SPR}. 

\subsubsection{Fluxes}\label{sec:Fluxes}

First, we examined the correlation between photometric and spectroscopic fluxes derived from simulations, as shown in the first row of Figure~\ref{fig:comparison-SPR}. The comparison shows that absorption fluxes for the silicate feature are overestimated, while those for the water ice feature remain nearly consistent, and the aliphatic hydrocarbon feature shows high consistency. A contributing factor to the deviations is the additional photon integration caused by the overlap between the filter throughputs and the absorption feature profiles. This is more pronounced for the silicate feature since we use a wide band filter. Additionally, asymmetries, width variations, and peak absorption shifts can influence peak absorption flux measurements.

The comparison of correlations between fluxes (first row) and optical depths (second row) indicates that the photometric method tends to systematically underestimate optical depths due to the smoothing of spectral features caused by the use of filters. Additionally, depending on the selected filter set, the blending of the wings or shoulders of absorption features with the filter profiles can further reduce the estimated continuum fluxes and, consequently, the derived optical depths (see Figure~\ref{fig:OD-Data-FilterSets}).

The resulting optical depth measurements exhibit some scatter due to additional spectral features intrinsic to individual lines of sight, as the spectral set spans a wide range of environments (from quiescent clouds to YSOs). This effect is more pronounced in the NIR region, especially along icy sight-lines, where the presence of the CO${2}$ (4.2--$\mu$m) and CO (4.7--$\mu$m) ice absorption features can affect continuum fitting. There are also H$_{2}$ emission lines at 2.12--$\mu$m and 3.23--$\mu$m, as well as H emission lines at 4.05--$\mu$m, which may contribute to the photometric fluxes. In the MIR region, although the scatter is primarily due to variations in absorption fluxes, the presence of PAH emission features at 7.7--$\mu$m and 11.3--$\mu$m, the O-C-O bending absorption feature at 15.3--$\mu$m, and the broad Si-O-Si bending absorption feature at 18.0--$\mu$m can cause deviations in optical depths.

\subsubsection{Optical Depths}\label{sec:OpticalDepths}

We then compared $\tau_{p}$ with $\tau_{s}$ values (the second row of Figure~\ref{fig:comparison-SPR}) to examine the influence of filter use, and $\tau_{p}$ with $\tau_{r}$ values (the third row of Figure~\ref{fig:comparison-SPR}) to reveal methodological discrepancies. As expected, the correlation between $\tau_{p}$ and $\tau_{r}$ was significantly weaker than that between $\tau_{p}$ and $\tau_{s}$, highlighting the impact of these methodological differences, including more accurate continuum and absorption constraining approximations unavailable to the photometric method due to the limited resolution.

We calculated absolute differences in optical depths ($\Delta \tau_{ps} = |\tau_p - \tau_s|$ and $\Delta \tau_{pr} = |\tau_p - \tau_r|$) and normalized these differences with reference optical depths ($\tau_{\text{s}}$ or $\tau_{\text{r}}$) to evaluate their relative magnitudes. We provided the absolute differences and the relative absolute differences ($\Delta \tau_{ps}/\tau_{\text{s}}$ and $\Delta \tau_{pr}/\tau_{\text{r}}$) in Table~\ref{tab:deviation-PS} and Table~\ref{tab:deviation-PSR}. 

For the water ice, aliphatic hydrocarbon, and silicate features, the average $\Delta \tau_{ps}/\tau_{\text{s}}$ values are 0.16, 0.14, and 0.18, respectively, with an overall average of 0.16 across all measurements, indicating consistency. The corresponding average $\Delta \tau_{pr}/\tau_{\text{r}}$ values are 0.28, 7.56, and 0.33, respectively, and are considerably larger, particularly for the aliphatic hydrocarbon feature. These discrepancies are primarily due to the methodological procedures used for continuum fitting and absorption constraint. The $\Delta \tau_{\text{sr}}/\tau_{\text{s}}$ values are also found to be substantial, with values of 0.24, 8.31, and 0.21 for the water ice, aliphatic hydrocarbon, and silicate features, respectively. This further supports the conclusion that the differences mainly arise from methodological factors, whereas the impact of filter usage appears to be relatively minor.

There are unavoidable deviations in the optical depth values as the photometric method is limited by three photometric data points. The $\Delta \tau_{pr}$ values are substantially higher for the aliphatic hydrocarbon feature, as the contamination from the long-wavelength wing of the water ice absorption feature is corrected in \cite{Gibb2004}. The relatively small average $\Delta \tau_{pr}$ values for the water ice and silicate features are primarily due to linear continuum fitting and uncaptured variations in the spectral profile. The differences are more pronounced for saturated sources, such as W33A and AFGL 7009S, which were analyzed using a different approximation in \cite{Gibb2004}. The saturated features in the spectra of these sources are modeled using analytical function fitting based on full spectroscopic information, and thus show the largest discrepancies from the photometric optical depths (see Figure~\ref{fig:comparison-SPR}). If we exclude the data from saturated sources in the observational spectra set, the average $\Delta \tau_{pr}/\tau_{\text{r}}$ is reduced to 0.25 for water ice and 0.29 for the silicate feature, based on the cleaned data set, thereby revealing the significant impact of observational data quality on the reliability of the analysis.

Therefore, the comparisons indicate that large differences from the reported values primarily stem from methodological approaches used in spectroscopy to improve results. With limited resolution, capturing the full profile of absorption features or accurately modeling them is not feasible. Also, the accuracy of continuum modeling depends on both spectral coverage and resolution. Therefore a complete consistency between the different methods cannot be expected.

\subsubsection{Total Absorptions}\label{sec:TotalAbsorptions}

Additionally, we investigated the correlation between $\tau_{p}$ and total absorption (integrated absorption, $\mathcal{A}$, cm$^{-1}$) (the fourth row of Figure~\ref{fig:comparison-SPR}). To obtain total absorption measurements for each spectrum \citep{Gibb2004}, we used Equation~\ref{eq:IntegratedAbsorption}, where optical depth spectra were derived based on a linear continuum approximation for consistency. This approach minimized deviations caused by the different approximations used in the spectroscopic method, such as more accurate continuum fitting and absorption constraining through model fitting. We found that the photometric optical depths align more closely with the total absorption values than with the reported optical depth values, although the quality of some observational spectra in the data set influences the photometric measurements to some extent. This comparison shows that the primary source of the discrepancies is the use of various methodological approximations applied to improve spectroscopic results.

\subsubsection{Methodologies}\label{sec:Methodologies}

The reported optical depths in the literature are obtained using the full spectroscopic information, allowing for more accurate absorption measurements and continuum estimation. However, a challenge in observational studies is the wide variation in optical depths reported for the same source in the literature, often due to differences in instrumentation, data quality, analysis methods, and continuum fitting approximations (as discussed in \citealt{Moultaka2004, Godard2012, Gunay2020, Gunay2022}). For example, \cite{Chiar2002} and \cite{Moultaka2004} reported different 3.4 $\mu$m optical depth values for the same sources (GCIRS 12N, GCIRS 1W, GCIRS 3, GCIRS 7). For GCIRS 7, optical depths at 3.4--$\mu$m were reported as 0.15 by \citet{Chiar2002} and 0.41 by \citet{Moultaka2004}. The difference between these two measurements, $\Delta \tau = 0.26$, corresponds to a relative deviation of $\Delta \tau/\tau_{0} = 1.73$ or $0.63$, depending on the selected reported value ($\tau_{0}$ = $\tau_{r}$). Moreover, \cite{Moultaka2004} showed that optical depths vary depending on the different approximations used to estimate the continuum. For example, for the source IRS 16C, the aliphatic hydrocarbon optical depths can be 0.14 or 0.49, depending on the continuum fitting \citep{Moultaka2004}. This results in $\Delta \tau = 0.35$, and $\Delta \tau/\tau_{0} = 0.71$ or $2.50$, depending on the reference optical depth. \cite{Godard2012} also discussed challenges in detection of real continuum and they preferred local linear continuum for the 3.4--$\mu$m feature.  

Consequently, we demonstrate that the differences between the photometric optical depths obtained in this study and the reported values fall within the range of variations observed among independent spectroscopic studies. As noted, optical depth values from different projects are often affected by instrumental and methodological variations. Simultaneous photometric measurements of all sources within the same frame mitigate these discrepancies by ensuring a more accurate relative optical depth distribution in the FoV through consistent data acquisition and processing.

\section{Calibrations} \label{sec:Calibrations}

In this section, we demonstrate that \textit{spectral models} can be used to derive calibration equations to minimize methodological biases in the photometric optical depths. 

\subsection{Spectral Models for Calibration \label{sec:CalibrationModels}}

To identify the best spectral models (optimal models) representing the data set ([G04]), we generated a test set of model spectra (test models) with varying parameters (CWs and FWHMs) and a range of reference optical depths (the average spectroscopic optical depth levels obtained in this study). We introduced flexibility to the FWHM values while constraining the CW parameter range to the commonly reported values in the literature, as discussed in Section~\ref{sec:ModelingSpectra}. We used 10 CW values and 10 FWHM values within the specified ranges, resulting in a total of 100 unique combinations for each feature. The parameter ranges are presented in Table~\ref{tab:model-parameter-ranges}. 

Then using these spectra set, we derived calibration equation sets that link photometric optical depths ($\tau_{p}$) to spectroscopic optical depths ($\tau_{s}$), as the true optical depths ($\tau_{0}$) of the observational spectra are unknown and reported optical depths ($\tau_{r}$) are influenced by various methodological approximations. 

We then tested all equations on the observational spectra set to determine the optimal model parameters (CWs $\&$ FWHMs) that represent the data set by yielding the best calibration. Since a single spectral model cannot perfectly represent the diverse feature profiles of the observational spectra, we applied the Least Squares Method to identify models that minimize the average $\Delta\tau_{ps}$, thereby providing the most effective calibration equations for the data set \citep{Gibb2004}. 

The parameters (CWs and FWHMs) of the optimal models identified for the calibration process are summarized in Table~\ref{tab:model-parameter-ranges}. In Figure~\ref{fig:calibration-models}, the absorption feature templates used in the standard and optimal models are shown in relation to the transmission wavelengths (half-power points) of the filters. For all models, the CWs are close to the frequently reported values, as the variations were constrained, but adjusted FWHM value for the water ice is slightly larger than the expected values (see Section~\ref{sec:ModelingSpectra}). We note that other combinations in the vicinity of these values may also yield approximately effective results.

\subsection{Calibration Equations \label{sec:CalibrationEquations }}

The best calibration equations for mitigating methodological biases arising from filter use have been identified through the determination of optimal models. These polynomial equations ($P_{ps}(x)$), obtained from the correlation between $\tau_{p}$ and $\tau_{s}$, are presented in Table~\ref{tab:coeffs}. 

We also used optimal models to derive correlations that link photometric optical depths to true optical depths. These polynomial equations ($P_{p0}(x)$), obtained from the correlation between $\tau_{p}$ and $\tau_{0}$ are also presented in Table~\ref{tab:coeffs} and can minimize overall methodological biases, including those arising from limited resolution.

We emphasize that the calibration equations depend on the characteristics of the observational spectral set modeled here \citep{Gibb2004} but they can be applicable to similar sightlines (dense sightlines) with some degree of success. However, calibration equations can be improved in future studies by incorporating larger data samples include high-resolution spectra obtained with JWST.

\subsection{Calibrated Results \label{sec:CalibratedOpticalDepths}}

\begin{figure*}[ht]
\begin{center}
\begin{tabular}{ccc}
    {\includegraphics[angle=0,scale=0.21]{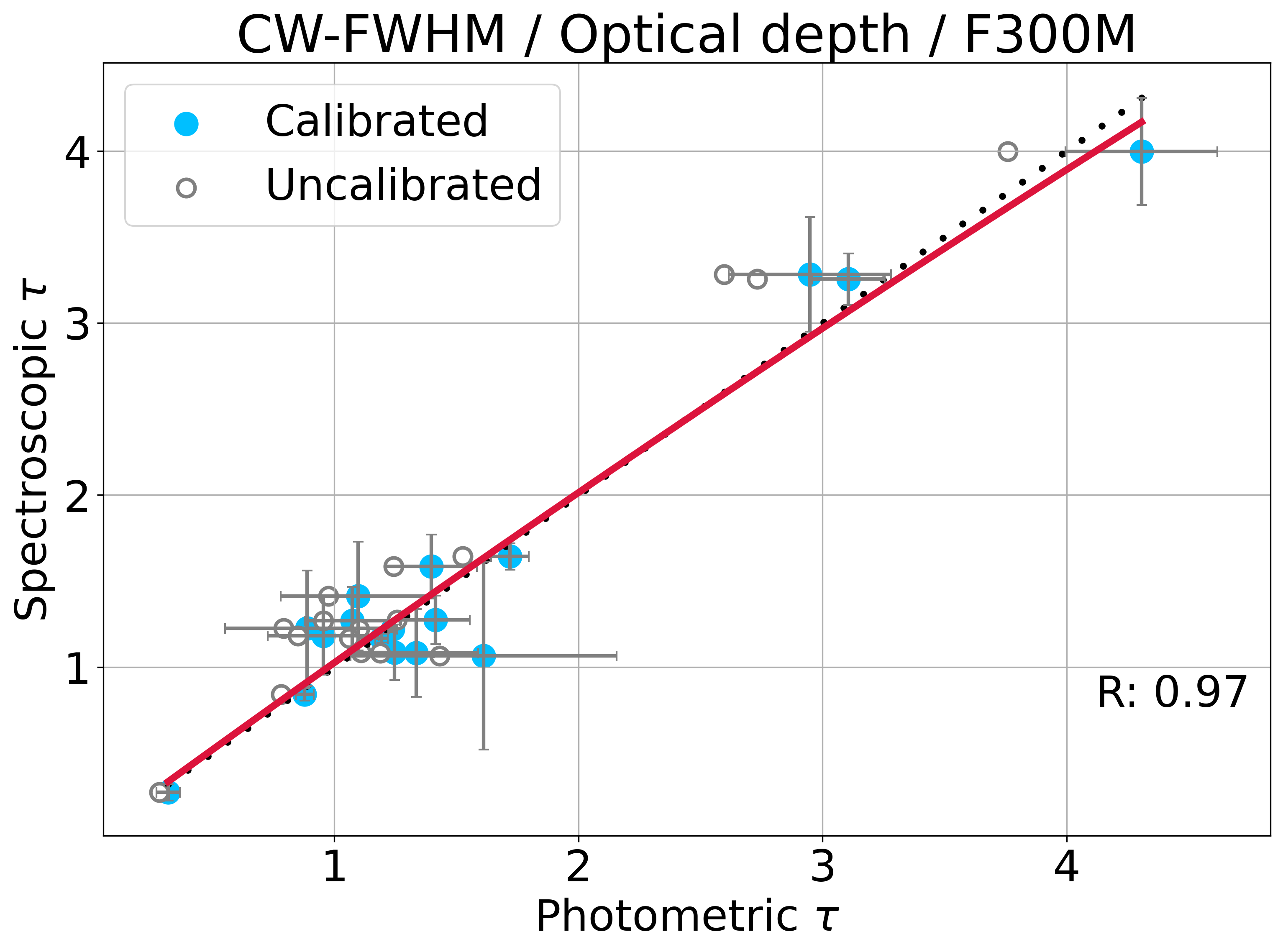}} &
    {\includegraphics[angle=0,scale=0.21]{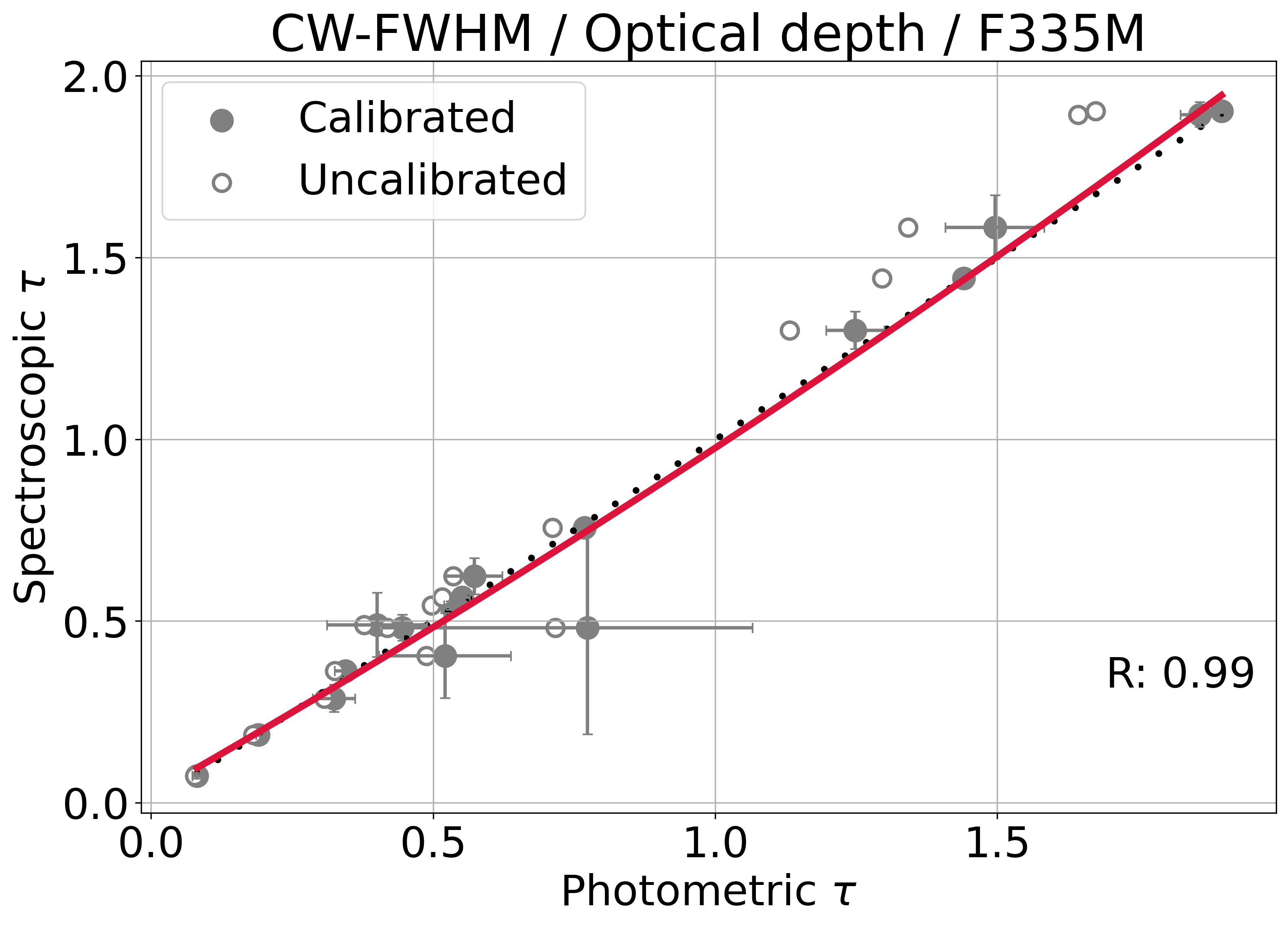}} &
    {\includegraphics[angle=0,scale=0.21]{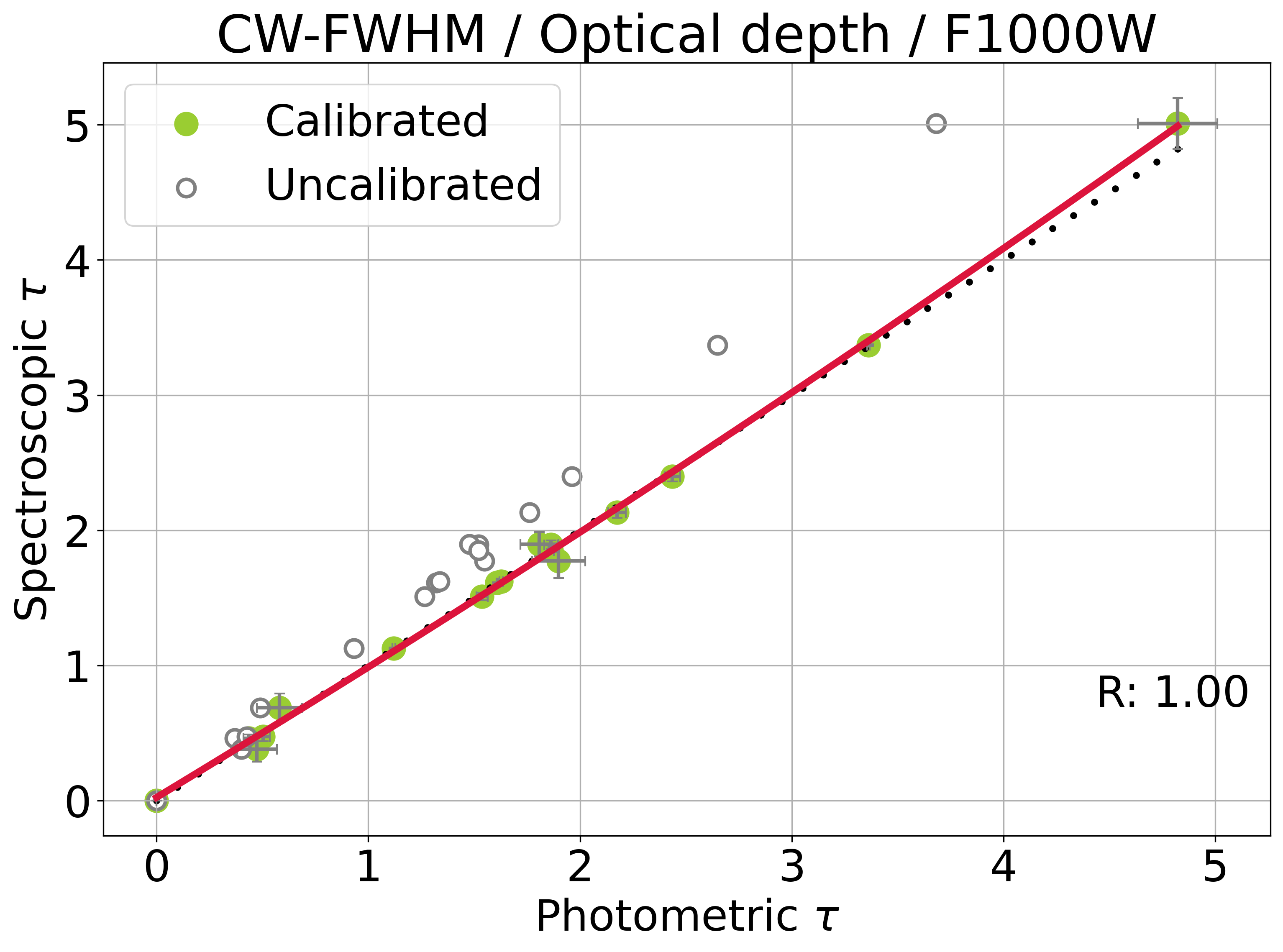}}\\
\end{tabular}
\caption{The photometric optical depths compared with the spectroscopic optical depths. In this comparison, calibrated optical depths are represented by filled dots, while uncalibrated optical depths are shown with unfilled dots. The polynomial fits show the trend between $\tau_{p}$ and $\tau_{s}$ values and the dotted line indicates equal $\tau_{p}$ and $\tau_{s}$ values.}
\label{fig:od-calibratations}
\end{center}
\end{figure*}

After calibration, the agreement between the photometric optical depths and those derived from spectroscopic simulations improved, indicating an effective correction of methodological biases introduced by the use of filters. The calibrated photometric optical depths ($\tau_{c}$) are compared with the spectroscopic optical depths in Figure~\ref{fig:od-calibratations}, demonstrating overall consistency. The $\tau_{c}$ values are presented in Table~\ref{tab:calibrated-improved}. The absolute differences between the calibrated and spectroscopic optical depths ($\tau_{cs}$), along with their relative values for each sightline, are presented in Table~\ref{tab:deviation-CSR}.

However, as anticipated, these calibrated optical depths still show unavoidable differences, stemming from the inherent, sightline-specific characteristics of the spectra. The average $\Delta \tau_{cs}/\tau_{\text{s}}$ values are found to be 0.15, 0.11, and 0.05 for the water ice, aliphatic hydrocarbon, and silicate features, respectively (see Table~\ref{tab:deviation-CSR}). 
The calibrated photometric optical depths are presented in Table~\ref{tab:calibrated-improved}. The absolute differences between calibrated and reported optical depths ($\tau_{cr}$) are presented in Table~\ref{tab:deviation-CSR}, together with their relative values for each sightline.

The calibrated results are compared with the reported values. For the silicate feature, the calibrated results generally became more consistent with the reported values, and the average differences were reduced. However, no improvement was observed for the water ice and aliphatic hydrocarbon features. The average $\Delta \tau_{cr}/\tau_{\text{r}}$ values are 0.28, 8.05, and 0.22 for the water ice, aliphatic hydrocarbon, and silicate features, respectively. The absolute differences between calibrated and reported optical depths are presented in Table~\ref{tab:deviation-CSR}, together with their relative values for each sightlines. If we exclude the data from saturated sources in the observational spectra set (see Section~\ref{sec:TestingwithObservationalData}), the average $\Delta \tau_{cr}/\tau_{\text{r}}$ is further reduced to 0.17 for silicate feature and also decreases to 0.25 for water ice feature, based on the cleaned data set. Additionally, the calibrated optical depths are potentially subject to further improvement, as discussed in the following section.

\subsection{Possible Improvements \label{sec:Possible Improvements}}

The calibrated photometric optical depths could be possibly improved to minimize overall methodological biases, continuum related differences, and contributions from other features, based on the approximations presented below.

\textit{Overall biases:}  We used the polynomial equations ($P_{p0}(x)$) that link $\tau_{p}$ values to estimated $\tau_{0}$ values to minimize overall biases (see Section~\ref{sec:MethodologicalBiases}). The optical depths calibrated to correct for overall biases ($\tau_{c0}$) are presented in Table~\ref{tab:calibrated-improved}. The results for the silicate feature became more consistent with previously reported values for some sightlines (see Table~\ref{tab:calibrated-improved}), and the average differences were reduced, whereas no improvement was observed in general for the water ice and aliphatic hydrocarbon features (the $\Delta \tau_{c0}/\tau_{\text{r}}$ values are 0.30, 8.27, and 0.19 for the water ice, aliphatic hydrocarbon, and silicate features, respectively). If the cleaned data set is used, the average $\Delta \tau_{cr}/\tau_{\text{r}}$ is further reduced to 0.15 for silicate feature and also decreases to 0.27 for water ice feature.

\textit{Continuum uncertainties:} The uncertainties due to a linear continuum fit may be minimized based on blackbody continuum approximation, as explained in Section~\ref{sec:Variations-continuum}. These uncertainties are more pronounced for NIR features, as the continuum in the MIR region is comparatively flatter for most of the spectra in the data set. The BB difference adjusted calibrated optical depths ($\tau_{BB}$) are presented in Table~\ref{tab:calibrated-improved}. Following this correction, the average differences are reduced for some sightlines (see Table~\ref{tab:deviation-CSR}), resulting in an overall decrease in the average relative differences for all features (the $\Delta \tau_{BB}/\tau_{\text{r}}$ values are 0.22, 7.19, and 0.18 for the water ice, aliphatic hydrocarbon, and silicate features, respectively). If the cleaned data set is used, the average $\Delta \tau_{cr}/\tau_{\text{r}}$ is further reduced to 0.14 for silicate feature and 0.19 for water ice feature.

\textit{Contributions from other features:} We estimated the contribution of the water ice feature's extended wing ($\sim$ 18.75\% of its peak optical depth) to the optical depth of the aliphatic hydrocarbon feature ($\tau_{3.35}$) using the water ice feature model as described in Section~\ref{sec:SpectralModels}). Following this correction, the $\Delta \tau_{pr} / \tau_{r}$ value decreased from 8.05 to 4.11. When combined with continuum correction and the exclusion of the three largest outliers, the differences are further reduced, and $\Delta \tau_{pr} / \tau_{r}$ approaches approximately 2, although it still remained substantially larger than those of the water ice and silicate features.

Actually, improving the accuracy of the results requires more realistic spectral models of absorption and continuum profiles. However, a comprehensive modeling study is beyond the scope of this paper. Therefore, the possible improvement approximations discussed are presented as preliminary examples and are excluded in the subsequent stages of the analysis. In the following section, we examine how the applied calibrations and residual discrepancies affect the accuracy of the photometric optical depth maps.

\section{Mapping Applications} \label{sec:MappingApplications}

In this section, we explore the feasibility of generating optical depth maps using photometric measurements. We also examine how applied calibrations and residual discrepancies affect the spatial consistency and accuracy of the resulting maps.

For the mapping applications, we generated a \textit{synthetic FoV} (scene) using 15 sources from \cite{Gibb2004} that were studied in this work. We selected sources with a complete set of NIR $\&$ MIR spectra (see Section~\ref{sec:ModelingFoV}). We employed optical depths derived from photometric measurement simulations and reported optical depth values in the literature (see Table~\ref{tab:comparison-all}). To model a FoV, we reused each point multiple times to fill x-y grid (for details, see Section~\ref{sec:ModelingFoV} and Figure~\ref{fig:key-map}). The data were categorized by optical depth intensity into groups and arranged into sub-regions, forming an optical depth gradient (column density gradient) that increases from the upper right to the lower left corner. Then, by using linear interpolation between data points, photometric optical depth maps obtained.

\begin{figure*}[htp]
  \begin{center}
    \begin{tabular}{ccc}
      {\includegraphics[angle=0,scale=0.4]{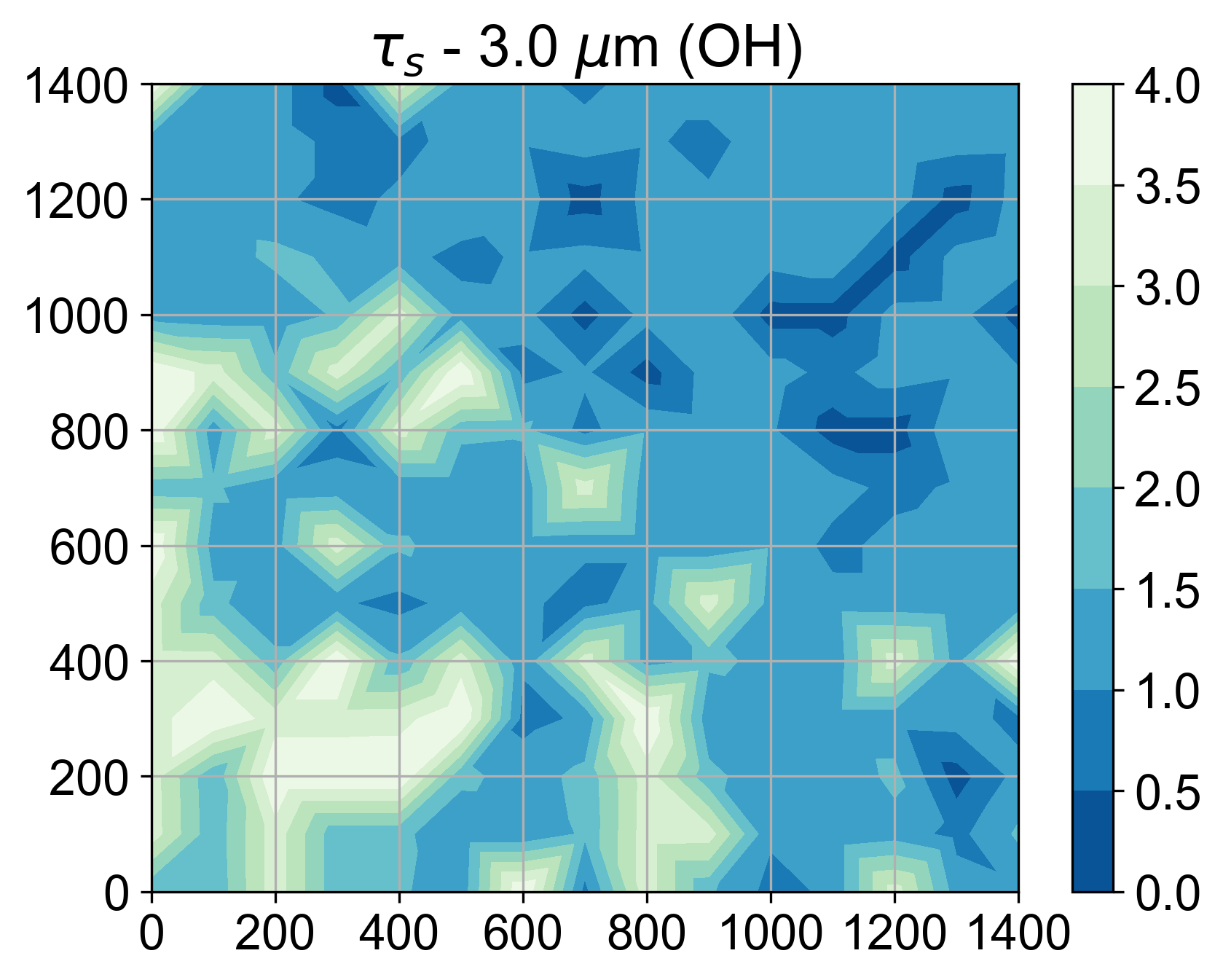}} &
      {\includegraphics[angle=0,scale=0.4]{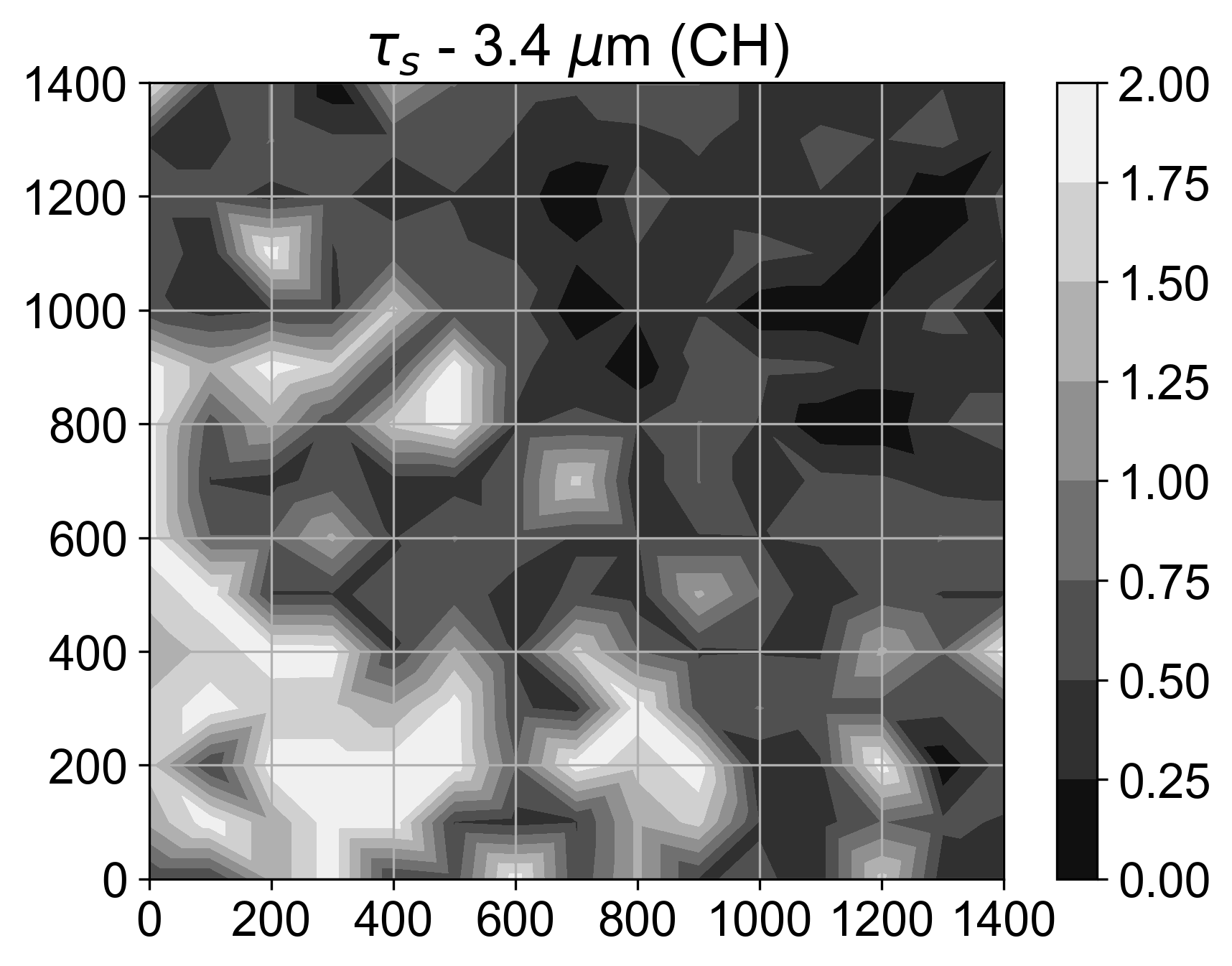}} &
      {\includegraphics[angle=0,scale=0.4]{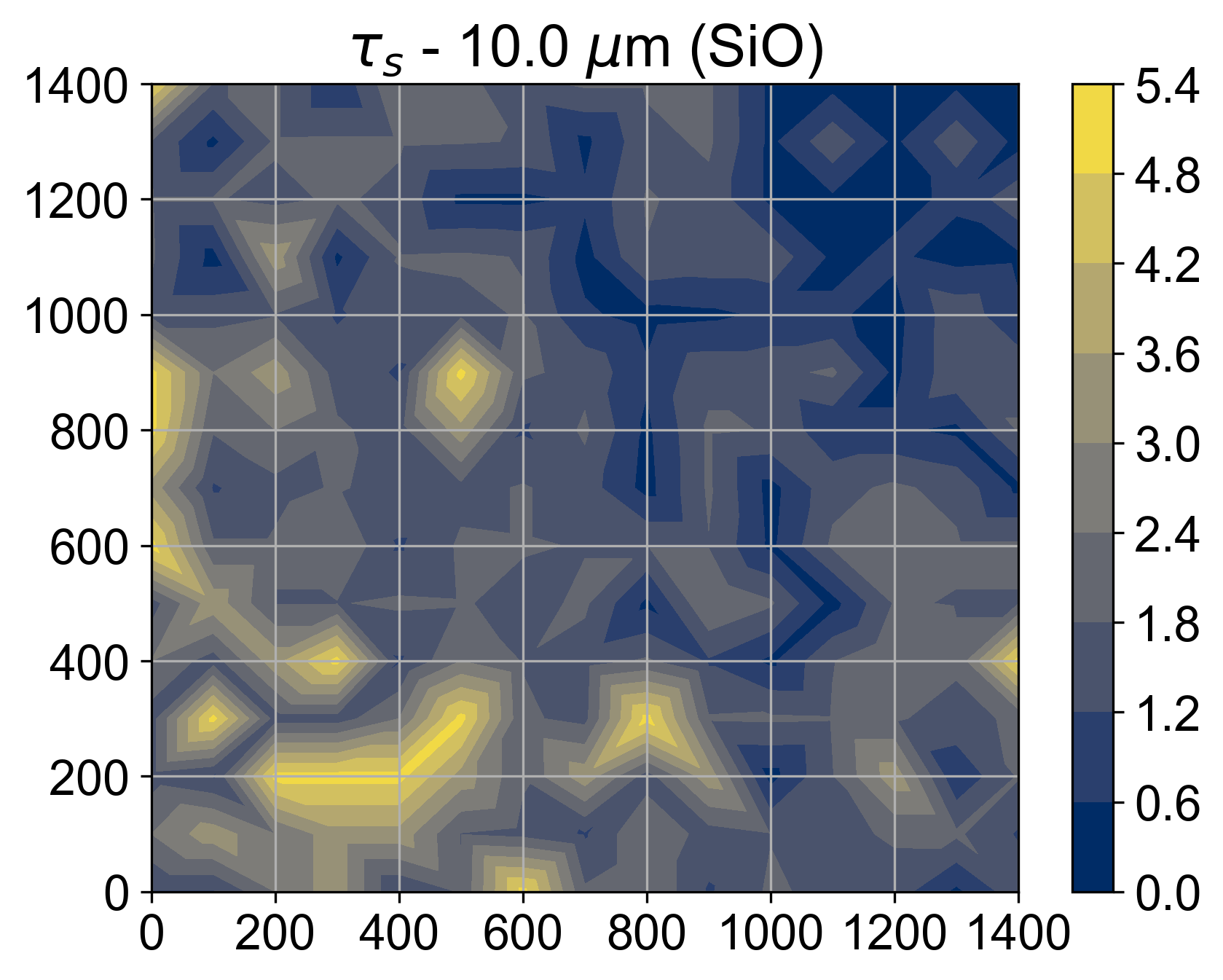}} \\
      {\includegraphics[angle=0,scale=0.4]{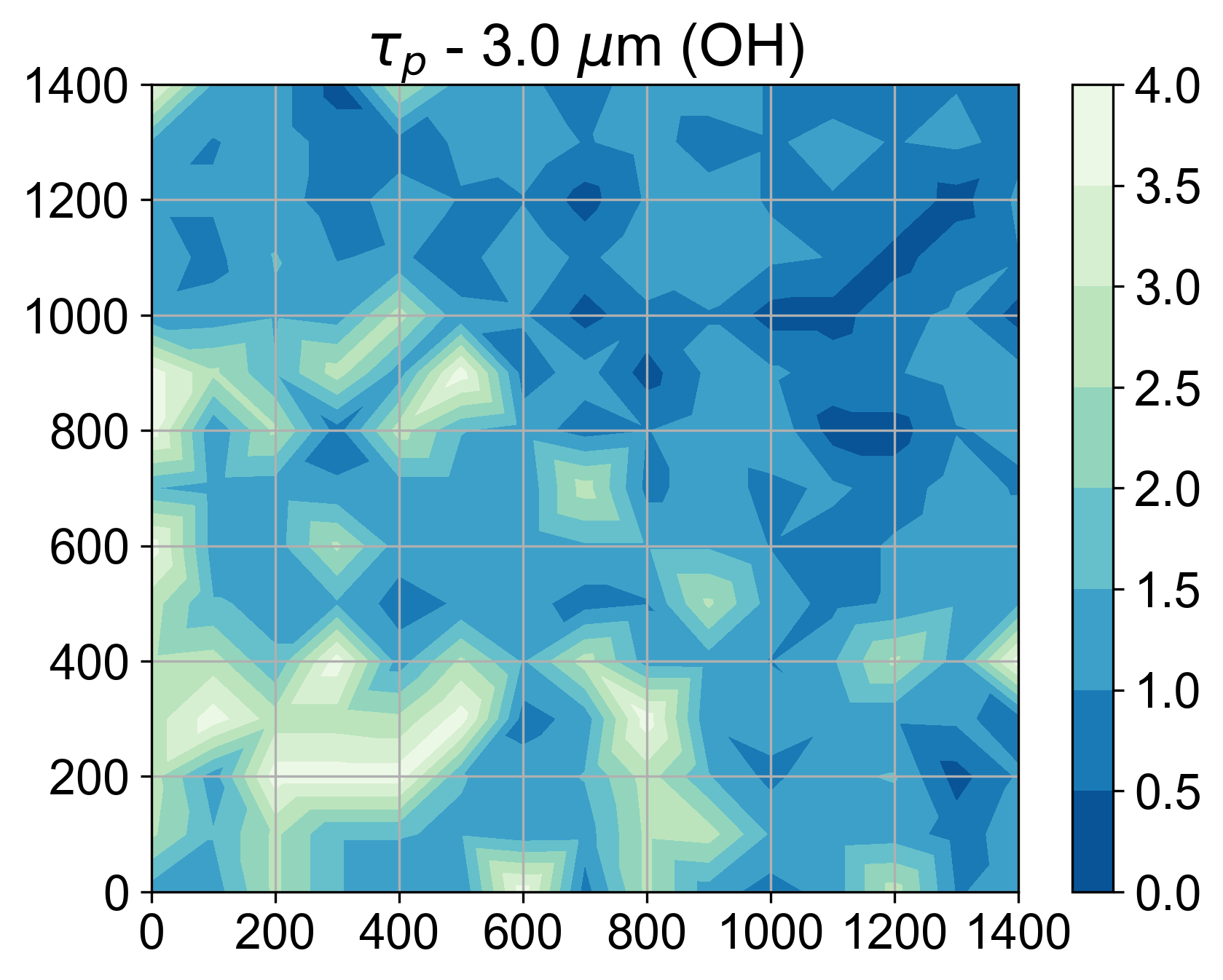}} &
      {\includegraphics[angle=0,scale=0.4]{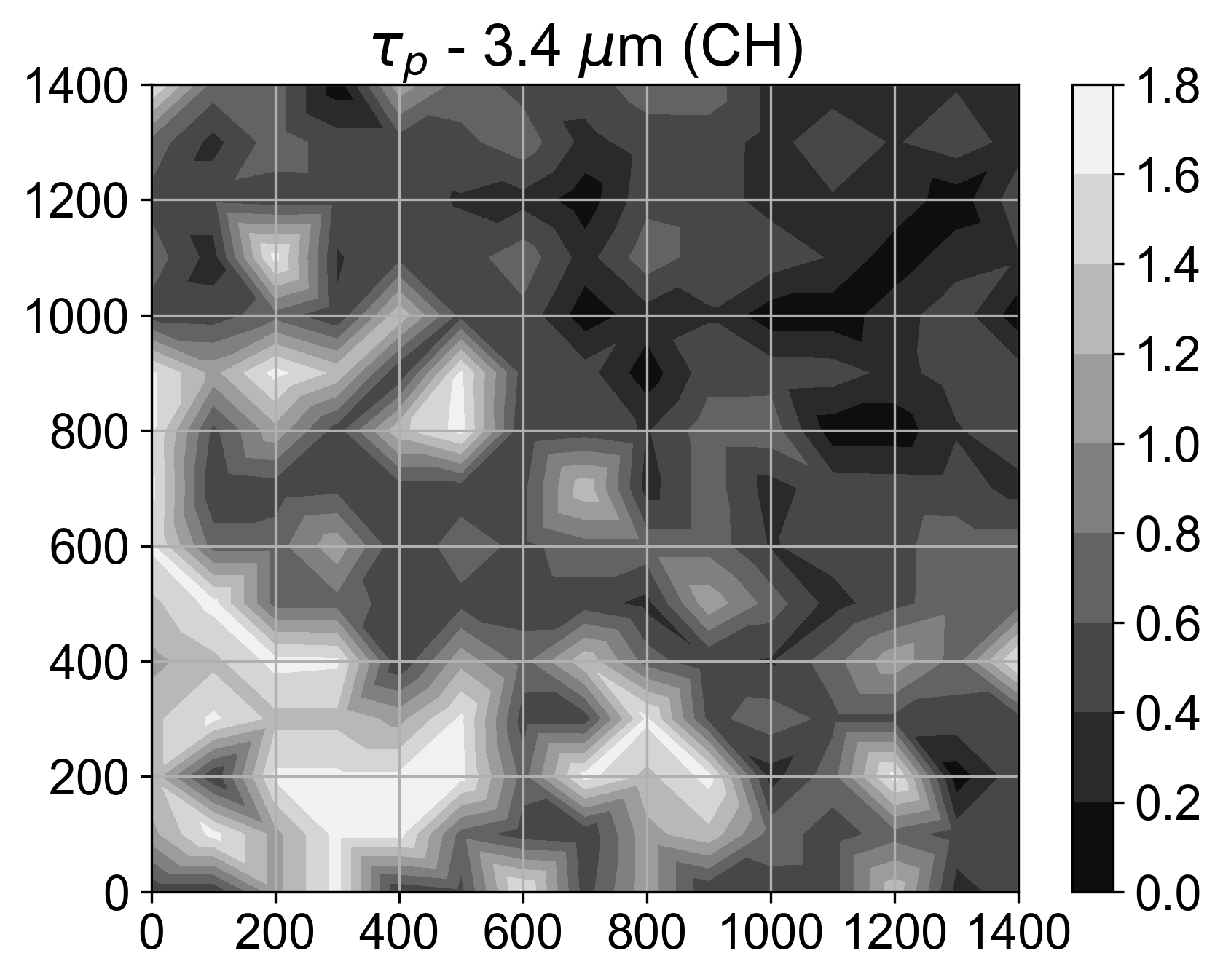}} &
      {\includegraphics[angle=0,scale=0.4]{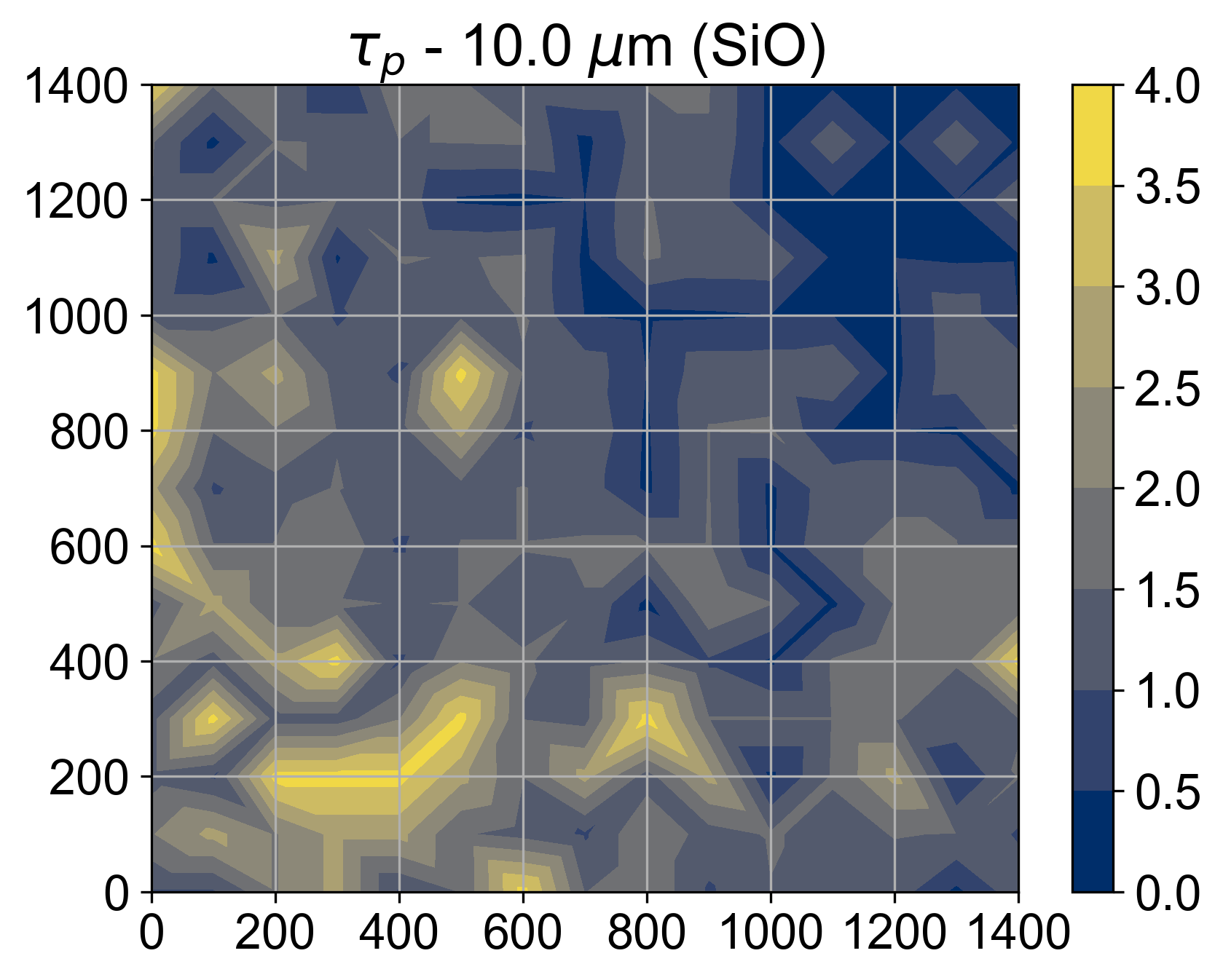}} \\
      {\includegraphics[angle=0,scale=0.4]{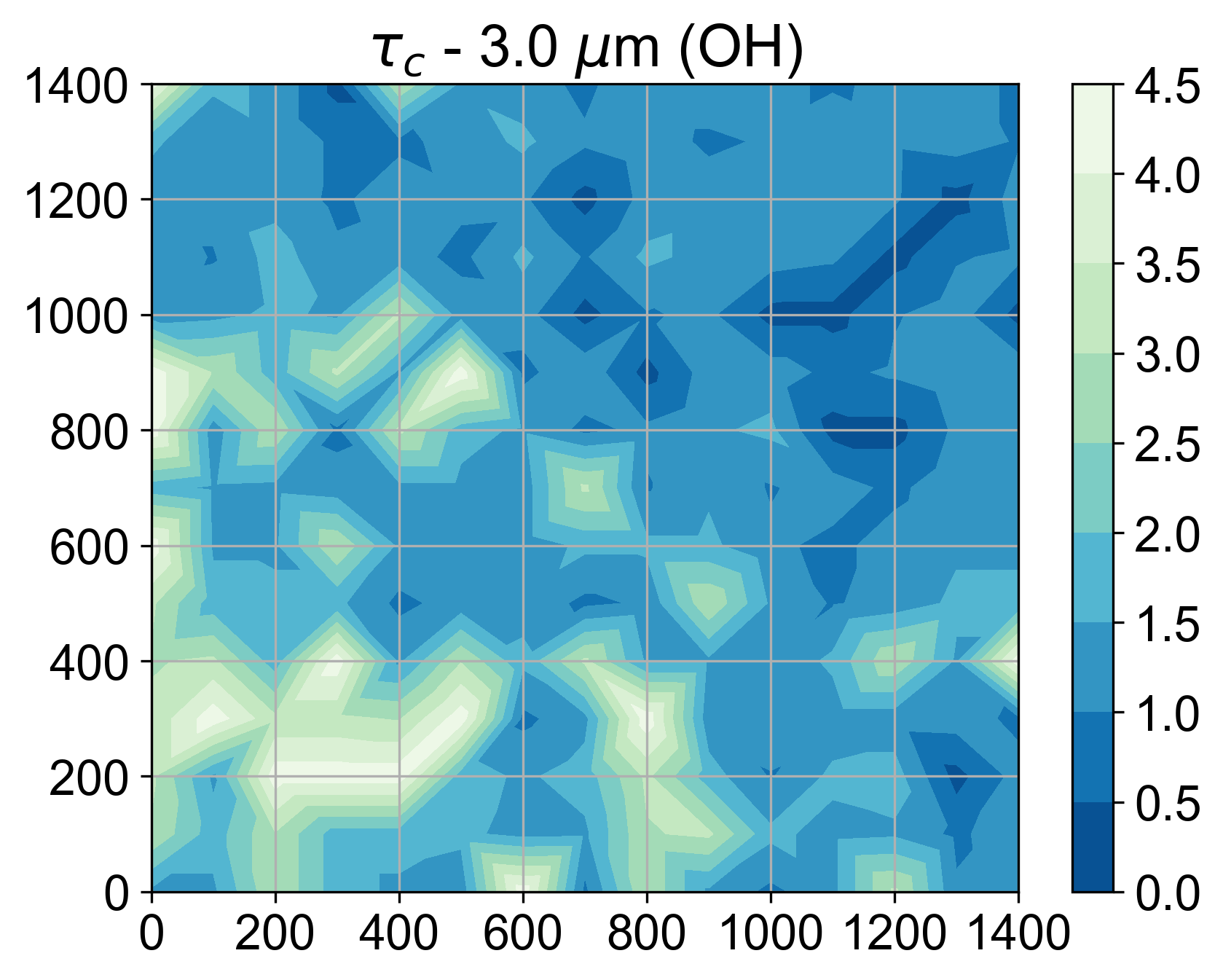}} &
      {\includegraphics[angle=0,scale=0.4]{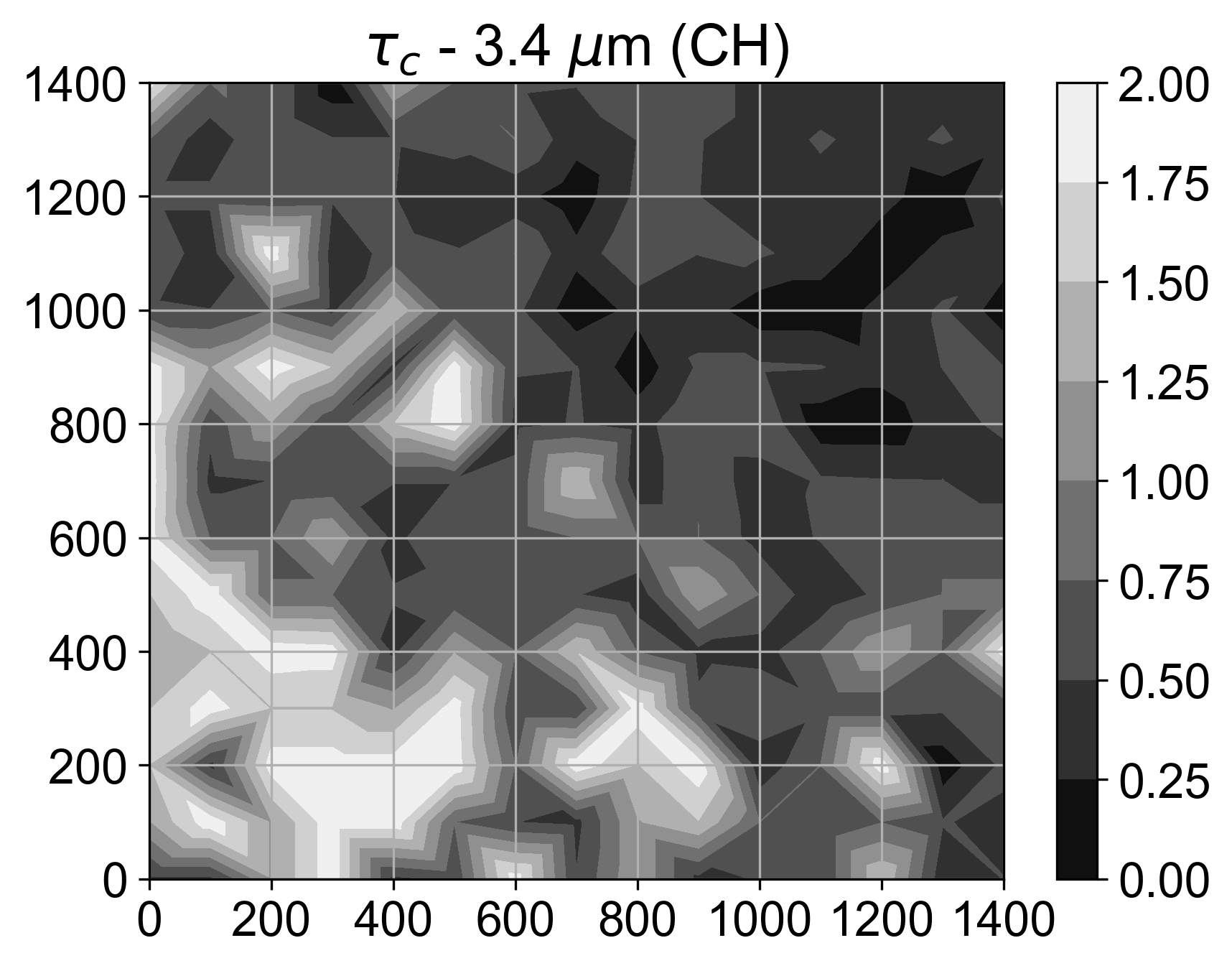}} &
      {\includegraphics[angle=0,scale=0.4]{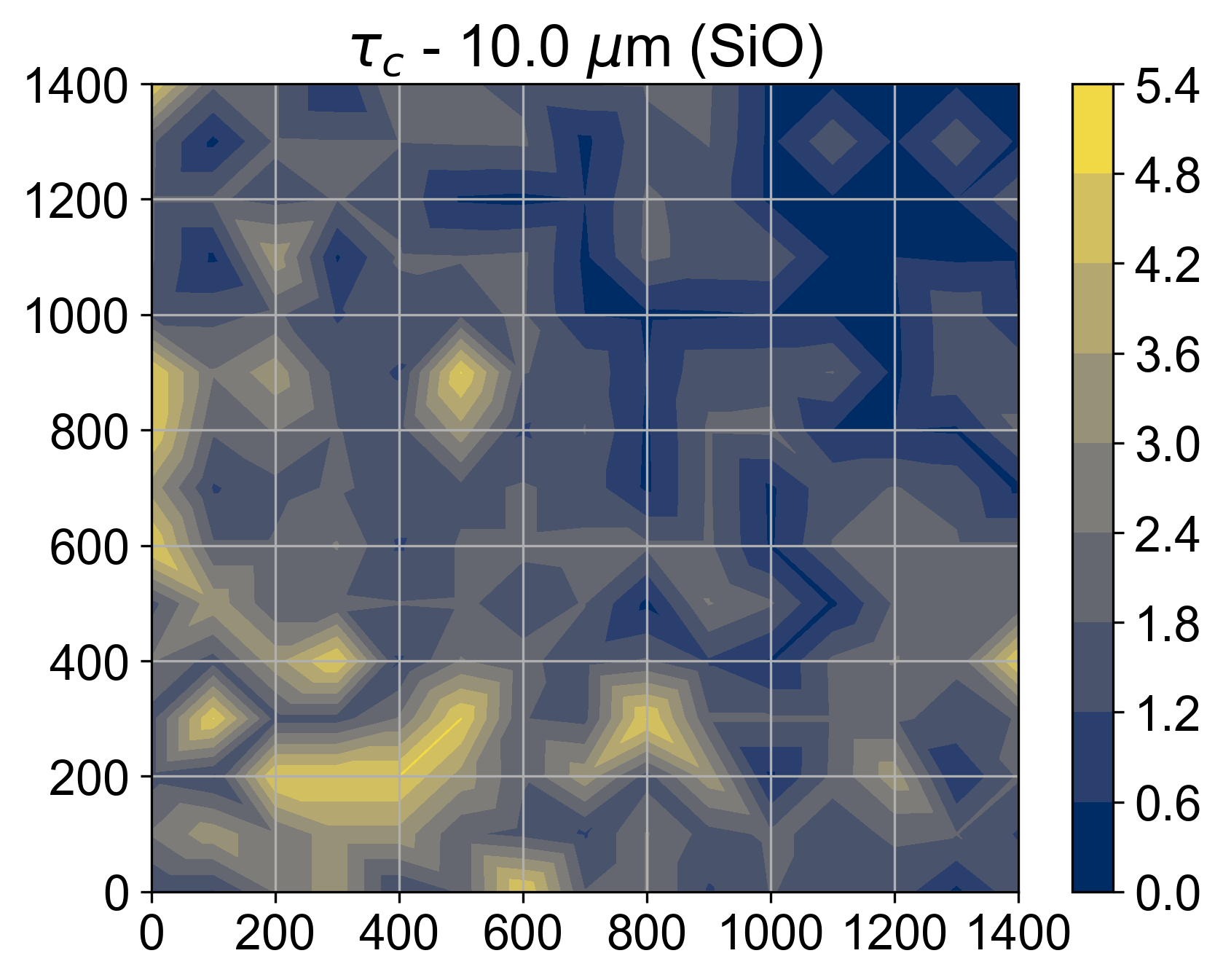}} \\
      {\includegraphics[angle=0,scale=0.4]{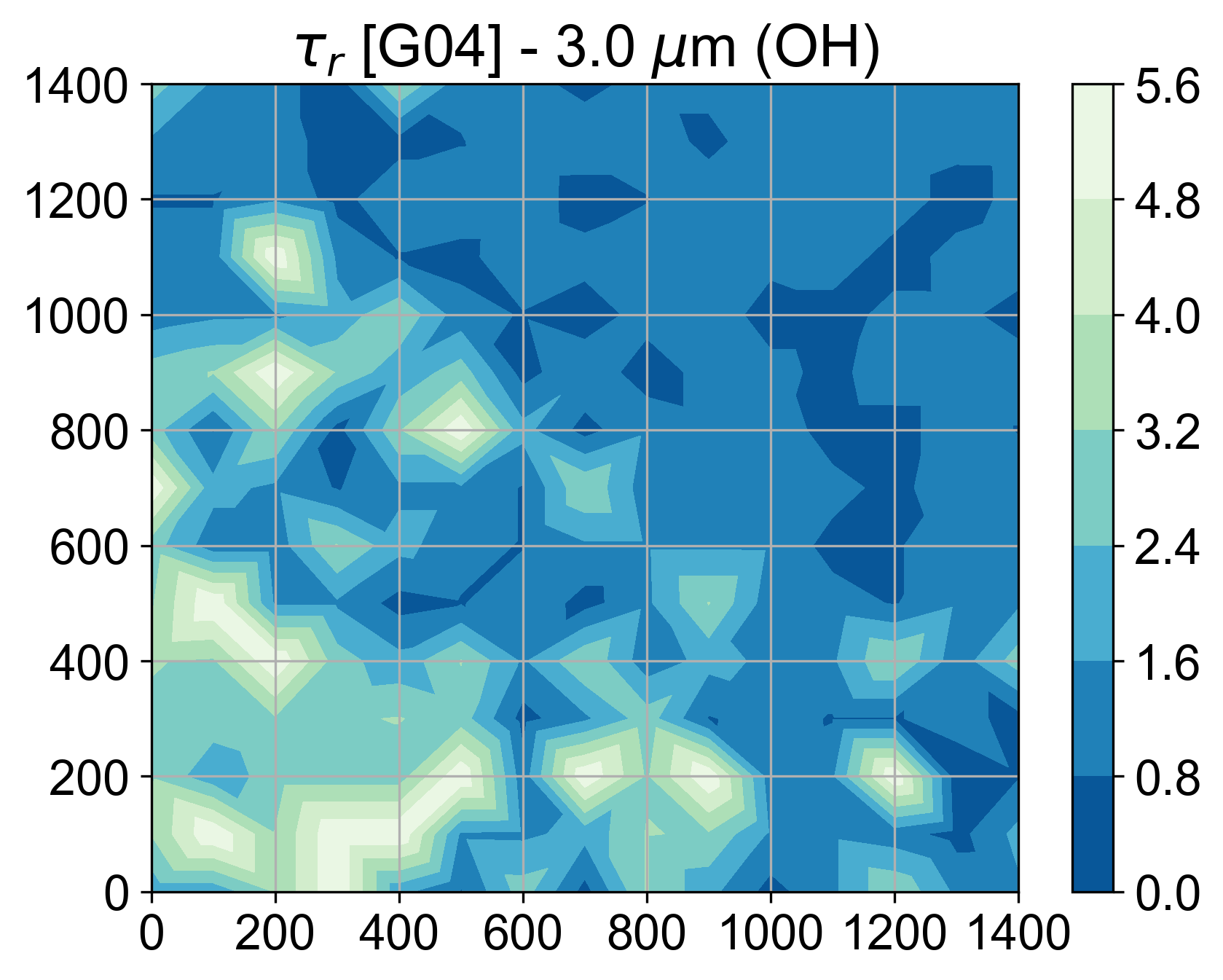}} &
      {\includegraphics[angle=0,scale=0.4]{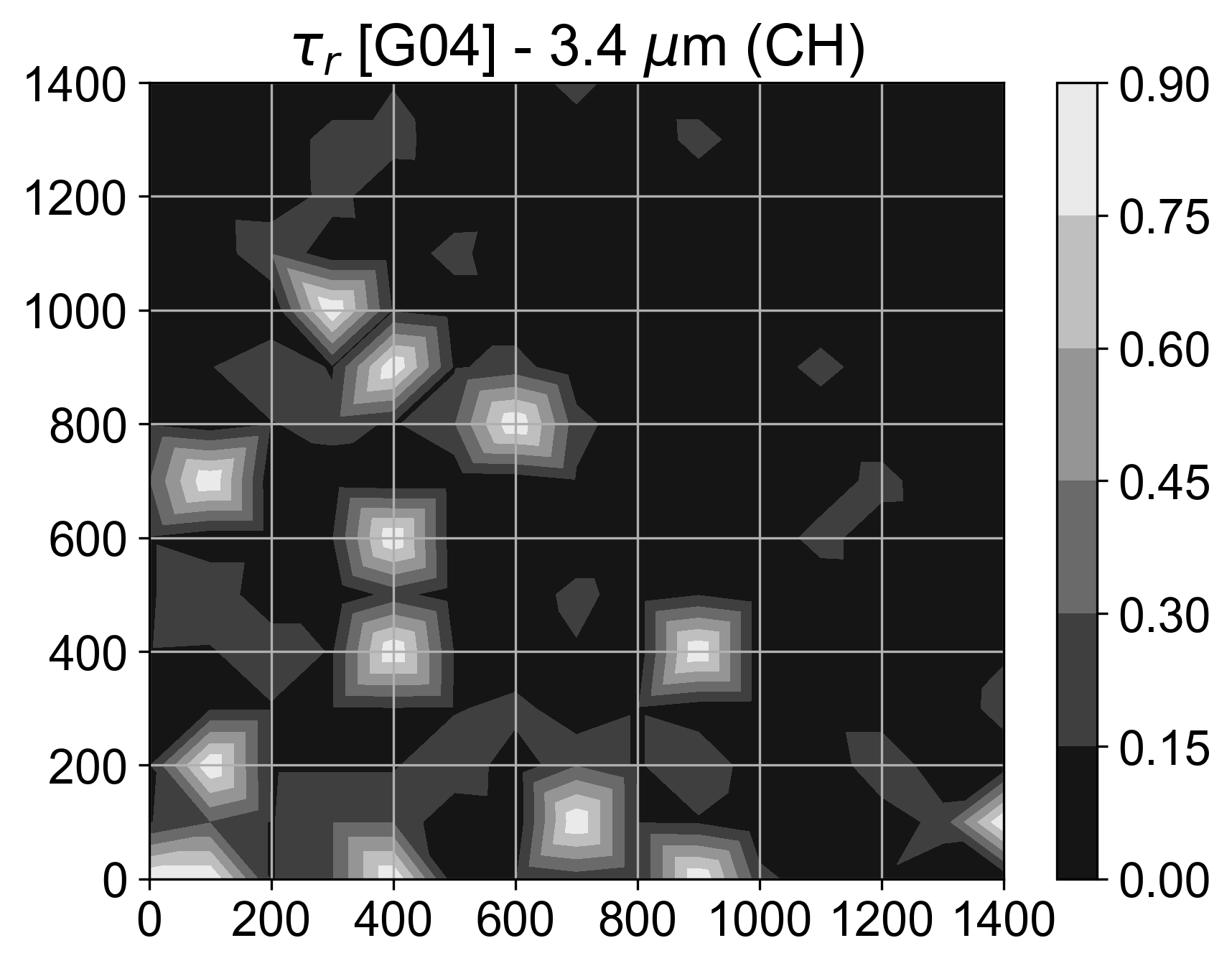}} &
      {\includegraphics[angle=0,scale=0.4]{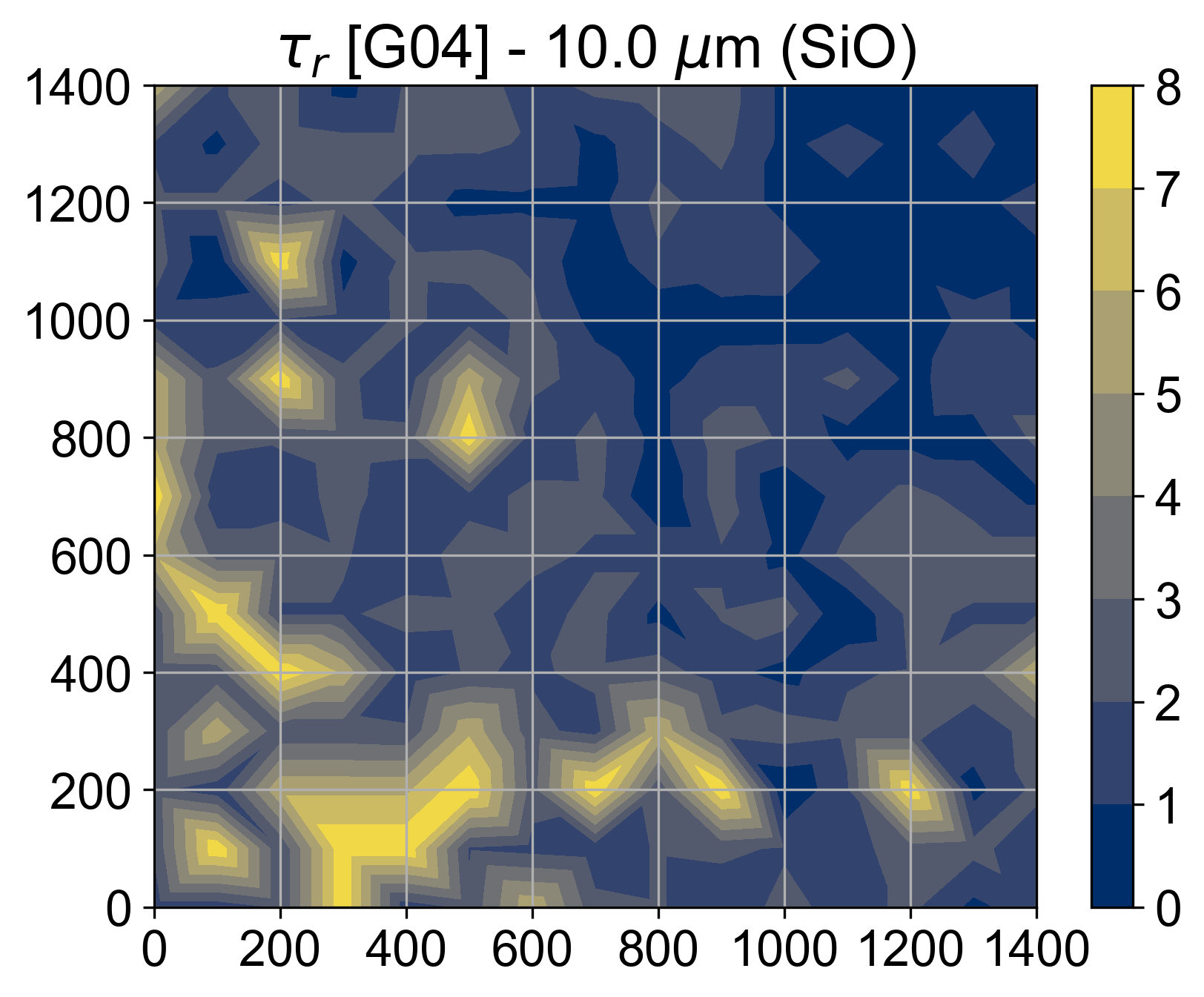}} \\
        \end{tabular}
     \caption{The maps illustrate the spatial gradient distribution of optical depths for water ice, aliphatic hydrocarbons, and silicates across \textit{the modeled FoV}. The first row shows spectroscopic optical depths ($\tau_{s}$), the second row presents photometric optical depths ($\tau_{p}$), and the third row displays calibrated photometric optical depths ($\tau_{c}$). The fourth row is constructed using reference optical depths ($\tau_{r}$) from the literature \citep{Gibb2004}. } 
    \label{fig:maps}
      \end{center}
\end{figure*}

The resultant optical depth maps of the 3.0--$\mu$m water ice $-$OH feature, the 3.4--$\mu$m aliphatic hydrocarbon $-$CH feature and the 10.0--$\mu$m silicate $-$SiO feature are presented in Figure~\ref{fig:maps}. The upper panels show optical depth maps generated via simulation of the spectroscopic method, the second row shows maps generated via simulation of the photometric method, the third row shows maps generated via calibrated optical depths of the photometric method, while the bottom maps are created directly using optical depth values reported in the literature \citep{Gibb2004}.

Figure~\ref{fig:maps} shows the gradual distribution of optical depths, representing a column density gradient from denser to less dense regions. We found similarities between the optical depth distribution patterns obtained from photometric and spectroscopic simulations and those reported in the literature. This suggests that, for a FoV with an adequate number density of background sources, photometric measurements with JWST are capable of tracing the bulk properties of ISM grains. Photometric optical depth maps can effectively reveal changes in the relative abundance distributions of water ice, aliphatic hydrocarbon, and silicate components.

The calibrated photometric and spectroscopic optical depth maps show agreement, although there is unavoidable scatter in the measurements due to intrinsic variations in each sightline. The maps obtained with values reported in the literature tend to align with similar optical depth levels for water ice feature and silicate feature, with exceptions for certain sightlines primarily resulting from data quality issues and different methodology. While the photometric method cannot fully disentangle the contribution of the water ice feature to the aliphatic hydrocarbon feature, it remains effective for tracing variation patterns of aliphatic hydrocarbons along dense sightlines in the ISM. The optical depths used in these maps are potentially subject to further improvement, as discussed in Section~\ref{sec:Calibrations}. Therefore, photometric measurements with JWST have the potential to reveal the distinct spatial abundance distributions of major dust components in the ISM.

\section{Summary and Discussion} \label{sec:SummaryandDiscussion}

In this study, we demonstrated that large-scale mapping of the optical depths of the 3.0--$\mu$m, 3.4--$\mu$m, and 10.0--$\mu$m absorption features can be achieved through photometric measurements using suitable NIRCam and MIRI filter combinations.

First, we explored the uncertainties arising from filter use and limited spectral information in three-band photometry by employing model spectra in Section~\ref{sec:TestingwithModelSpectra}. We analyzed the impact of potential variations in the continuum and absorption features separately for a range of reference optical depths (Figure~\ref{fig:Variations-Abs} and Figure~\ref{fig:Variations-Cont}). We found that the normalized differences ($\Delta \tau/\tau_{\text{0}}$) arising from the expected variations in spectra remain below \( \sim 15\% \). 

We then tested the accuracy of the photometric method using observational data in Section~\ref{sec:TestingwithObservationalData}. We compared the photometric optical depths with both the spectroscopic simulations and the reported reference values. We found that the optical depths derived from photometric and spectroscopic measurement simulations show a notable level of consistency, even prior to calibration. The average $\Delta \tau_{ps}/\tau_{\text{s}}$ values are 0.16, 0.14, and 0.18 for the water ice, aliphatic hydrocarbon, and silicate features, respectively (Table~\ref{tab:deviation-PS}). These values are interpreted as a methodological bias, primarily associated with the use of filters.

We found that the optical depths derived from photometric measurements and spectroscopic simulations employing a linear continuum fit differ significantly from those obtained using spectroscopic methods, especially for the aliphatic hydrocarbon feature. The average $\Delta \tau_{pr}/\tau_{\text{r}}$ values are 0.28, 7.56, and 0.33 for the water ice, aliphatic hydrocarbon, and silicate features, respectively (Table~\ref{tab:deviation-PSR}). The corresponding average $\Delta \tau_{sr}/\tau_{\text{s}}$ values are 0.24, 8.31, and 0.21 for the same features. This suggests that the discrepancies from the reported values are primarily due to missing spectral information, which limits both continuum estimation and the accurate characterization of absorption features, while the impact of filter usage appears to be relatively minor. 

We also showed that the photometric optical depths correlate more strongly with the total absorption measurements calculated using the linear fit approximation than with the reported values (Figure~\ref{fig:comparison-SPR}). This comparison reveals the critical importance of accurate continuum estimation and the need for further refinements to improve optical depth measurements, which are not fully feasible with three-band photometry due to its limited capacity to resolve detailed absorption profiles. 

Actually, complete consistency between the methods cannot be expected since spectroscopic studies using full spectral information and involve different steps that we were not able to apply in this study. While three-band photometry simplifies the spectral information, low-resolution spectra still yield reasonably accurate values for the water ice and silicate features, with uncertainties of approximately 30\%, even prior to calibration when compared to the reported values Table~\ref{tab:deviation-PSR}.

We investigated the methodological biases, employing model spectra and derived polynomial equations to correct method-related biases to calibrate photometric optical depths in Section~\ref{sec:Calibrations}. While model spectra require further improvement to match with more realistic spectral profiles (Figure~\ref{fig:calibration-models}), we have demonstrated that methodological biases arising from the use of filters can be mitigated using models (Figure~\ref{fig:od-calibratations}), and that the results can potentially be improved through further refinements.

The calibrated photometric optical depths ($\tau_{c}$) became highly consistent with those derived from spectroscopic simulations, especially for the silicate feature. The average $\Delta \tau_{cs}/\tau_{\text{s}}$ values decreased to 0.15, 0.11, and 0.05 for the water ice, aliphatic hydrocarbon, and silicate features, respectively. These residual differences in the NIR region primarily arise from spectral variations that cannot be represented by the uniform spectral models used for calibration.

Nevertheless, after calibration, the differences with the reported values remained significant, especially for the aliphatic hydrocarbon feature. The $\Delta \tau_{cr}/\tau_{\text{r}}$ values were 0.28, 8.05, and 0.22 for the water ice, aliphatic hydrocarbon, and silicate features, respectively (see Tables~\ref{tab:deviation-PSR} and \ref{tab:deviation-CSR}). If the problematic (saturated) data are excluded from the analysis (see Section~\ref{sec:TestingwithObservationalData}), the average $\Delta \tau_{cr}/\tau_{\text{r}}$ is reduced to 0.25 for the water ice feature and 0.17 for the silicate feature, based on the cleaned data set. This highlights the impact of observational data quality on the analysis.

Following the additional refinements described in Section~\ref{sec:Calibrations}, and using the cleaned data set, the average $\Delta \tau_{cr}/\tau_{\text{r}}$ is further reduced, reaching 0.14 for silicate feature and 0.19 for water ice feature, thereby indicating improved agreement with the reference values. For the aliphatic hydrocarbon feature, after applying combined refinements and excluding outliers, $\Delta \tau_{cr}/\tau_{\text{r}}$ of $\sim 2$ is approached. It should be noted that further improvements in accuracy are needed for the refinements as discussed in Section~\ref{sec:Calibrations}, which may be achieved by employing more realistic spectral models. As a comprehensive modeling analysis is beyond the scope of this study, the proposed improvements are presented as illustrative approximations.

To evaluate the impact of unavoidable sightline-specific residual deviations in the calibrated optical depths, we compared their influence on the resultant maps (Figure~\ref{fig:maps}) in Section~\ref{sec:MappingApplications} using a synthetic FoV. Despite the inclusion of problematic data, as well as inherent scatter in the measurements, the resulting photometric and spectroscopic maps are in agreement with some differences between the ones obtained with literature values. This suggests that photometric measurements of the 3.0--$\mu$m, 3.4--$\mu$m, and 10.0--$\mu$m features with JWST can trace changes in the relative spatial abundance patterns of water ice, aliphatic hydrocarbons, and silicates in the ISM.

In practice, larger differences may arise due to the combined influence of sightline-specific variations. However, statistical methods enable the identification and exclusion of outliers that reflect properties unrelated to the extended characteristics of the intervening medium.

In conclusion, after calibration, the photometric method can provide reasonable estimates of $-$OH abundances in water ice and $-$SiO abundances in silicates, yielding optical depth values close to the reported ones, with discrepancies of approximately 20--25\% for the water ice and 15--20\% for the silicate feature, depending on the data set used and the extent of further refinements. While the method exhibits limited effectiveness for $-$CH abundances in aliphatic hydrocarbons due to the masking effect of the water ice feature, improvements may be achieved using more realistic spectral models for dense sightlines to disentangle overlapping contributions. Additionally, more accurate measurements and improved abundance estimates are expected for translucent and diffuse sightlines, where the contribution of water ice is minimal. Overall, the uncertainty levels in the improved optical depths of the water ice and silicate features remain comparable to those observed in spectroscopic measurements based on different methodological approximations, while potential improvements for the aliphatic hydrocarbon feature offer promise for more representative measurements. 

JWST not only opens new windows for observation from space, but also provides exceptional measurement stability and sensitivity, thanks to its low thermal backgrounds, enabling precise infrared measurements that are challenging to achieve from the ground. Additionally, JWST allows for high-resolution mapping of crowded fields, reaching distant sources with the sensitivity required to detect faint signals. In this work we showed that, while JWST has not been optimally designed for this kind of experiment (the filters were not specified for the purpose of our application) it does provide a unique facility for quantification of column densities of $-$OH, $-$CH, and $-$SiO groups in the solid grains of the ISM. JWST enables the optical depths for all three features to be measured simultaneously for each field, allowing the analysis of relative variability in column densities and chemical abundances. 

The photometric method with JWST has the potential to be applied to various fields/objects in the ISM of our Galaxy as well as the ISM of nearby galaxies. Through a comparative analysis of optical depth maps with other tracers in the electromagnetic spectrum, we can shed light on the interplay between gas, ice, and dust in large fields and investigate how elemental abundances of carbon and oxygen vary under different conditions, thereby contributing to our understanding of the chemical evolution of the ISM.

\begin{acknowledgments} 
We would like to thank Dr. Adwin Boogert and Dr. Erika Gibb for providing the spectra set. BG would like to thank T\"{U}B\.{I}TAK 2219 International Postdoctoral Research Fellowship Programme and Ege University. MD acknowledges support from the Research Fellowship Program of the European Space Agency (ESA). We are grateful to the STScI Director’s Research Fund for partial support of this work.

\end{acknowledgments}

\software{Astropy \citep{astropy1:2013, astropy2:2018, astropy3:2022}, SciPy \citep{SciPy-NMeth2020}, NumPy \citep{NumPy-harris2020array}, Matplotlib \citep{Matplotlib-Hunter2007}, seaborn \citep{Seaborn-Waskom2021}}. BG acknowledges the use of the \textit{OpenAI GPT-4 model} (ChatGPT, \url{https://openai.com}) and Gemini for proofreading support.

\clearpage
\appendix

\section{Modeling Spectra} \label{sec:ModelingSpectra}

\subsection{Absorption Features} \label{sec:AbsorptionFeatures}

Using analytical equations with a range of parameters, we modeled absorption features. We used a range of central wavelengths (CWs) and widths (FWHMs) as parameters (see Table~\ref{tab:model-parameter-ranges}) to obtain absorption profile templates using the Gaussian function (Equation~\ref{eq:gaussian}) presented below ($A$: the amplitude is optical depth, $\mu$: CW ($\mu$m), $\sigma$: the standard deviation is a measure of width ($\mu$m), and \(\text{FWHM} = 2 \sqrt{2 \ln(2)} \times \sigma\) $\approx$ 2.355 $\times$ $\sigma$). 

\begin{equation}
\label{eq:gaussian}
\tau = A \times \exp\left(-\frac{(x - \mu)^2}{2\sigma^2}\right) 
\end{equation}

For the standard spectral models, we preferred to set CWs of absorption features at the \(\lambda_{0}\) of filters (3.00--$\mu$m, 3.35--$\mu$m and 10.00--$\mu$m) and adopted FWHM values based on the average of reported values in the literature. Then, we altered the CWs and FWHMs to study the effects of variations in the absorption features. For this we used frequently reported CW and FWHM values in the literature, as we describe below. We present the model parameters we used to model the spectra in Table~\ref{tab:model-parameter-ranges}. 

\textit{Water Ice Feature}: The water ice feature in the ISM can vary based on factors such as temperature, morphology, structure and size, besides impurities. In general, the CW of the water ice absorption feature is around 2.95--3.05 $\mu$m and the FWHM is commonly reported to be in the range of approximately 0.36--0.43 $\mu$m (for details, see \citealt{Maldoni1998, Chiar2002, Gibb2004, Dartois2004, Dartois2024, Bergner2024, Rocha2024}).

\textit{Aliphatic Hydrocarbon Feature}: The aliphatic hydrocarbon feature can vary depending on the types of hydrocarbon groups and their ratios (such as CH$_{2}$/CH$_{3}$). The CW of the aliphatic hydrocarbon absorption feature complex (including sub-features) is around 3.35--3.45 $\mu$m and the FWHM is typically reported to be around 0.14--0.17 $\mu$m (for details, see \citealt{Mennella2002, Chiar2002, Dartois2004, Gibb2004, Moultaka2004, Godard2012, Chiar2013, Gadallah2015, Gunay2018, Gerber2025}).

\textit{Silicate Feature}: The silicate feature can vary depending on the specific type of silicate minerals (such as pyroxene and olivine) and physical properties of grains (in particular morphology, such as crystalline or amorphous). Generally, the CW of the silicate absorption feature is around 9.7--10.0--$\mu$m and the FWHM is typically reported to be in the range of approximately 1.50--3.20 $\mu$m  (for details, see \citealt{Jaeger1994, Dorschner1995, Fabian2001, Gibb2004, Kemper2004, Min2007, Speck2011, Fogerty2016, Shao2024, Gordon2021, Decleir2025}).

\subsection{Continuum} \label{sec:Continuum}
\begin{figure*}[htbp]
  \begin{center}
    \begin{tabular}{c}
        {\includegraphics[angle=0,scale=0.45]{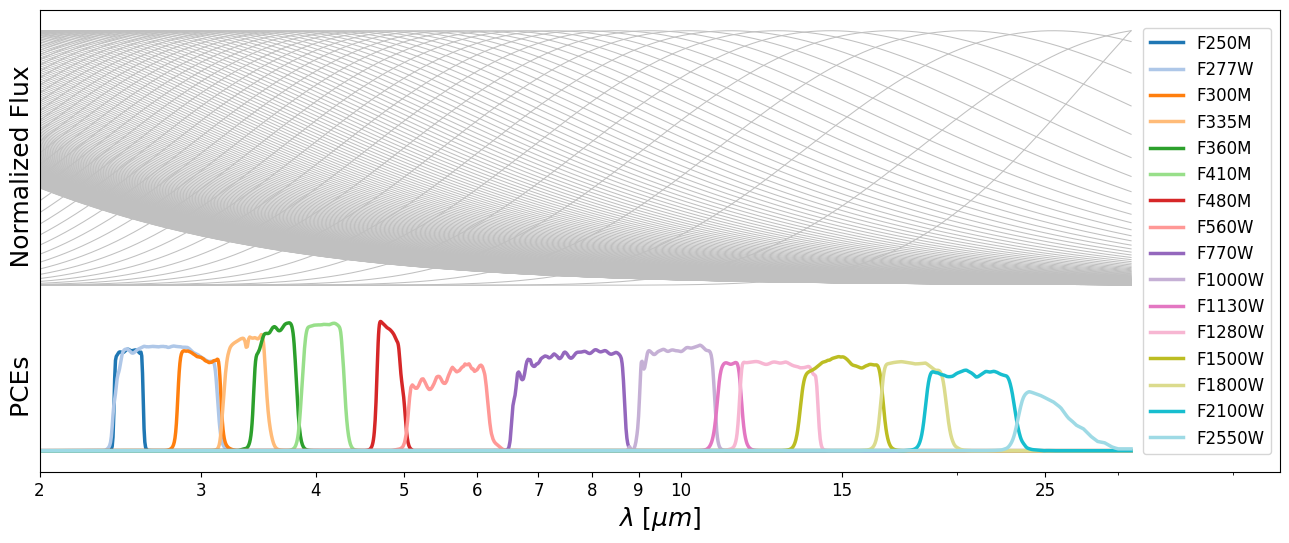}} 
        \end{tabular}
    \caption{The set of continuum profile templates: \textit{normalized blackbody curves} (BBs) with temperatures ranging from $T = 100$--$10000\,\text{K}$, used for continuum modeling, is illustrated. The throughput functions of the NIRCam and MIRI filters are also presented with the color code at the bottom.}
    \label{fig:BB}
      \end{center}
\end{figure*}

The continuum profiles of spectra in the IR region can primarily vary depending on the thermal radiance of background stellar sources \citep{Rieke2008, Gordon2022Calibration}. For dense sightlines, observable background sources are relatively cool IR-bright stars at various evolutionary stages (from YSOs to AGBs), whose continua peak at longer wavelengths (Wien's displacement law), causing the IR absorption features to fall within the slope of the spectrum (e.g., \citealt{Chiar2002}; \citealt{Moultaka2004}; \citealt{Gibb2004}), in contrast to diffuse sightlines where the IR absorption features typically fall within the flatter Rayleigh-Jeans tail. In the presence of shells, envelopes, and disks (e.g., \citealt{Evans2003, Crapsi2008, Robitaille2017, Richardson2024}) continuum profiles are shaped by circumstellar dust emission (dependent on the temperature gradient of grains), which enhances long-wavelength flux and extinction (dependent on grain size distribution), which reduces short-wavelength flux. As a result, the slope of the spectral shape further deviates from that of a blackbody: it can become steeper or relatively flattened across the wavelength region of absorption features, which can be described using the slope of the spectra (or spectral index, such as \citealt{Evans2009}). We employed analytical functions to model continuum with a slope range. We generated a range of continuum profiles using \textit{normalized blackbody curves} (BBs) spanning $T = 100$--$10000\,\text{K}$ (Figure~\ref{fig:BB}) to represent a variety of conditions, including extreme cases where the continuum is dominated primarily dominated hot stellar radiance and by cool dust radiance. Additionally, to generate more realistic continuum profiles affected by dust, we derived templates by fitting polynomial functions to observational spectra set \citep{Gibb2004}. We also used a flat continuum as an ideal case for comparison.

\section{Modeling Field of View} \label{sec:ModelingFoV}

\begin{figure*}[h]
  \begin{center}
    \begin{tabular}{ccc}
      {\includegraphics[angle=0,scale=0.4]{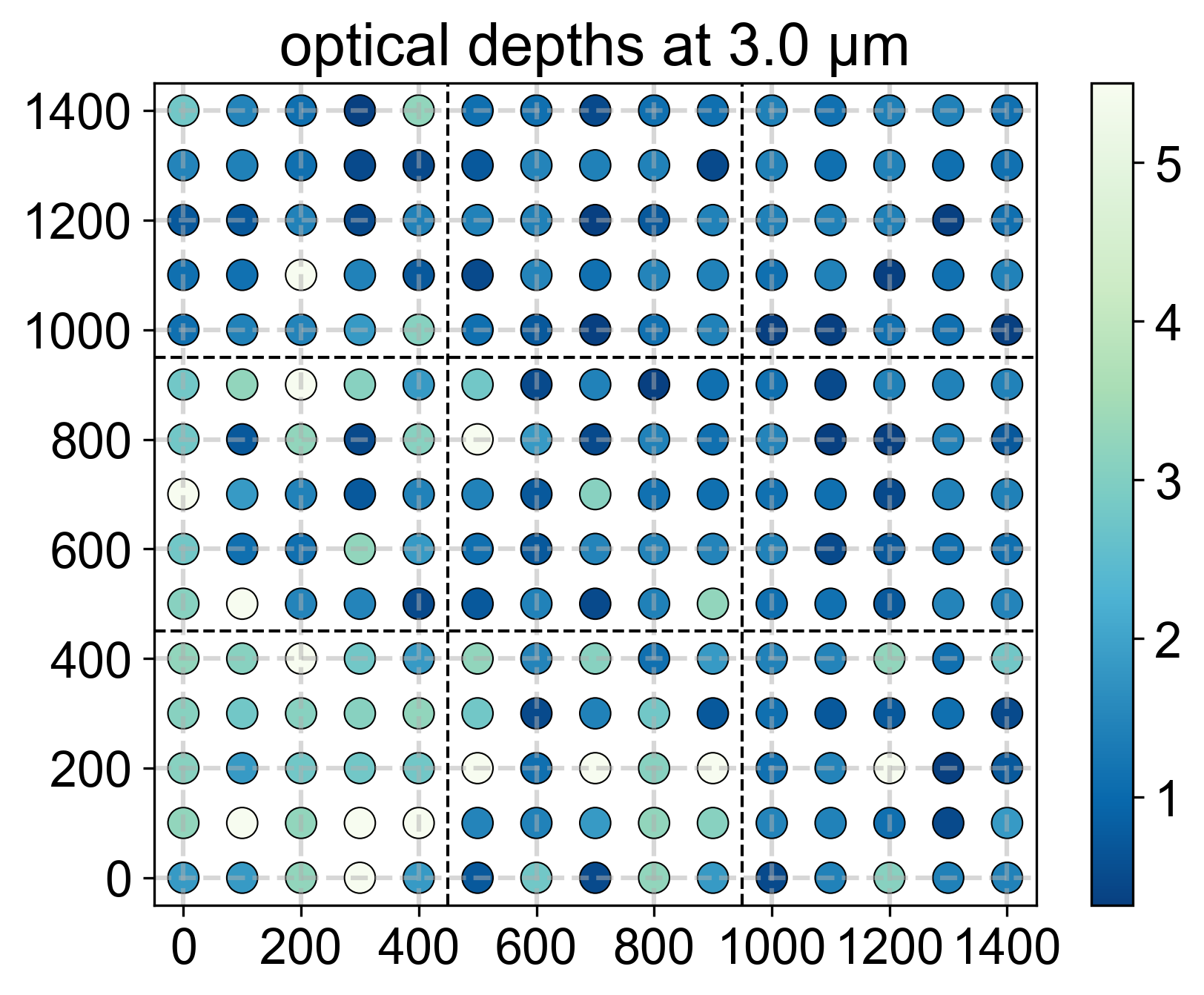}} &
      {\includegraphics[angle=0,scale=0.4]{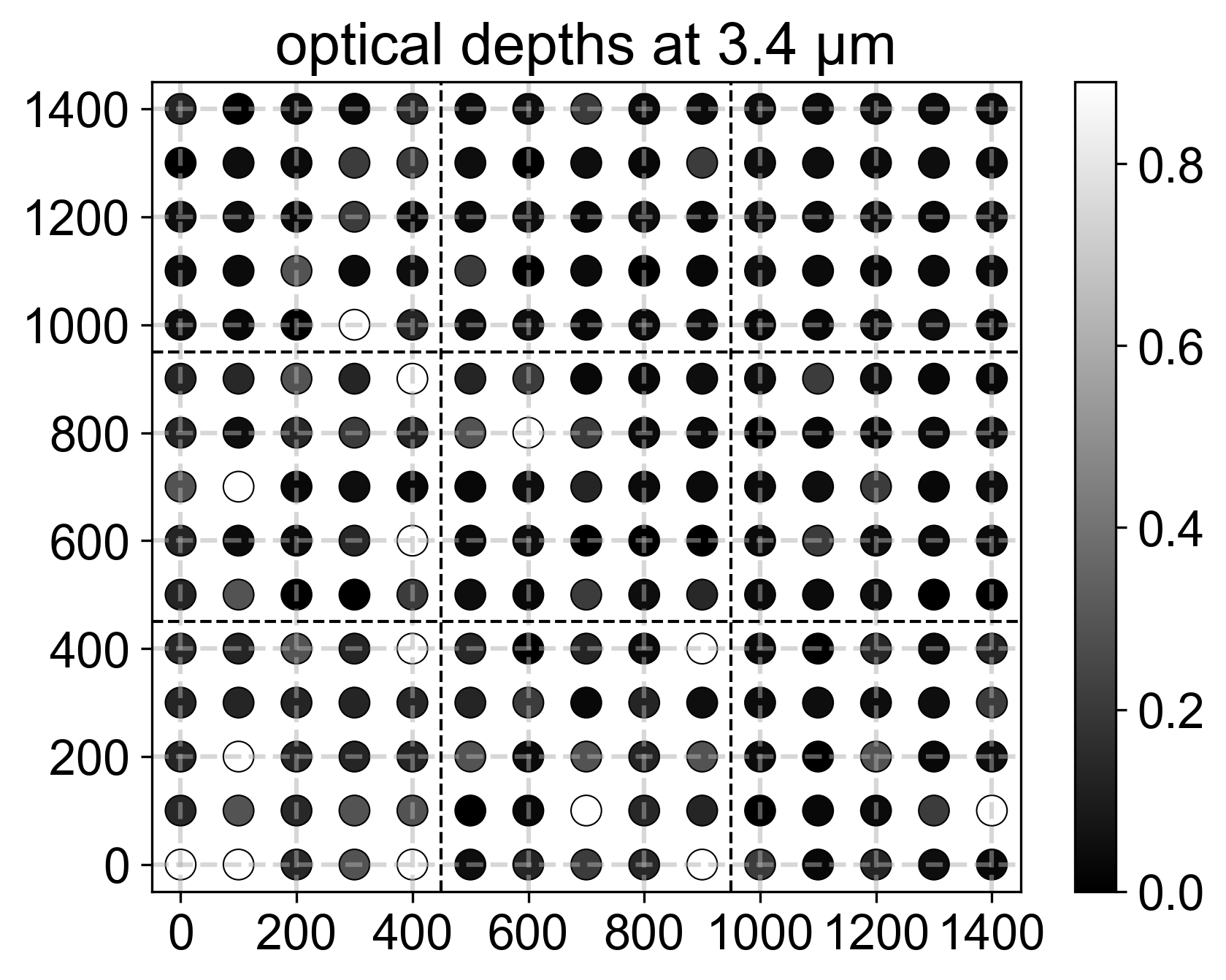}} &
      {\includegraphics[angle=0,scale=0.4]{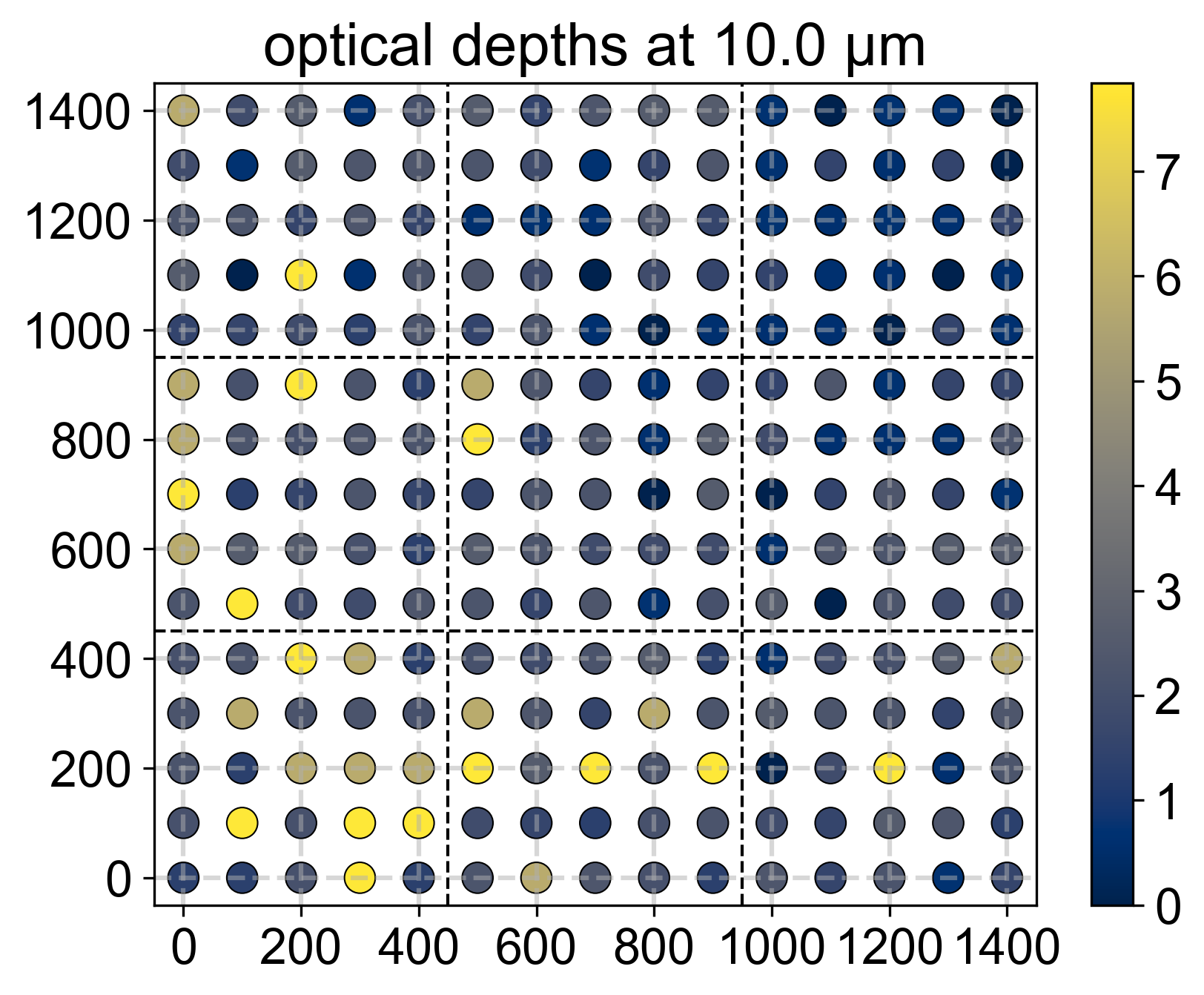}} \\
        \end{tabular}
    \caption{The maps show coordinates for each data distribution of the 3.0-$\mu$m water ice, the 3.4-$\mu$m aliphatic hydrocarbon, and the 10.0-$\mu$m silicate optical depth values in the scene (modeled FoV). The color of the dots (as indicated by the color bar) represents the optical depth values reported in the literature \citep{Gibb2004}. The sub-regions (tiles) are also indicated with dotted lines. }   
    \label{fig:key-map}
      \end{center}
\end{figure*}

To simulate a FoV, the data points were assigned coordinates evenly on an x-y grid spanning from 0 to 1400 elements (detector elements: pixels). The data were categorized into three groups based on the average optical depth intensity of the three features: Group-A (W33 A, W3 IRS 5, AFGL 2136, NGC 7538 IRS 9, Elias 29), Group-B (Mon R2 IRS 3, Orion IrC2, Orion BN, AFGL 2591, Sgr A$^{*}$), and Group-C (S140, AFGL 989, R CrA IRS 1, R CrA IRS 2, AFGL 490). Then, these groups were repeatedly used with gradual mixing to form hybrids, resulting in a total of five optical depth groups: Group-1: AAAAA, Group-2: AAABBB, Group-3: ABBBC, Group-4: BBBCC, and Group-5: CCCCC, each containing 25 data points (5$\times$5). Then we divided FoV into 3$\times$3 = 9 sub-region, each spanning 500 elements (x1:0-400, x2:500-900, x3:1000-1400 and y1:0-400, y2:500-900, y3:1000-1400) to form 9 tiles (x1y1, x1y2, x2y1, x1y3, x2y2, x3y1, x2y3, x3y2, x3y3). The data points were then \textit{randomly} assigned to the coordinates of \textit{each tile} (Group-1$\to$x1y1, Group-2$\to$x1y2, Group-2$\to$x2y1, Group-3$\to$x1y3, Group-3$\to$x2y2, Group-3$\to$x3y1, Group-4$\to$x2y3, Group-4$\to$x3y2, Group-5$\to$x3y3) to generate 3$\times$3 = 9 distinct sub-region scenes, which were subsequently used to generate an optical depth gradient (column density gradient) in the FoV that increases from the upper right to the lower left corner (Figure~\ref{fig:key-map}). The resultant frame has 225 source in 1500$\times$1500 element size frame with a resolution of 100$\times$100, comprising 15$\times$15 data points, which matches the number density of sources in previous studies ($\sim$ 200 sources in a frame with coordinates spanning from 0 to 1400 in pixel units. (FoV = 167 arcsec), (see \citealt{Gunay2020, Gunay2022}).

\section{Supporting Figures and Tables} \label{sec:SupportingFiguresandTables}

\begin{figure*}[htbp]
\begin{center}
\begin{tabular}{ccc}
    {\includegraphics[angle=0,scale=0.2]{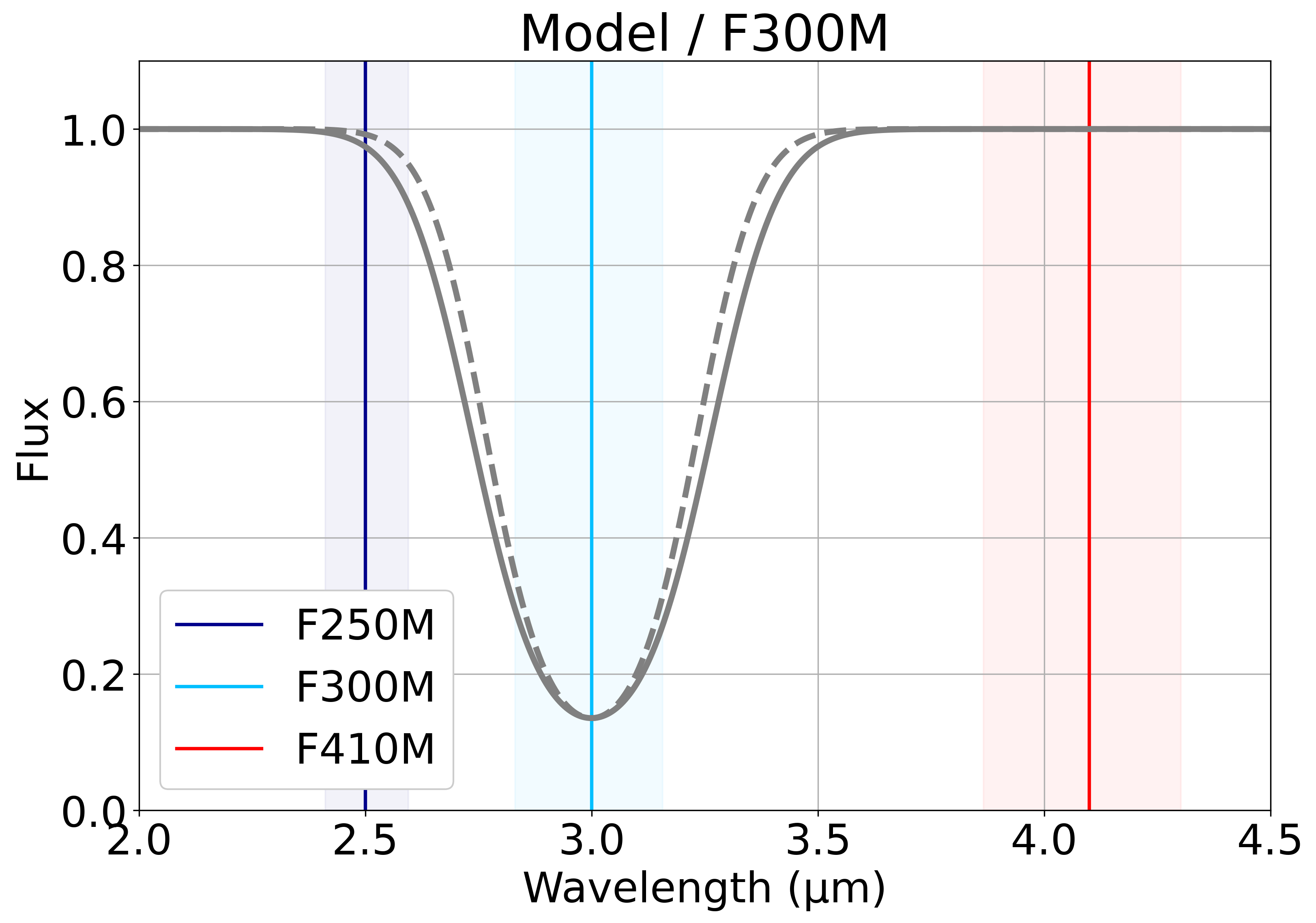}} &
    {\includegraphics[angle=0,scale=0.2]{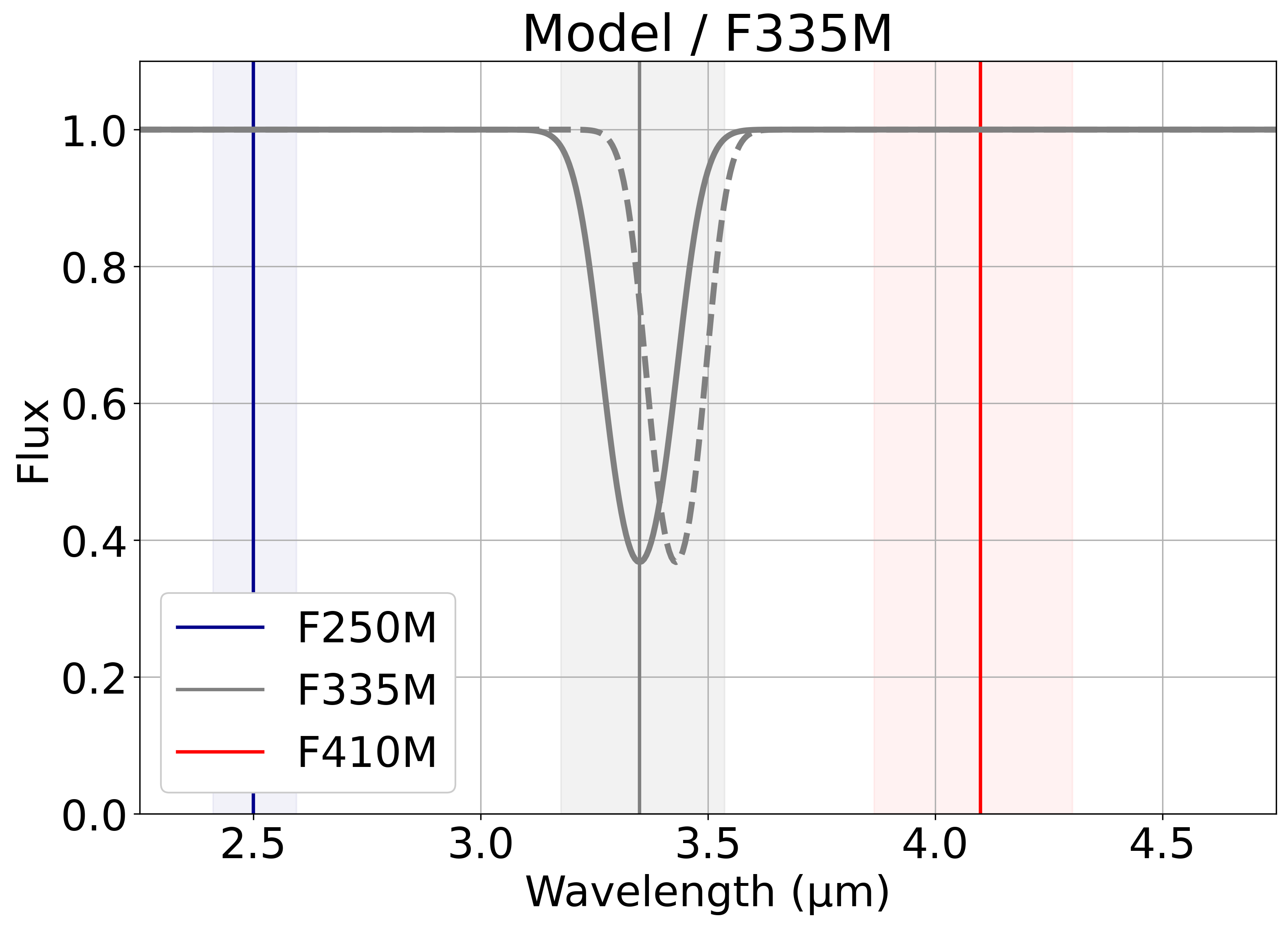}} &
    {\includegraphics[angle=0,scale=0.2]{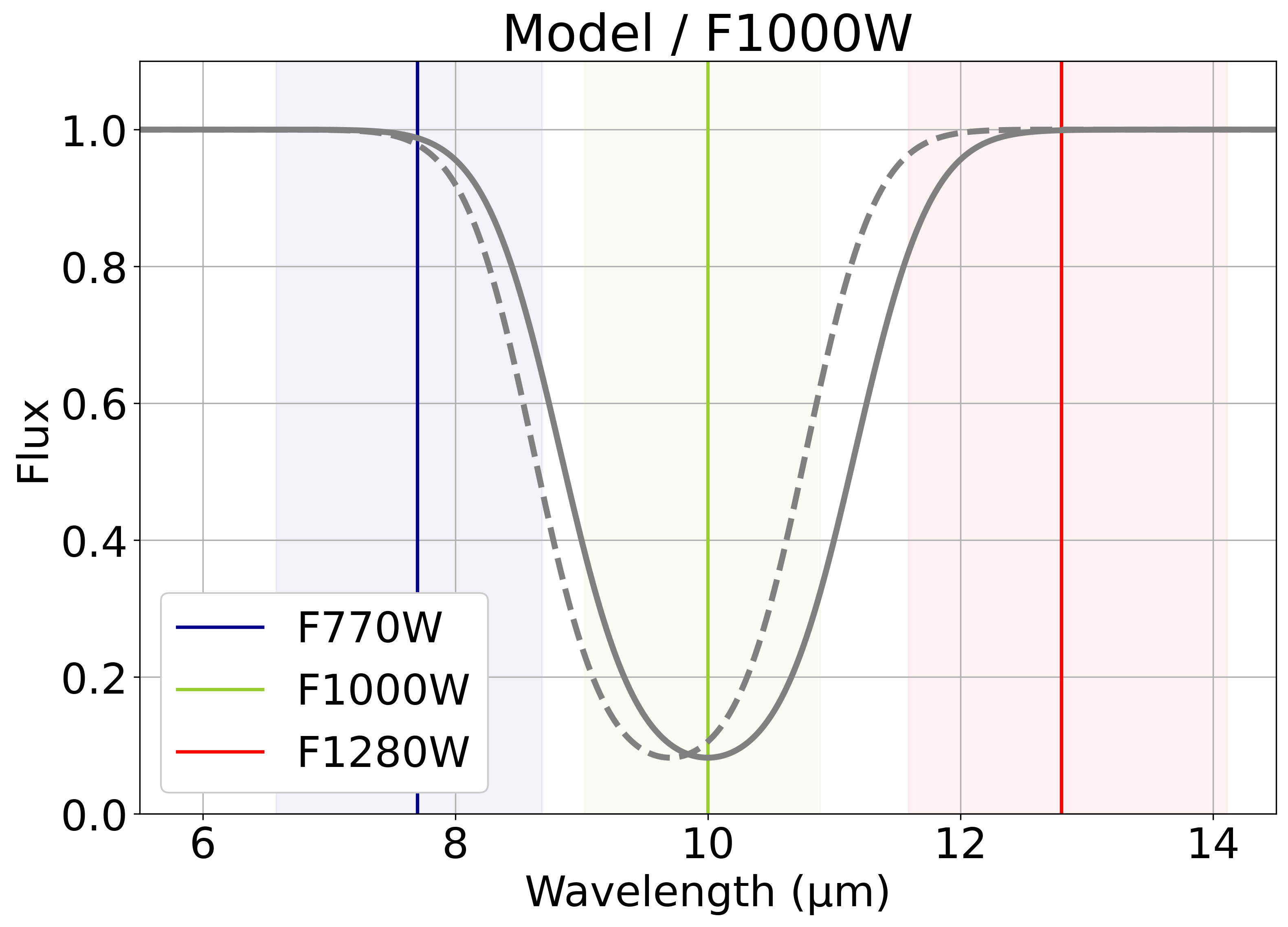}}\\
\end{tabular}
\caption{The absorption feature templates at 3.0--$\mu$m, 3.35--$\mu$m, and 10.0--$\mu$m used in the \textit{standard model} spectra (gray curves) are shown alongside those in the \textit{optimal models} (gray dashed curves), which were obtained by adjusting the CW and FWHM parameters to achieve optimal calibration for the observational spectra set from \citet{Gibb2004}. The central wavelengths (vertical lines) and the half power pass-band wavelengths of the filters are also indicated. 
 }    
\label{fig:calibration-models}
\end{center}
\end{figure*}

\begin{deluxetable*}{c|cc|cc|cc|cc}[htbp]
\tabletypesize{\small} 
\tablecaption{List of the parameters used for the \textit{Spectral Models}. The CW ($\mu$m) and FWHM (2.355 $\times$ $\sigma$) ranges and values for the analytical functions (Gaussian functions) used for absorption features are presented, along with the types of continua. The Standard Models serve as references, the Exploratory Models are used to explore uncertainties in photometric optical depths arising from variations in the spectra, the Test Models are employed to investigate the optimal models that represent each absorption feature in the spectra set from \cite{Gibb2004}, and the Optimal Models are selected to calibrate the photometric optical depths derived from the spectra set ([G04]).}
\label{tab:model-parameter-ranges}
\tablehead{
\colhead{} & \multicolumn{2}{c}{Standard Models} & \multicolumn{2}{c}{Exploratory Models} & \multicolumn{2}{c}{Test Models} & \multicolumn{2}{c}{Optimal Models ([G04])} \\
\cline{2-9} 
\colhead{} & \colhead{CW} & \colhead{FWHM} & \colhead{CW} & \colhead{FWHM} & \colhead{CW } & \colhead{FWHM} & \colhead{CWs} & \colhead{FWHMs }
}
\startdata
$-$OH (3.0--$\mu$m) &   3.00--$\mu$m  & 0.40 $\mu$m & 3.00--3.10 $\mu$m& 0.35--0.45 $\mu$m  & 3.00--3.05 $\mu$m & 0.35--0.75 $\mu$m  & 3.00--$\mu$m & 0.50 $\mu$m \\
$-$CH (3.4--$\mu$m) &   3.35--$\mu$m  & 0.15 $\mu$m &  3.35--3.45 $\mu$m & 0.13--0.17 $\mu$m  & 3.40--3.45 $\mu$m & 0.13--0.20 $\mu$m &  3.43--$\mu$m & 0.17 $\mu$m\\
$-$SiO (10.0--$\mu$m) & 10.00--$\mu$m & 2.35 $\mu$m  & 9.70--10.00 $\mu$m & 1.50--3.20 $\mu$m &  9.70--10.00 $\mu$m & 1.50--3.50 $\mu$m &  9.70--$\mu$m  & 2.17 $\mu$m\\
\hline
Continuum & \multicolumn{2}{c|}{Flat Continuum} & \multicolumn{2}{c|}{Gaussians and Polynomials} & \multicolumn{2}{c|}{Flat Continuum} & \multicolumn{2}{c}{Flat Continuum} \\
\enddata
\end{deluxetable*}

\begin{figure*}[htbp]
  \begin{center}
    \begin{tabular}{ccc}
      {\includegraphics[angle=0,scale=0.25]{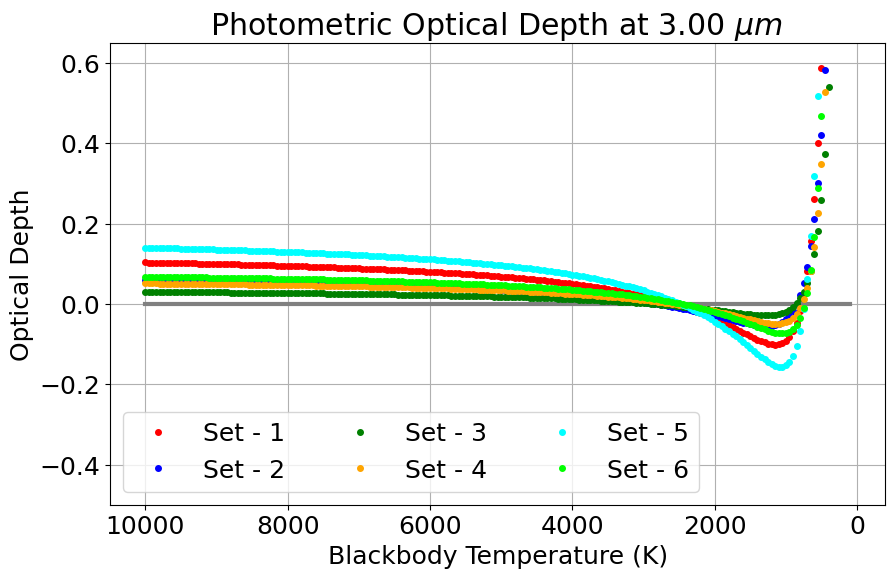}} &
      {\includegraphics[angle=0,scale=0.25]{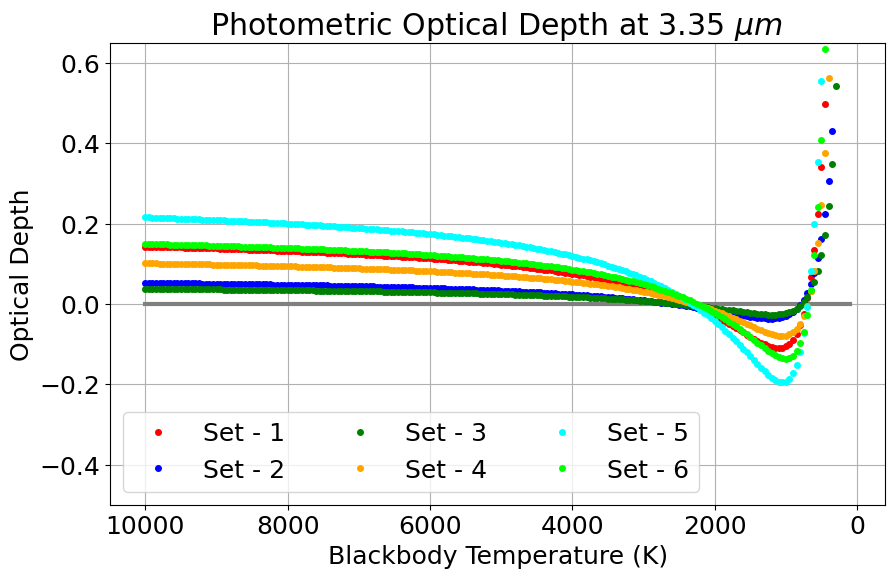}} &
      {\includegraphics[angle=0,scale=0.25]{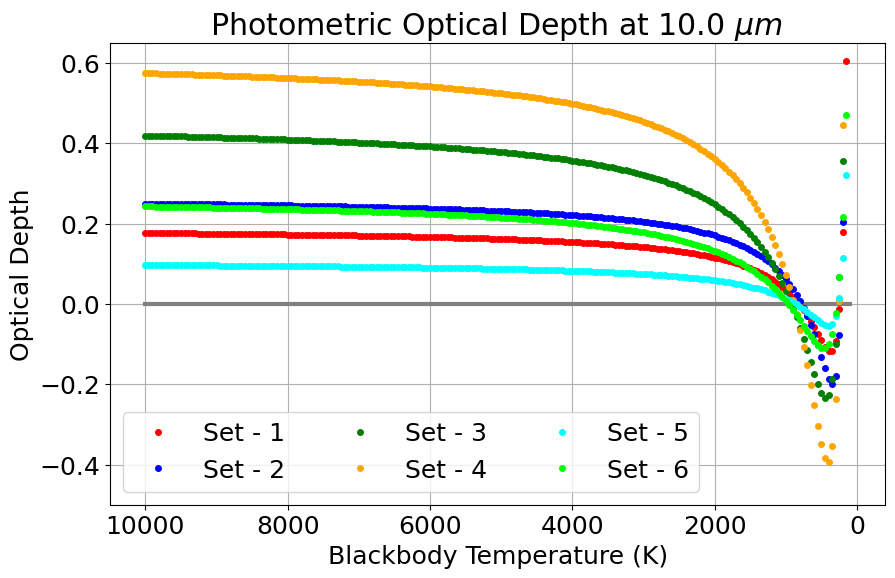}} \\
       {\includegraphics[angle=0,scale=0.25]{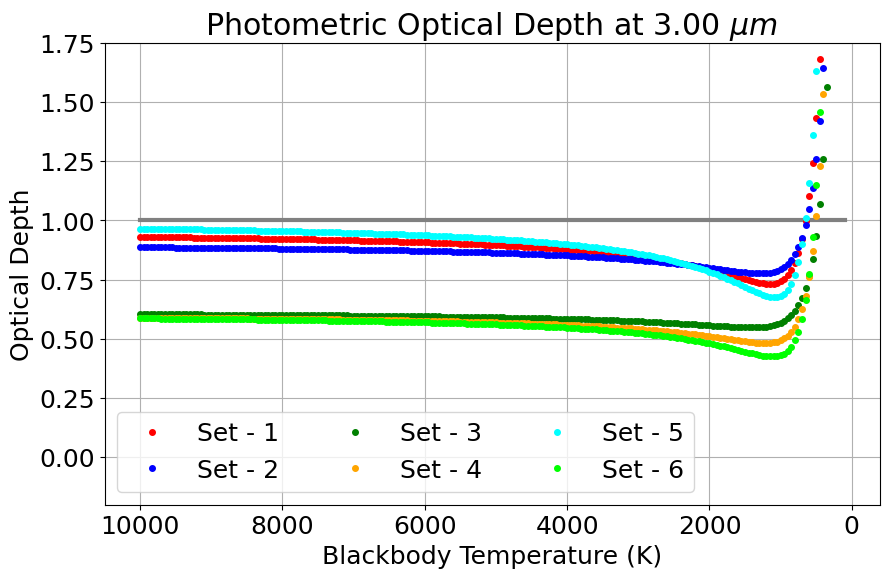}} &
      {\includegraphics[angle=0,scale=0.25]{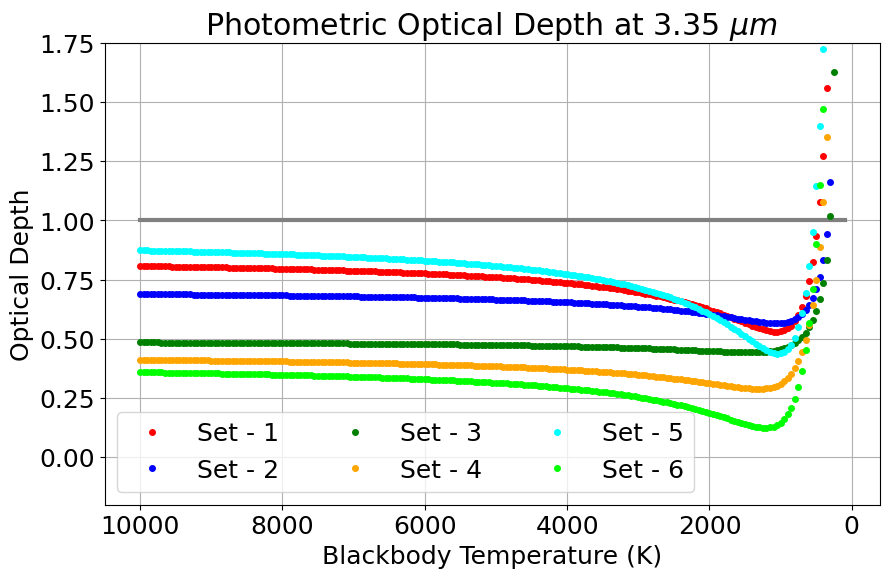}} &
      {\includegraphics[angle=0,scale=0.25]{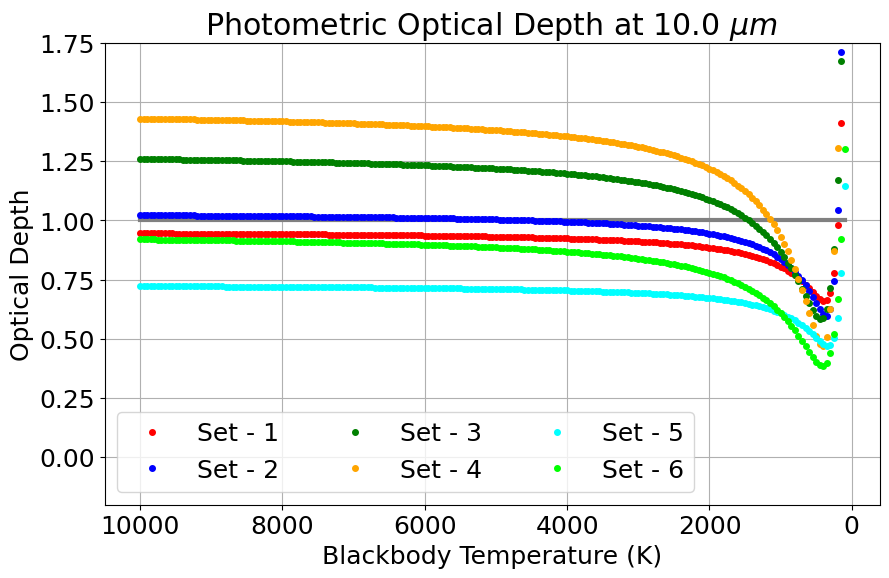}} \\
        \end{tabular}
    \caption{The model spectra are generated using normalized blackbody curves (BBs) with temperatures ranging from $T = 100$--$10000\,\text{K}$. Photometric optical depths at 3.0--$\mu$m, 3.35--$\mu$m, and 10.0--$\mu$m were obtained, employing the Filter Sets listed in Table~\ref{tab:filter-sets} and compared with the reference optical depths ($\tau_{0}$) to evaluate the accuracy of the photometric method with \textit{linear continuum fit} estimation. The efficiency of the Filter Sets in estimating continuum fluxes is tested with $\tau_{0} = 0$ (upper panels), while their ability to constrain absorption features is evaluated with $\tau_{0} = 1$ (lower panels).}  
    \label{fig:BB-OpticalDepths-AllFilterSets}
      \end{center}
\end{figure*}

\begin{figure*}[htbp]
  \begin{center}
    \begin{tabular}{ccc}
      {\includegraphics[angle=0,scale=0.20]{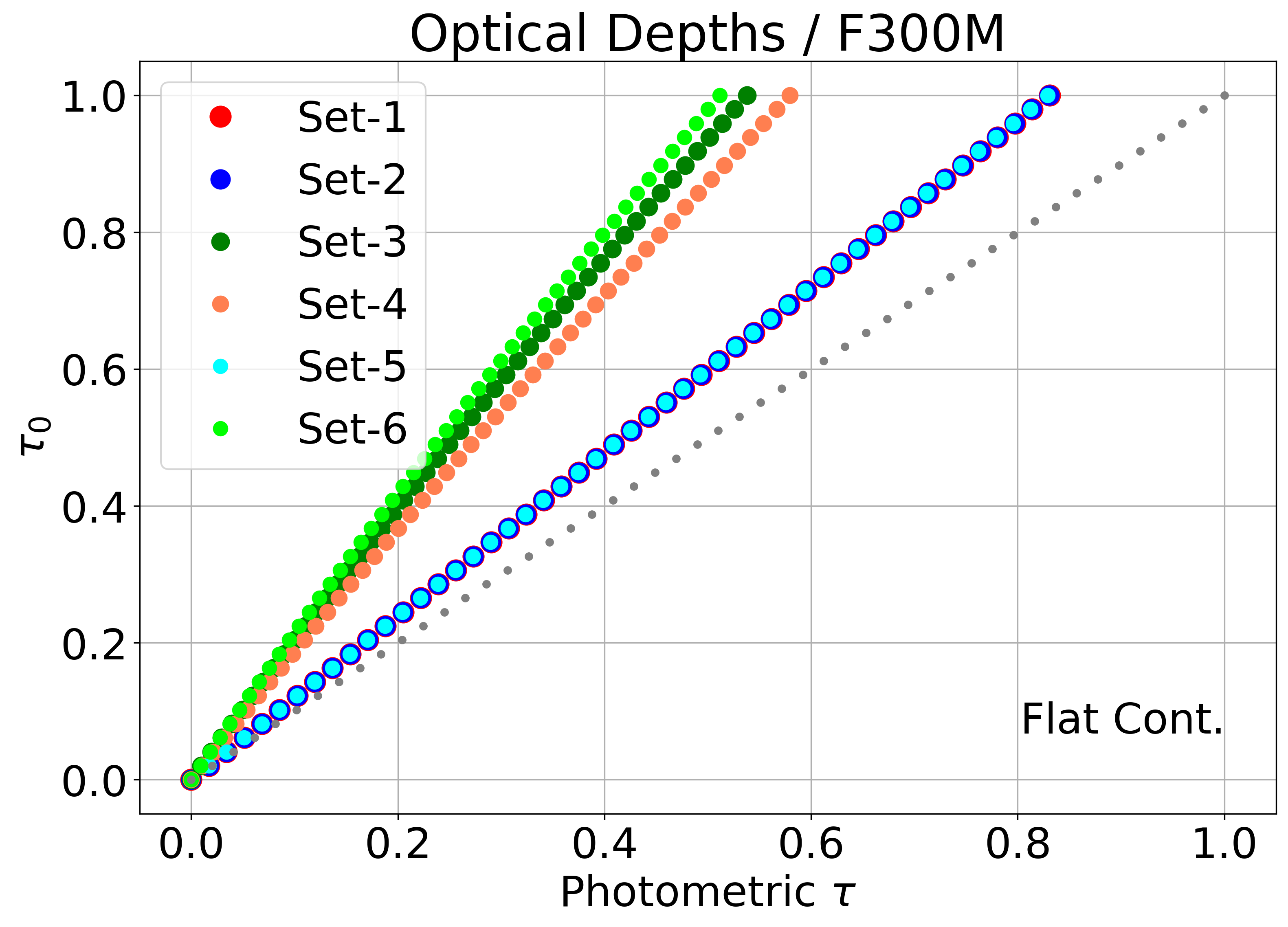}} &
      {\includegraphics[angle=0,scale=0.20]{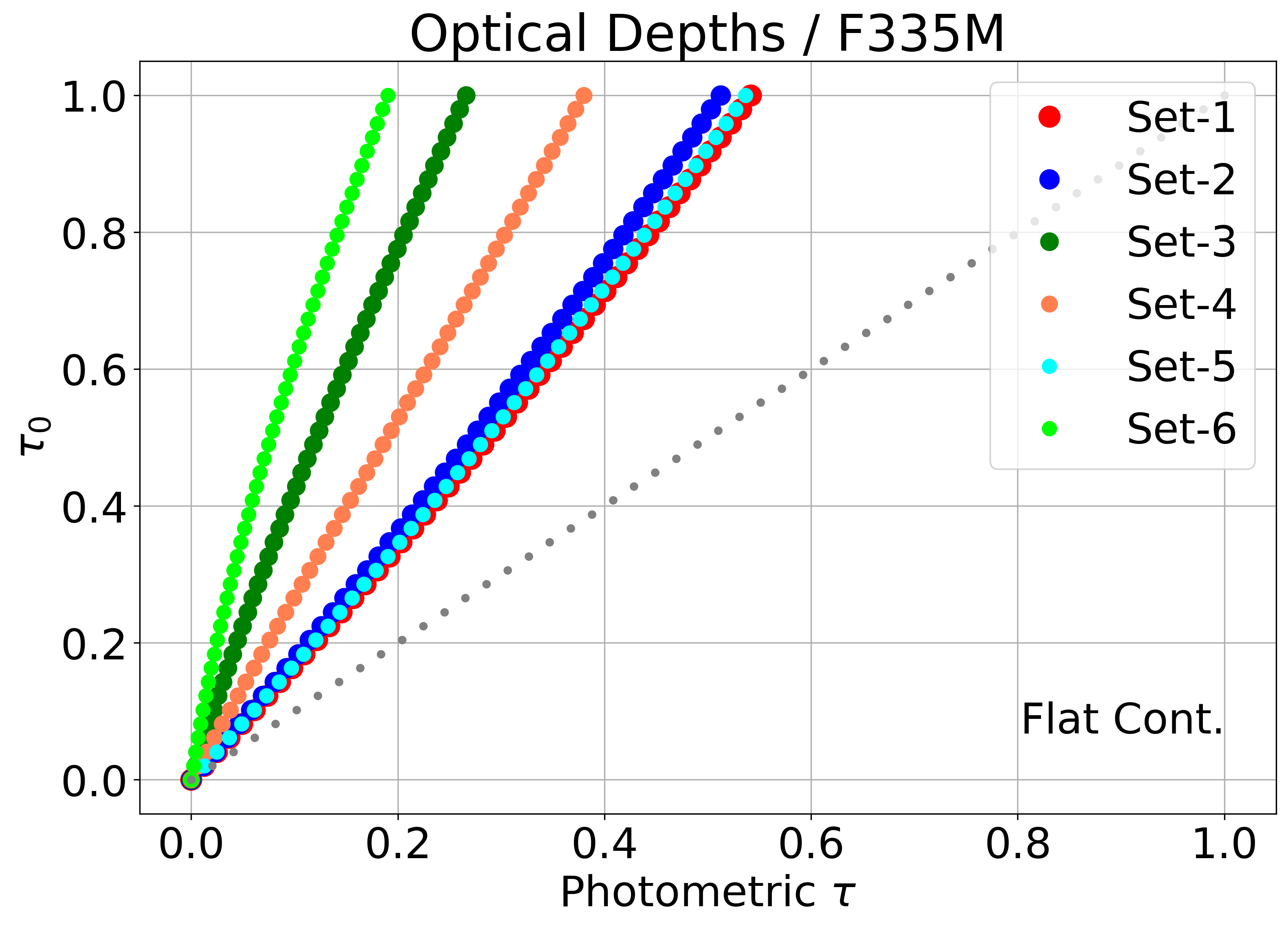}} &
      {\includegraphics[angle=0,scale=0.20]{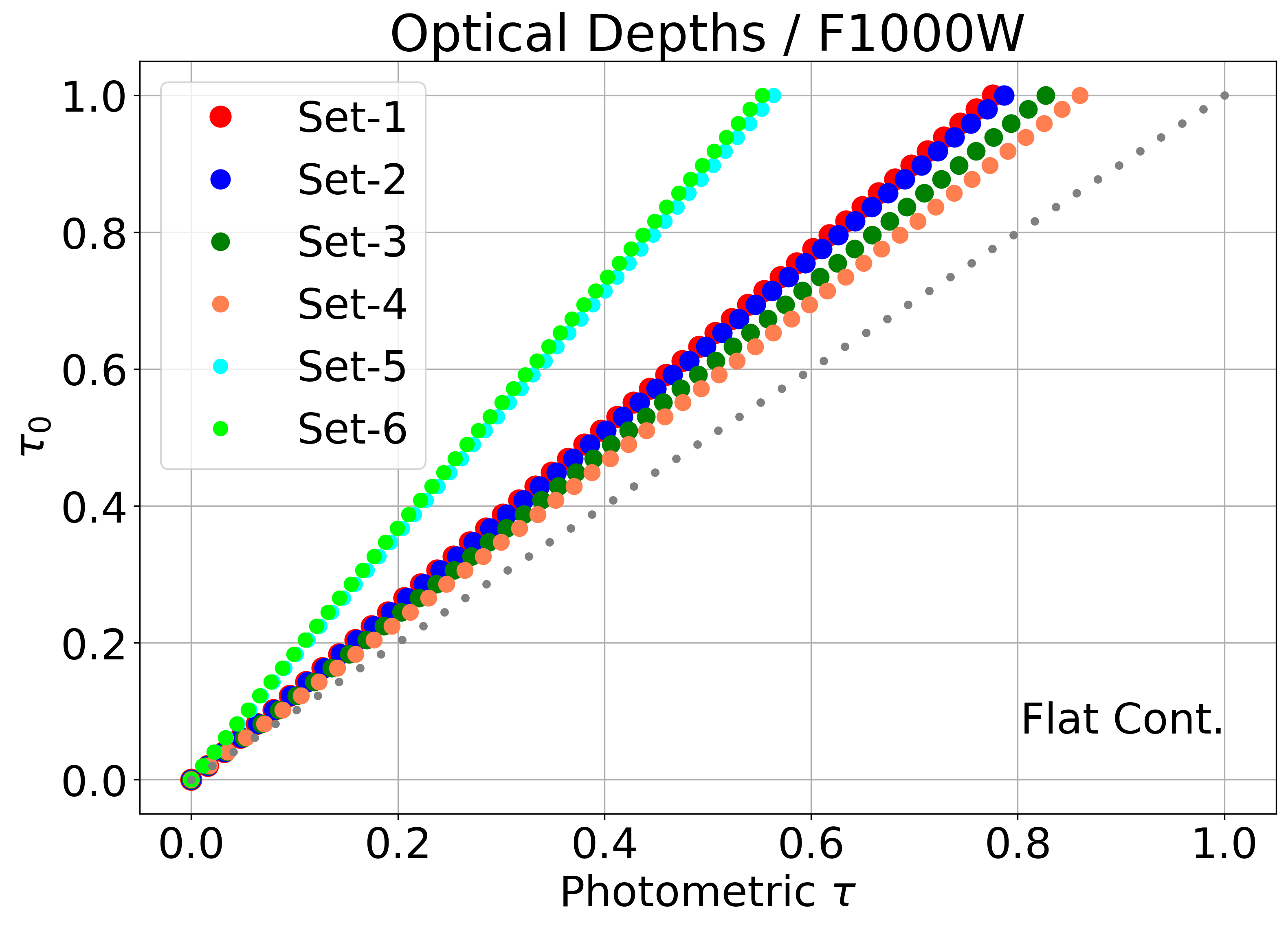}} \\
        \end{tabular}
    \caption{The photometric optical depths and spectroscopic optical depths at 3.0--$\mu$m, 3.35–$\mu$m, and 10.0--$\mu$m are measured using \textit{standard spectral models} (to represent ideal conditions) and compared to investigate methodological biases. The gray dotted line represents equal $\tau_{p}$ and $\tau_{s}$ values.} 
    \label{fig:Method-Biases}
      \end{center}
\end{figure*}

\begin{figure*}[htbp]
  \begin{center}
    \begin{tabular}{ccc}
      {\includegraphics[angle=0,scale=0.20]{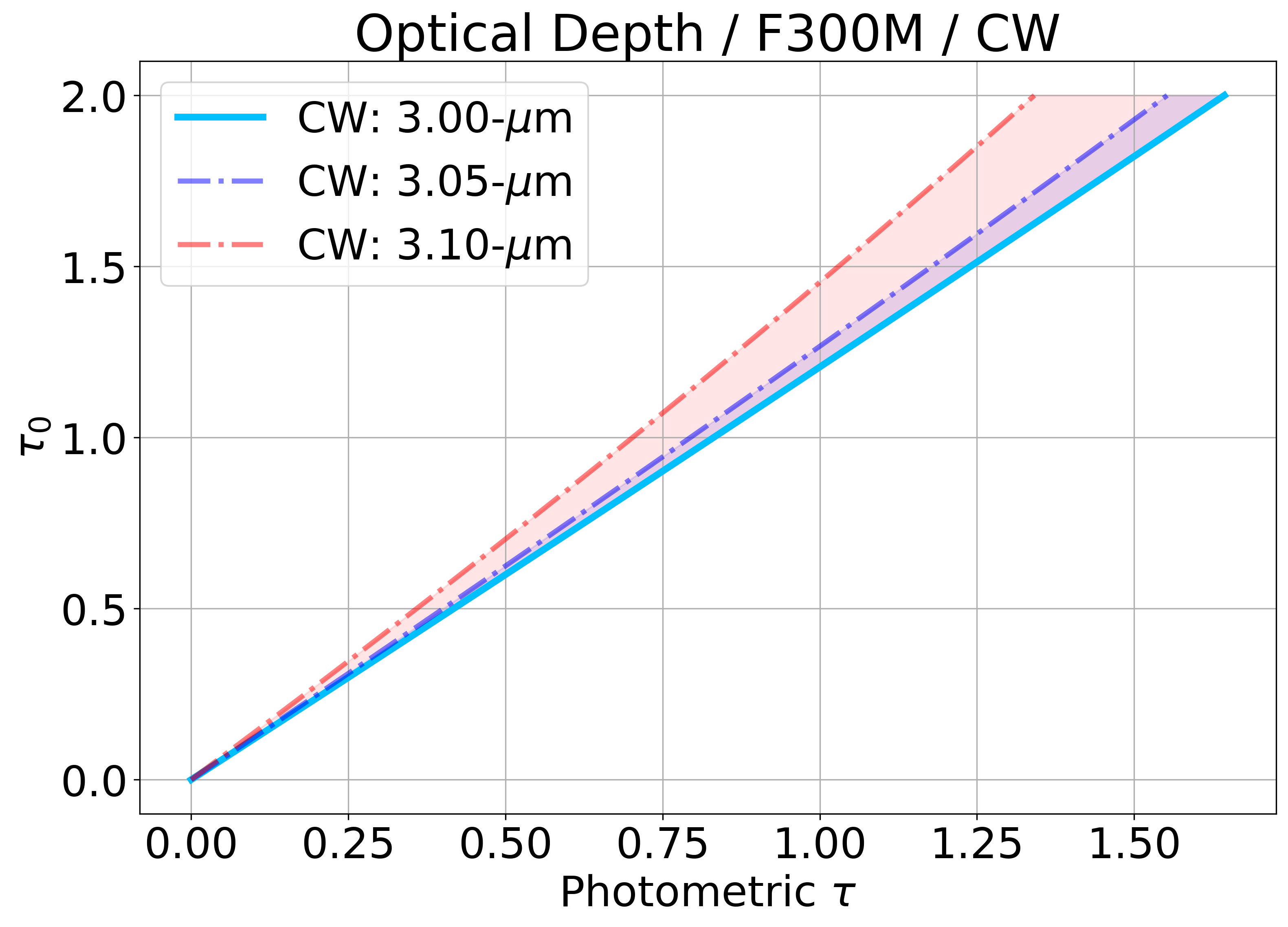}} &
      {\includegraphics[angle=0,scale=0.20]{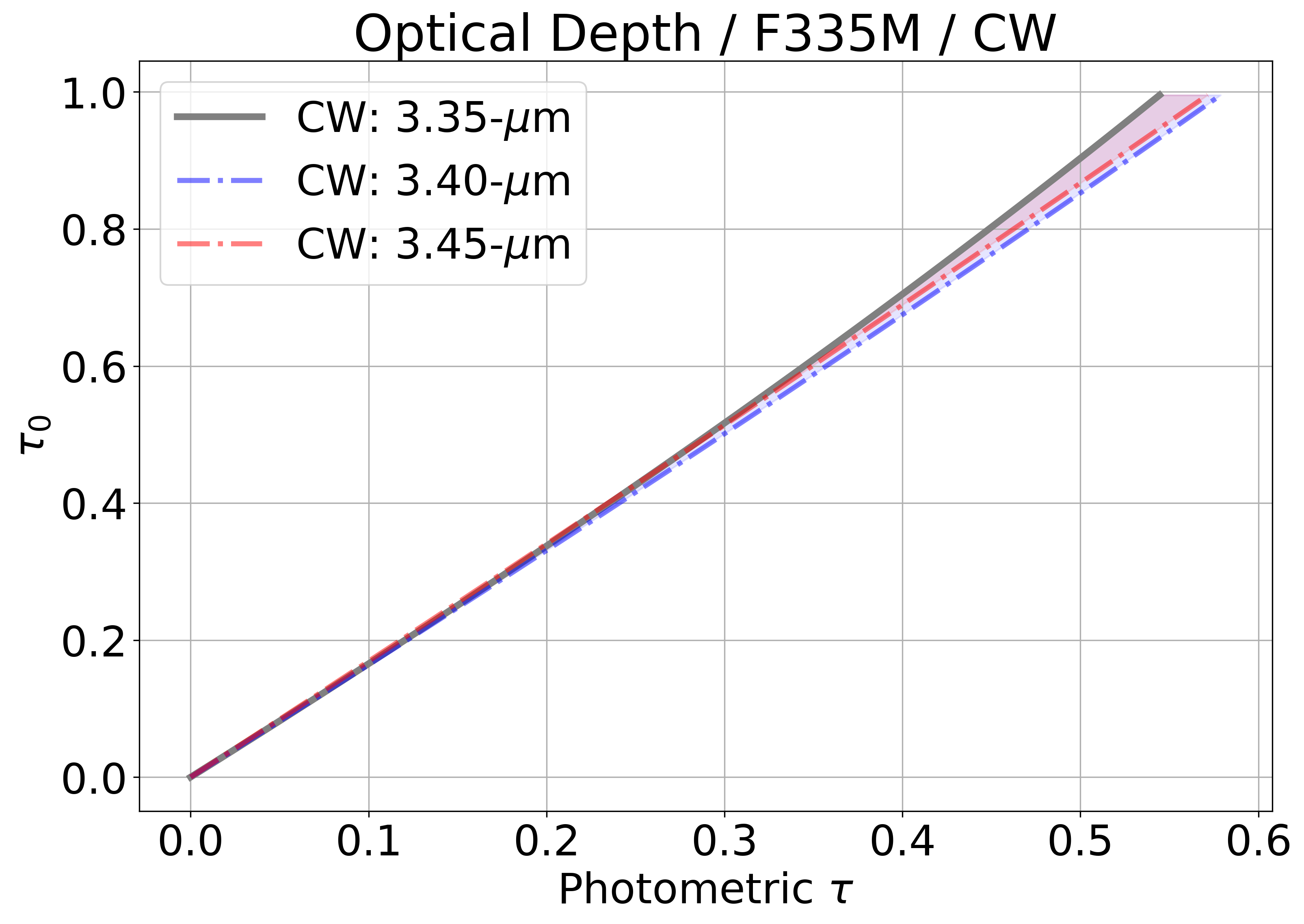}} &
      {\includegraphics[angle=0,scale=0.20]{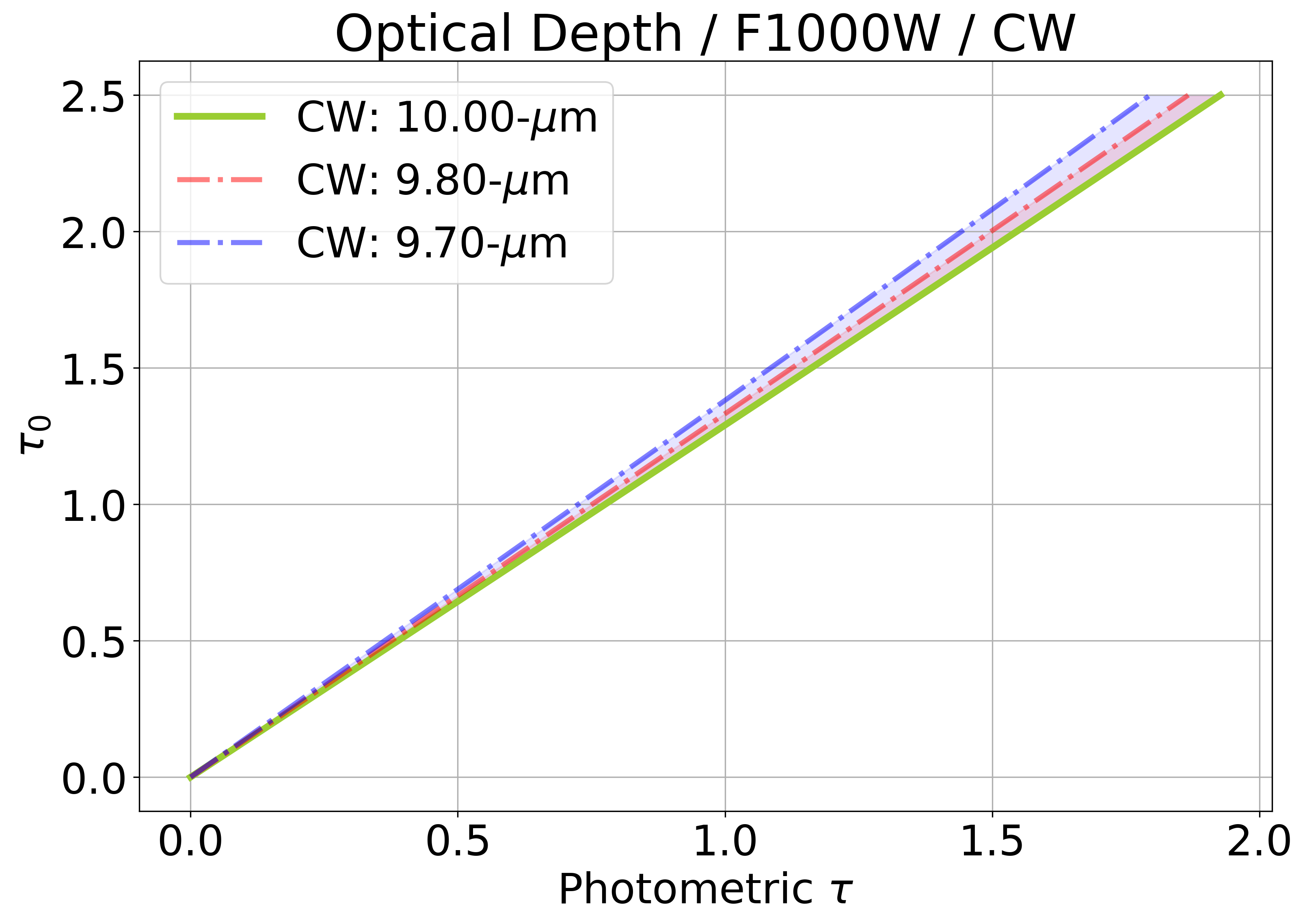}} \\
      {\includegraphics[angle=0,scale=0.20]{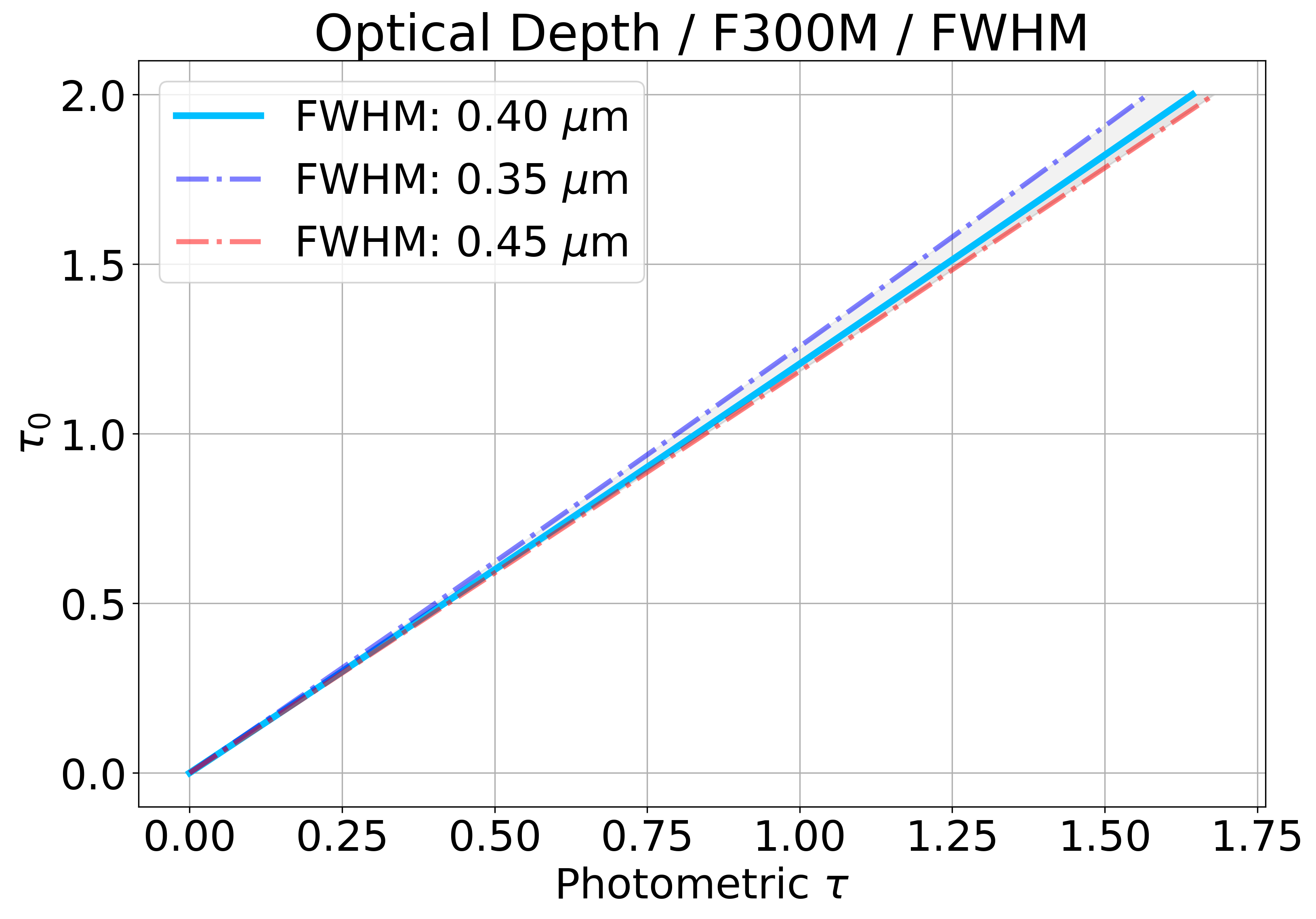}} &
      {\includegraphics[angle=0,scale=0.20]{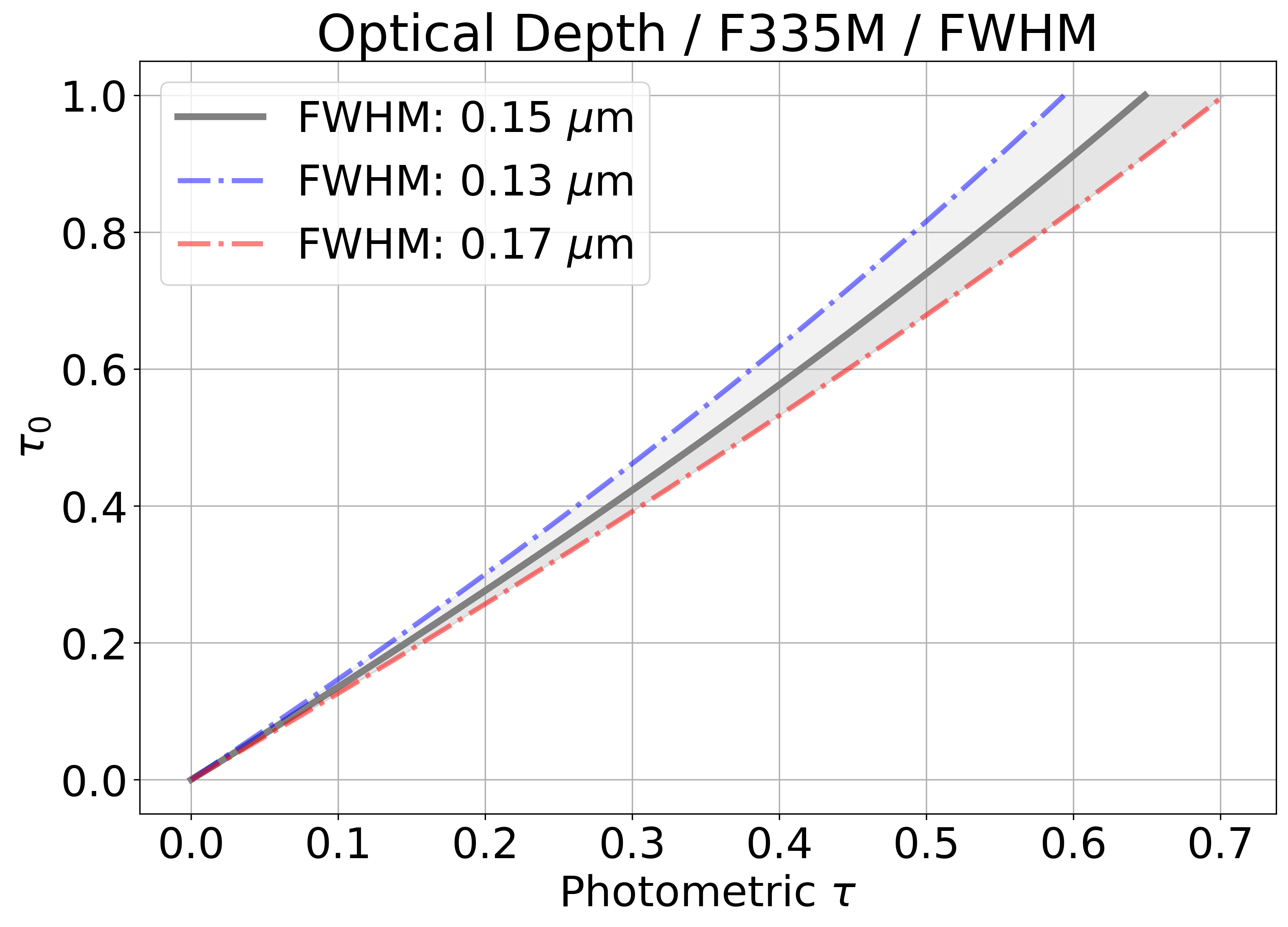}} &
      {\includegraphics[angle=0,scale=0.20]{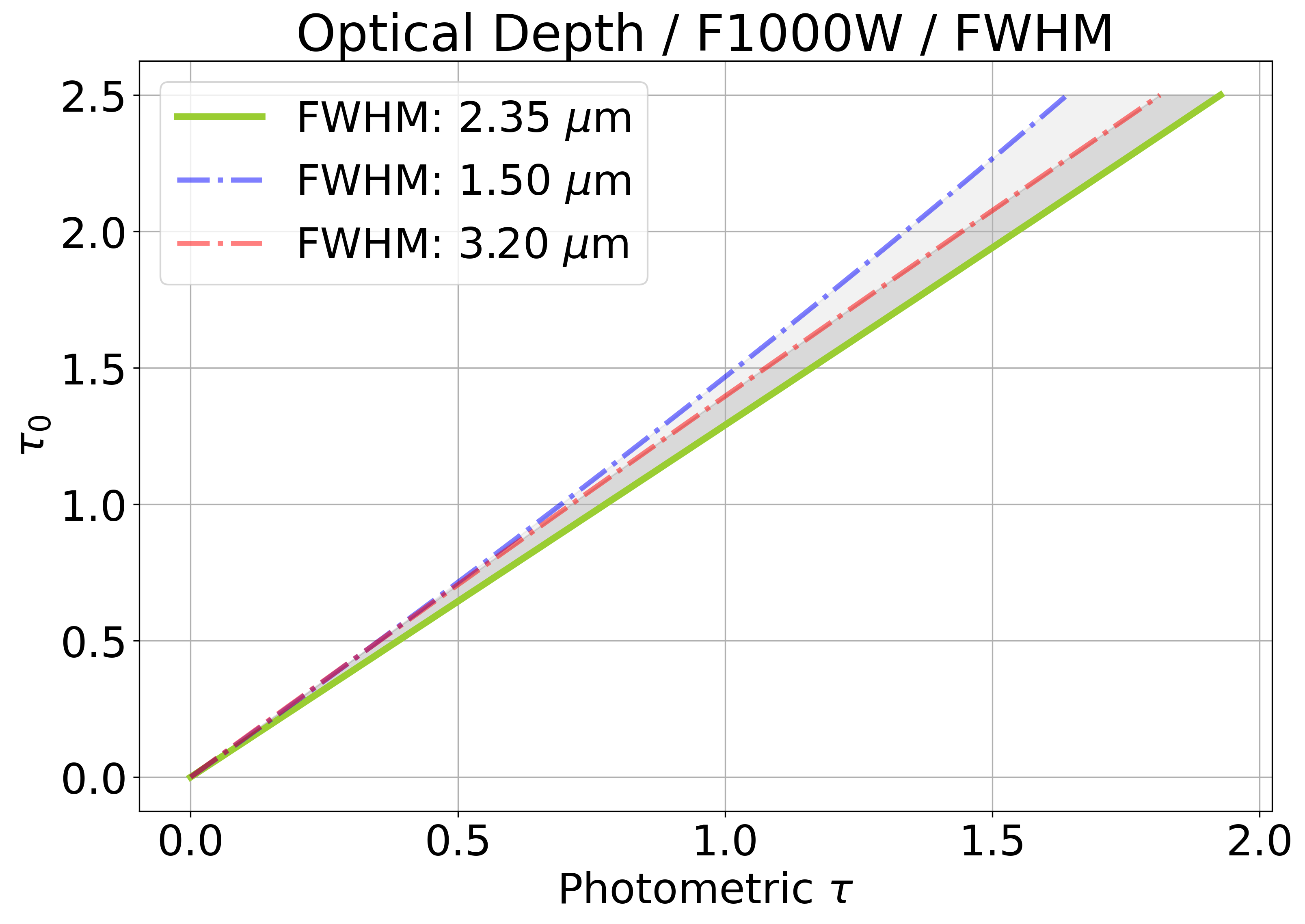}} \\
        \end{tabular}
    \caption{The photometric optical depths and spectroscopic optical depths are compared to investigate deviations across the applied optical depth ranges ($\tau_{\text{0[3.0]}}$ = 0–2, $\tau_{\text{0[3.4]}}$ = 0–1, and $\tau_{\text{0[10.0]}}$ = 0–2.5). The optical depths obtained with \textit{standard spectra model} parameters are indicated with blue, gray and green solid lines for the water ice, aliphatic hydrocarbons, and silicate features, respectively. \textit{Upper Panels}: Optical depths compared using exploratory spectral models created with CW shifts. The optical depths obtained with shifted CWs to shorter and longer wavelengths are indicated in blue and red respectively. \textit{Bottom Panels}: Optical depths compared using exploratory spectral models created with FWHM variations. The optical depths obtained with narrower and broader FWHMs are indicated in blue and red respectively. The CW and FWHM values applied in exploratory models are summarized in Table~\ref{tab:model-parameter-ranges}. }
    \label{fig:Variations-Abs}
      \end{center}
      \end{figure*}

\begin{figure*}
  \begin{center}
  \begin{tabular}{ccc}
    \includegraphics[angle=0,scale=0.20]{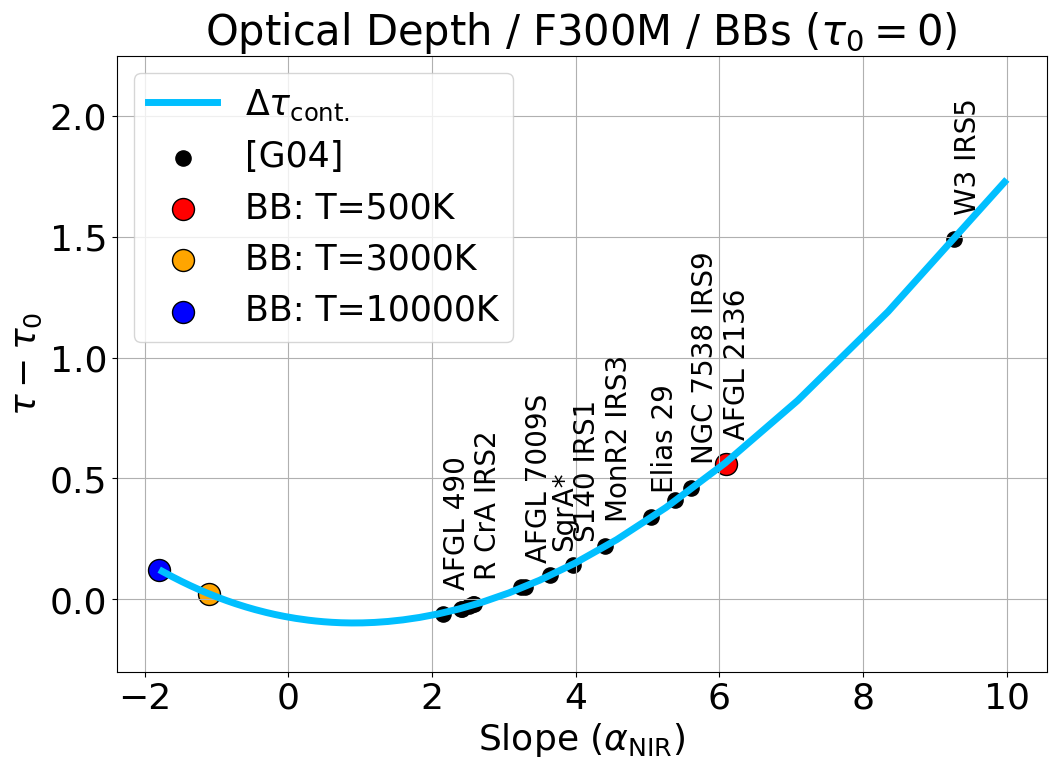} &
    \includegraphics[angle=0,scale=0.20]{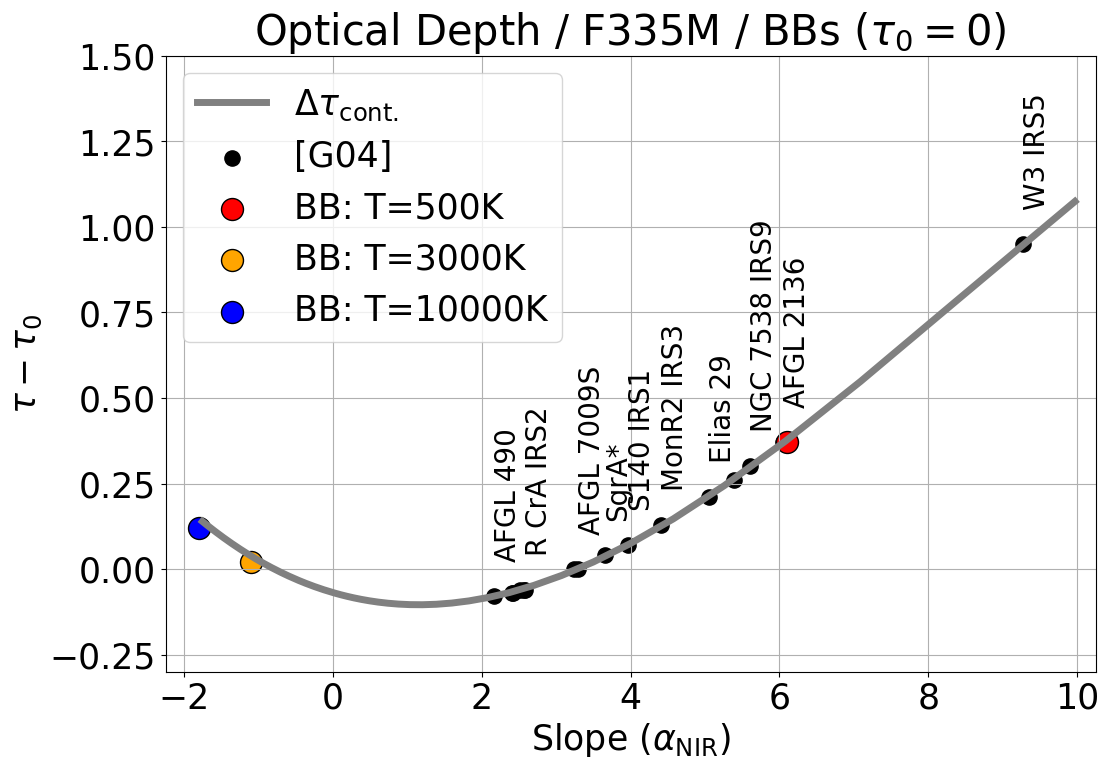} &
    \includegraphics[angle=0,scale=0.20]{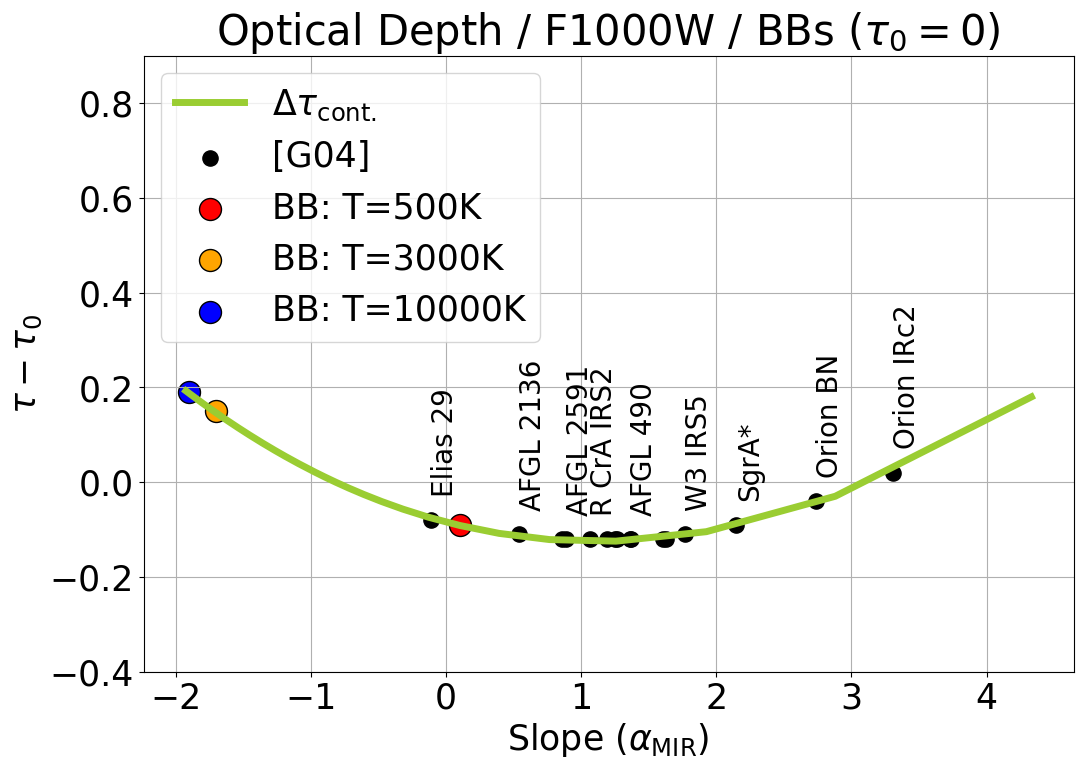} \\
  \end{tabular}
  \caption{The BBs (without absorption, $\tau_{0} = 0$) are used to derive correlations between spectral index values and differences arising from the linear continuum fit approximation. The optical depths at 3.0--$\mu$m, 3.35–$\mu$m, and 10.0–$\mu$ are measured. The model spectra set was generated using a polynomial fit to the observational spectra set ([G04]) that exhibit different slope values in the NIR ($\alpha_{\text{NIR}}$) and MIR ($\alpha_{\text{MIR}}$), covering $\alpha_{\text{NIR}}$ from $-2.2$ to $9.3$ and $\alpha_{\text{MIR}}$ from $-0.1$ to $3.3$. The slope ($\alpha$) of each spectrum was used to estimate linear continuum fit related differences ($\Delta \tau_{\text{cont.}}$) in the optical depth values based on the correlations between $\alpha$ and $\Delta \tau_{\text{cont.}}$ derived using BBs. Comparative models include BBs at $T = 500\,\text{K}$ ($\alpha_{\text{NIR}} = 6.1$, $\alpha_{\text{MIR}} = 0.1$), $3000\,\text{K}$ ($\alpha_{\text{NIR}} = -1.1$, $\alpha_{\text{MIR}} = -1.7$), $10000\,\text{K}$ ($\alpha_{\text{NIR}} = -1.8$, $\alpha_{\text{MIR}} = -1.9$) are also presented. The names of selected sources are labeled to indicate variations in slope levels.}
  \label{fig:SpectralIndex-Error}
  \end{center}
\end{figure*}

\begin{figure*}[htbp]
  \begin{center}
    \begin{tabular}{ccc}
      {\includegraphics[angle=0,scale=0.21]{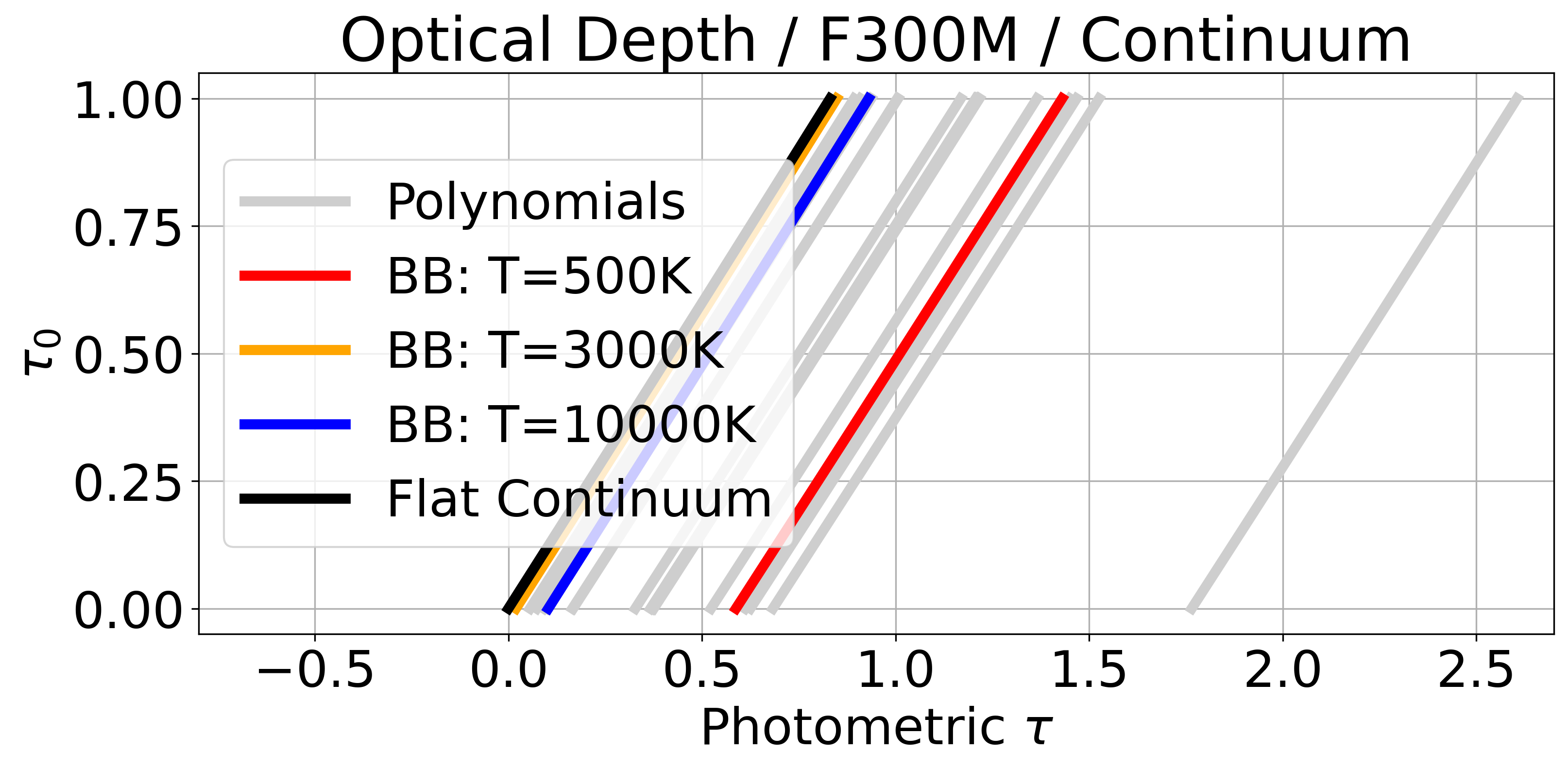}} &
      {\includegraphics[angle=0,scale=0.21]{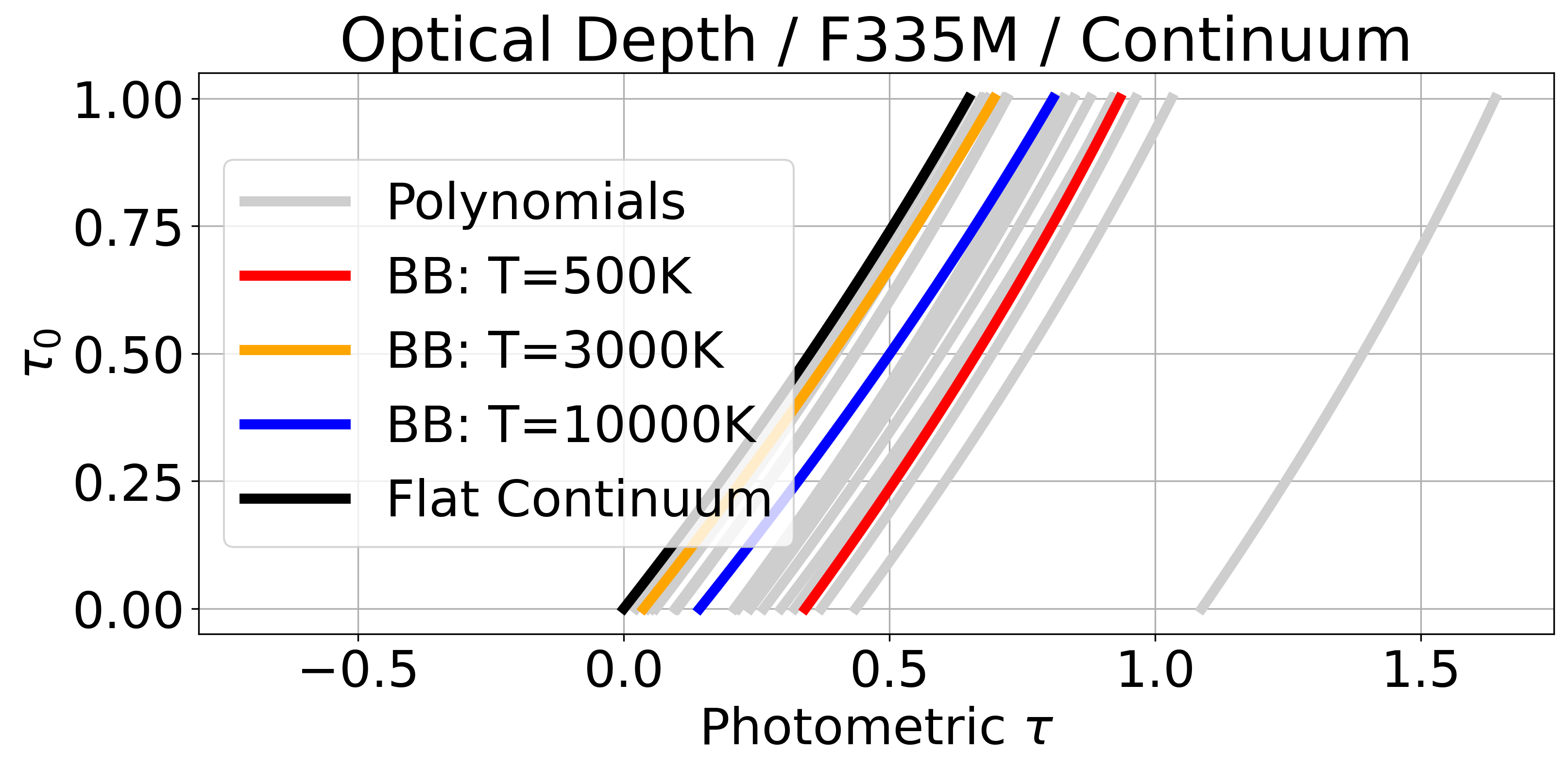}} &
      {\includegraphics[angle=0,scale=0.21]{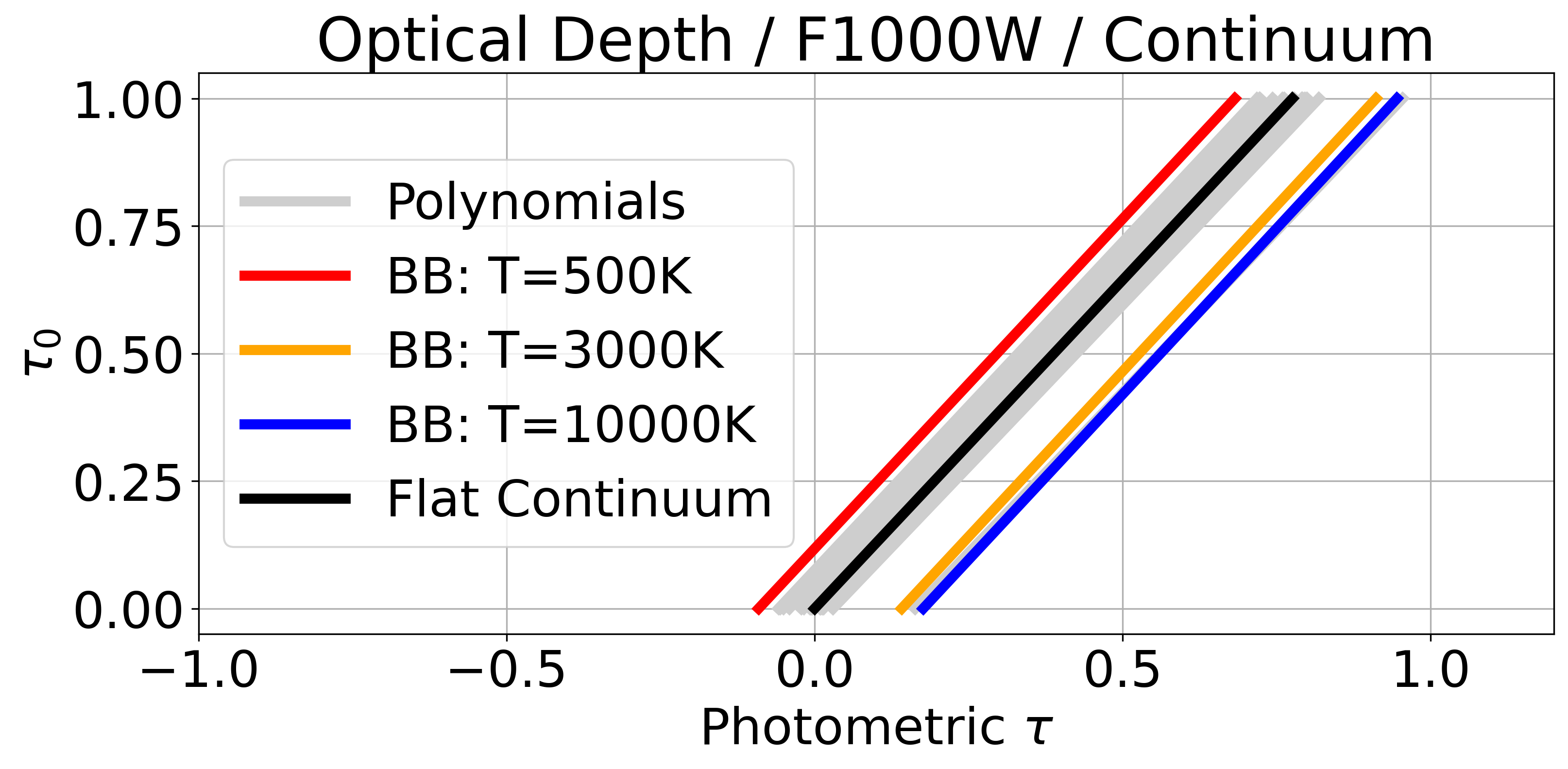}} \\
        \end{tabular}
    \caption{Spectral models were generated with continua having continuum profiles with different slopes. An optical depth range of $\tau_{0} = 0$–$1$ was applied. Photometric optical depths at 3.0--$\mu$m, 3.35–$\mu$m, and 10.0--$\mu$m were obtained and compared with the reference optical depths to explore continuum related differences ($\Delta \tau_{\text{cont.}}$). The optical depths obtained using models with polynomial continua derived from the observational spectra set ([G04]) are shown (in gray), alongside those obtained using models that include BBs at $T = 500\,\text{K}$, $3000\,\text{K}$, $10000\,\text{K}$ and a flat continuum. 
 } 
    \label{fig:Variations-Cont}
      \end{center}
\end{figure*}

\begin{figure*}[htbp]
  \begin{center}
    \begin{tabular}{cc}
      {\includegraphics[angle=0,scale=0.23]{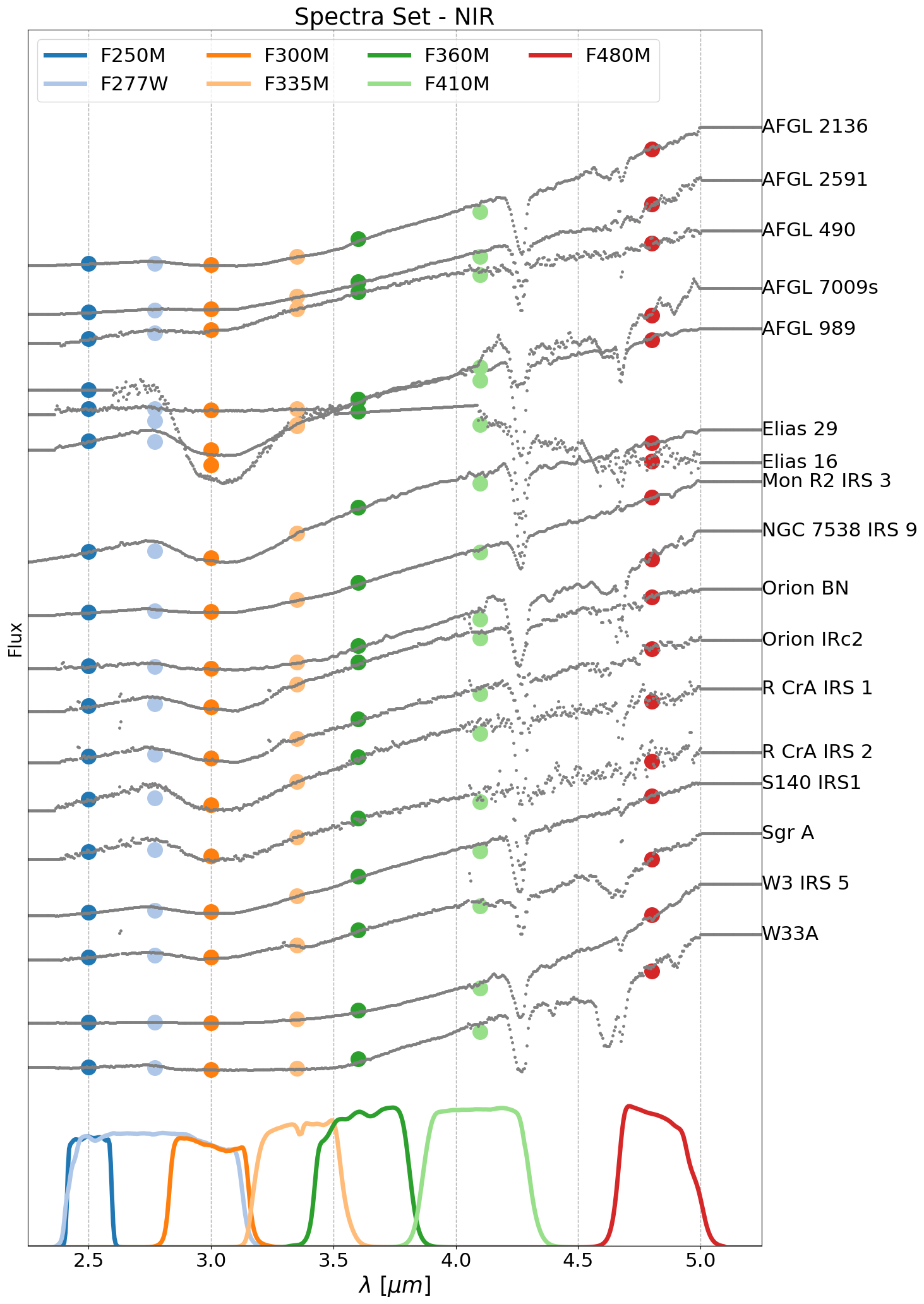}}  &
      {\includegraphics[angle=0,scale=0.23]{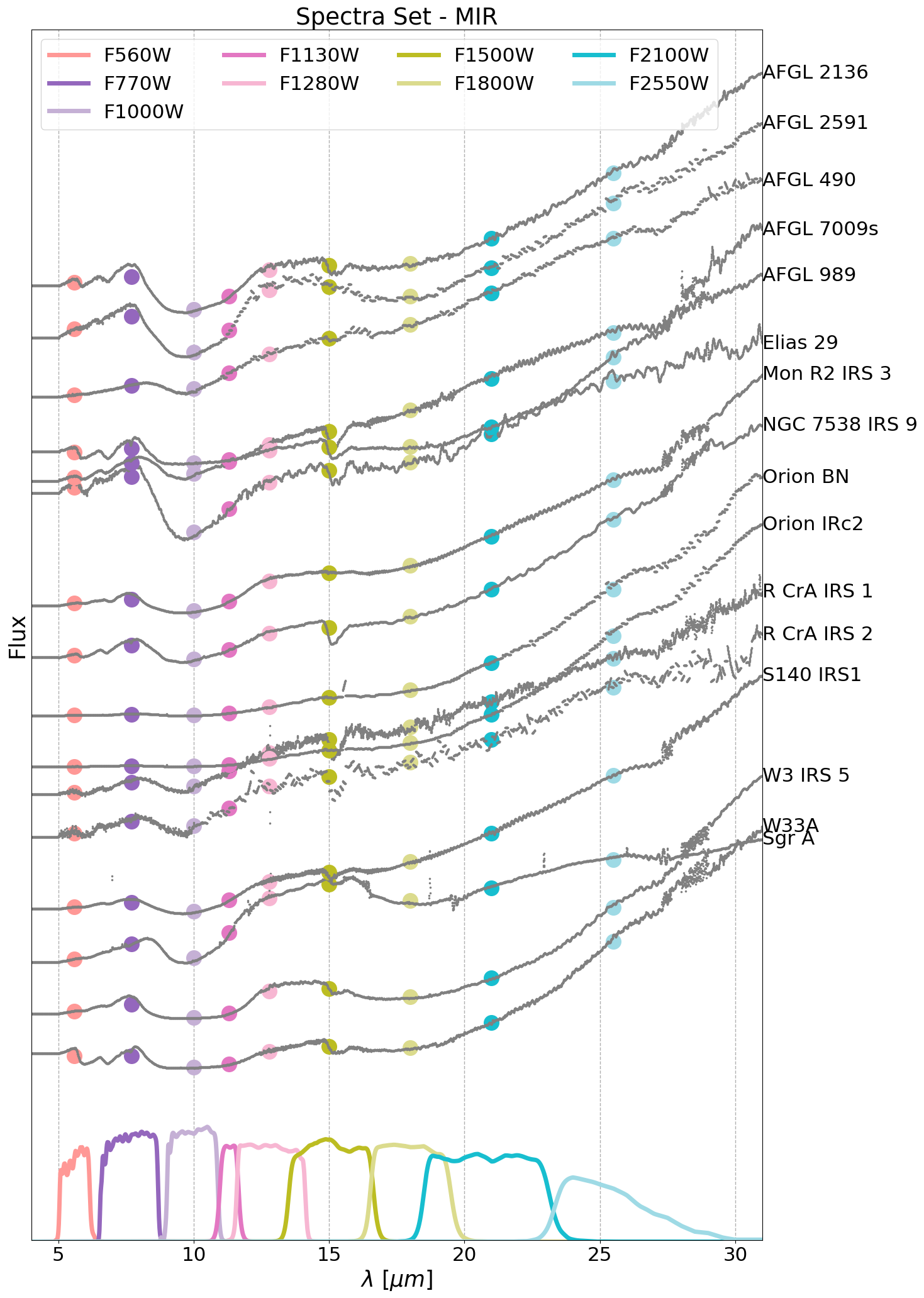}} 
        \end{tabular}
    \caption{The photometric fluxes are shown on the spectra set from the literature \citep{Gibb2004} with colored dots indicating the employed JWST filters. The throughput functions of the NIRCam and MIRI filters used for the photometric flux measurements are also presented at the bottom.}   
    \label{fig:AllSpectraG04}
      \end{center}
\end{figure*}

\begin{figure*}[htbp]
  \begin{center}
    \begin{tabular}{ccc}
      {\includegraphics[angle=0,scale=0.2]{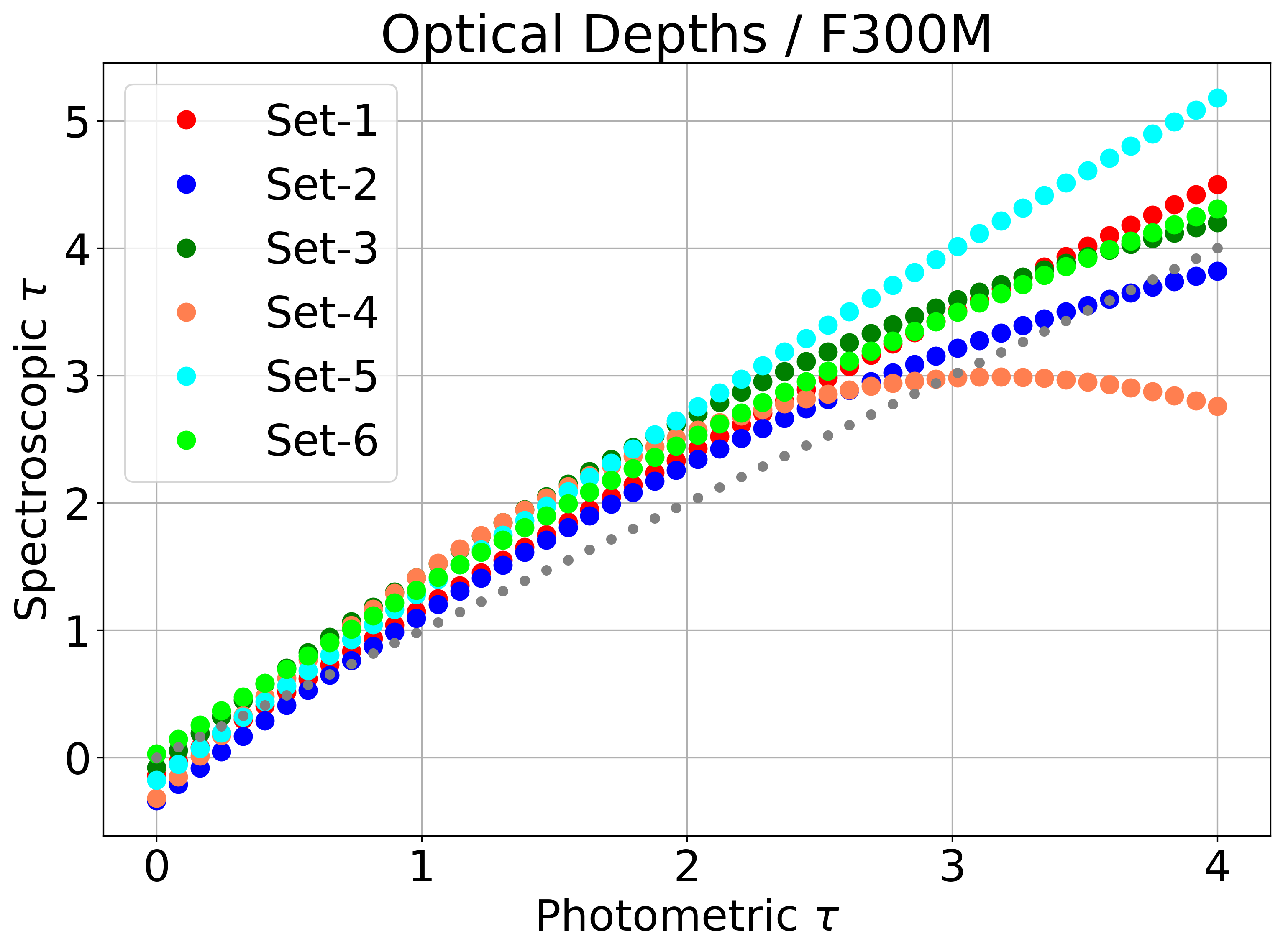}} &
      {\includegraphics[angle=0,scale=0.2]{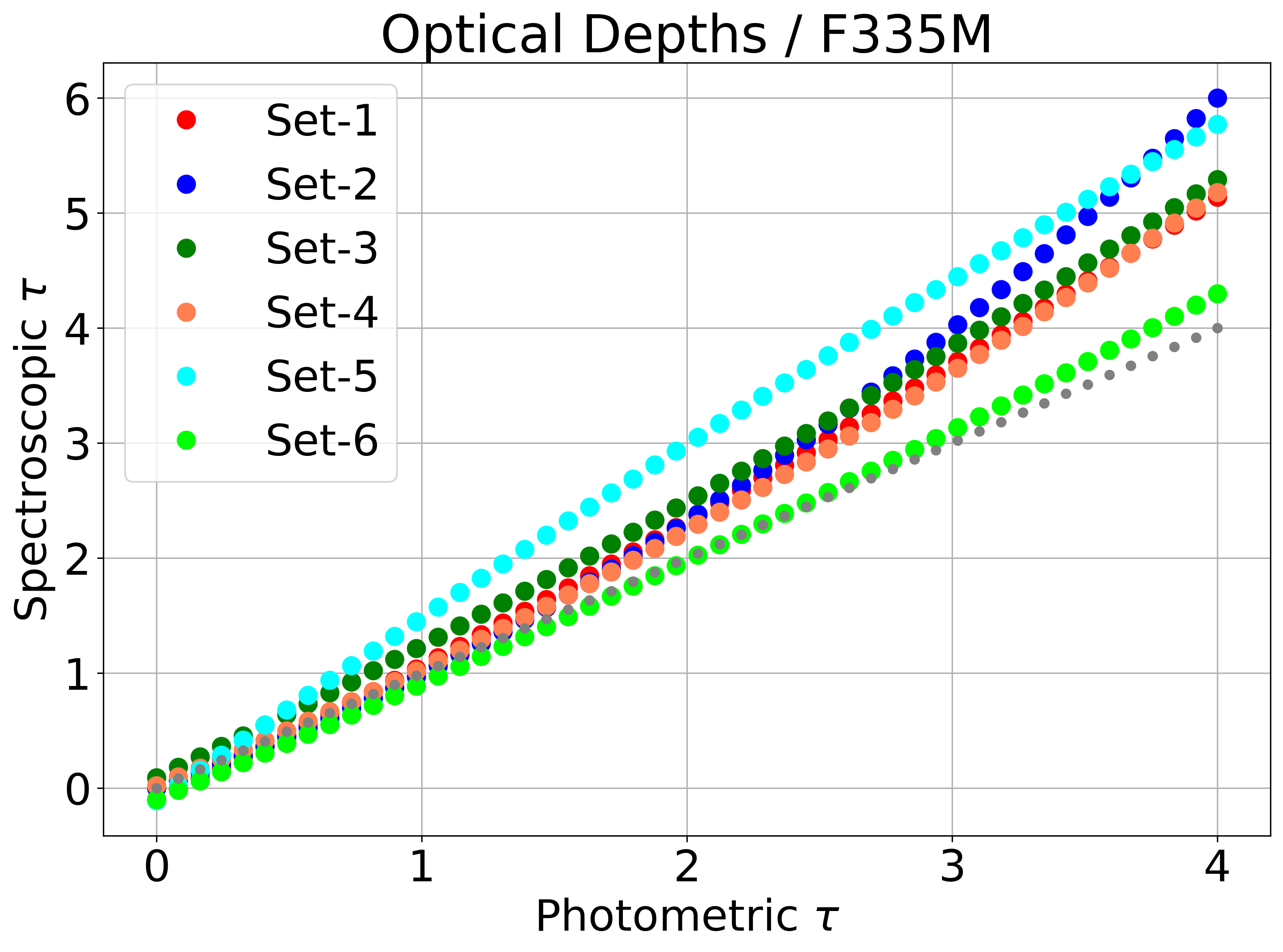}} &
      {\includegraphics[angle=0,scale=0.2]{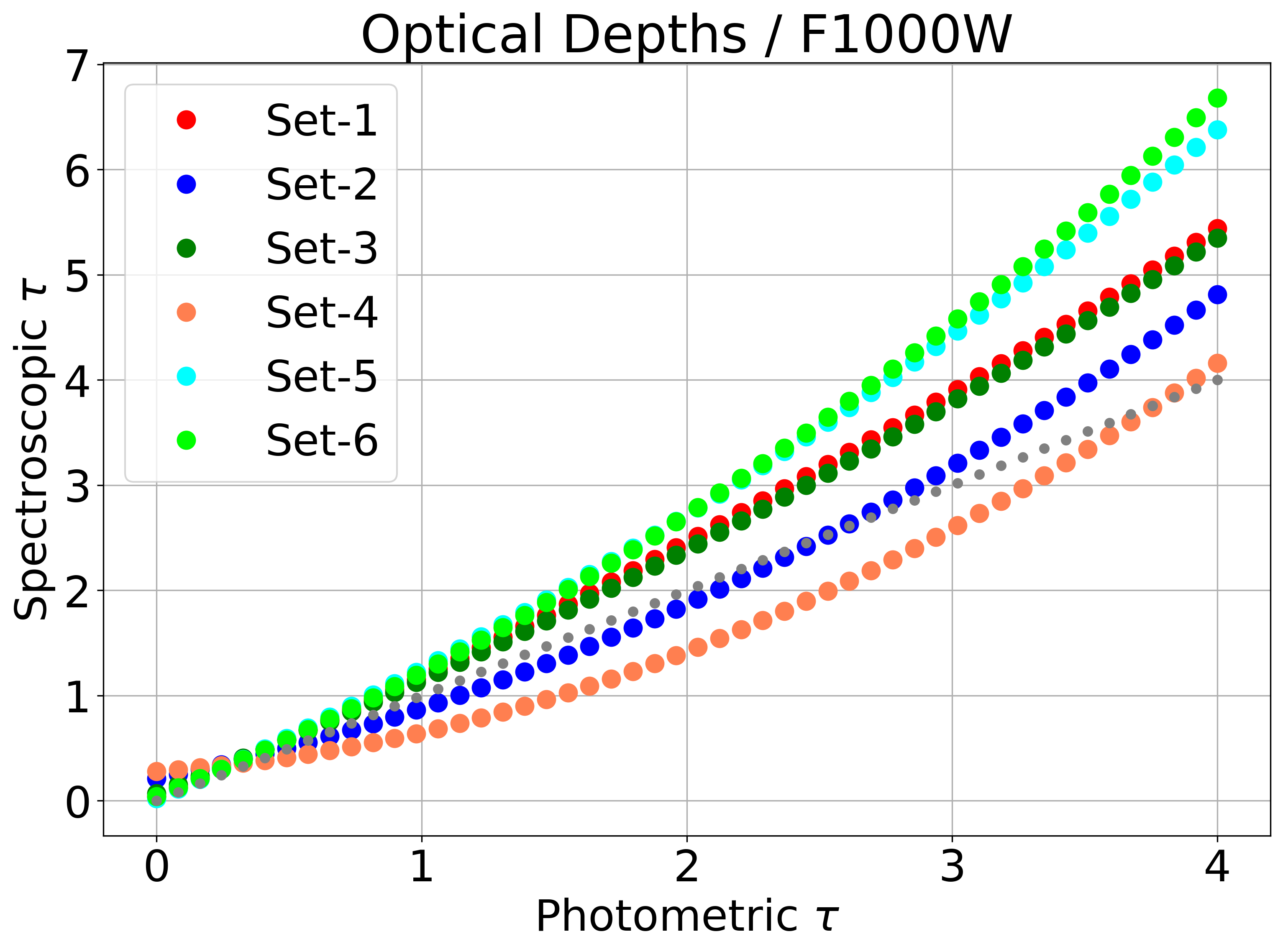}} \\
        \end{tabular}
    \caption{Optical depths at 3.0--$\mu$m, 3.35–$\mu$m, and 10.0--$\mu$m are derived from literature spectra \citep{Gibb2004}. Polynomial fits (colored dots) show the correlation between photometric and spectroscopic optical depths, with the gray dotted line representing equal $\tau_{p}$ and $\tau_{s}$ values.}
    \label{fig:OD-Data-FilterSets}
      \end{center}
\end{figure*}

\begin{deluxetable*}{cccccccccc}
\tabletypesize{\small}  
\setlength{\tabcolsep}{3pt}
\tablecaption{Photometric and spectroscopic optical depth measurements for the water ice $-$OH (3.0--$\mu$m), aliphatic hydrocarbon $-$CH (3.4--$\mu$m), and silicate $-$SiO (10.0--$\mu$m) features are compared with reported values in \citealt{Gibb2004} ([G04]).}
\label{tab:comparison-all}
\tablehead{
\colhead{} & \multicolumn{3}{c}{$-$OH (3.0--$\mu$m)}  & \multicolumn{3}{c}{$-$CH (3.4--$\mu$m)}  & \multicolumn{3}{c}{$-$SiO (10.0--$\mu$m)}  \\
\cline{2-10}
\colhead{} & \colhead{Photometric}  & \colhead{Spectroscopic} & \colhead{[G04]} & \colhead{Photometric} & \colhead{Spectroscopic} & \colhead{[G04]} & \colhead{Photometric} & \colhead{Spectroscopic} & \colhead{[G04]}
}
\startdata
Sgr A$^{*}$ & 0.78 & 0.84 & 0.50 & 0.50 & 0.54 & 0.21 & 1.48 & 1.90 & 2.31 \\
AFGL 2136 & 2.60 & 3.28 & 3.24 & 1.13 & 1.30 & 0.14 & 1.96 & 2.40 & 2.10 \\
NGC 7538 IRS 9 & 2.73 & 3.25 & 3.10 & 1.34 & 1.58 & 0.13 & 1.27 & 1.51 & 2.23 \\
Elias 29 & 1.24 & 1.58 & 1.85 & 0.42 & 0.48 & 0.89 & 0.93 & 1.13 & 1.32 \\
Orion BN & 1.19 & 1.08 & 1.44 & 0.49 & 0.40 & 0.03 & 1.32 & 1.61 & 1.59 \\
RCrA IRS 1 & 0.98 & 1.41 & 1.44 & 0.38 & 0.49 & 0.04 & 0.43 & 0.47 & 0.61 \\
AFGL 989 & 0.96 & 1.27 & 1.43 & 0.33 & 0.36 & 0.05 & 0.37 & 0.46 & 0.72 \\
Mon R2 IRS 3 & 1.26 & 1.27 & 1.12 & 0.71 & 0.76 & 0.04 & 1.52 & 1.89 & 2.55 \\
S140 & 1.10 & 1.22 & 1.12 & 0.52 & 0.57 & 0.05 & 1.34 & 1.62 & 1.51 \\
AFGL 2591 & 1.06 & 1.17 & 0.74 & 0.54 & 0.62 & 0.05 & 1.52 & 1.85 & 2.25 \\
AFGL 490 & 0.28 & 0.27 & 0.32 & 0.08 & 0.07 & 0.03 & 0.49 & 0.69 & 0.64 \\
W3 IRS 5 & 3.76 & 4.00 & 2.78 & 1.64 & 1.89 & 0.13 & 3.68 & 5.01 & 5.80 \\
R CrA IRS 2 & 0.85 & 1.18 & 1.14 & 0.31 & 0.29 & 0.04 & 0.40 & 0.38 & ...  \\
Elias 16 & 0.79 & 1.23 & 1.36 & 0.18 & 0.19 & 0.03 & -- & -- & -- \\
W33A & 1.53 & 1.64 & 5.50 & 1.67 & 1.90 & 0.29 & 2.65 & 3.37 & 7.84 \\
Orion IRc2 & 1.43 & 1.07 & 1.48 & 0.72 & 0.48 & ...  & 1.55 & 1.77 & 1.87 \\
AFGL7009s & 1.11 & 1.08 & ... & 1.30 & 1.44 & ... & 1.76 & 2.13 & 4.00 \\
\hline
Average	&	1.39	&	1.58	&	1.79	&	0.72	&	0.79	&	0.14	&	1.42	&	1.76	&	2.49	\\
\enddata
\tablecomments{The spectrum of R CrA IRS 2 in the MIR region is noisy and optical depths are not reported in [G04]. The spectrum of Elias 16 covers only the NIR region. For Orion IRc2, an absorption at 3.4--$\mu$m is not reported. The spectrum of W33A and AFGL 7009S are saturated around 3.0--$\mu$m $\&$ 9.0–$\mu$m. For AFGL 7009S, optical depths are not reported in G04 for the 3.0--$\mu$m $\&$ 3.4--$\mu$m features.}
\end{deluxetable*}

\begin{deluxetable*}{c|cc|cc|cc}
\tabletypesize{\small} 
\setlength{\tabcolsep}{20pt} 
\tablecaption{The absolute differences between photometric and spectroscopic optical depths ($\Delta \tau_{ps} = |\tau_p - \tau_s|$) that arise from the use of filters are presented for the water ice $-$OH (3.0--$\mu$m) feature, the aliphatic hydrocarbon $-$CH (3.4--$\mu$m) feature, and the silicate $-$SiO (10.0--$\mu$m) feature. The relative (normalized) absolute differences (${\Delta \tau_{ps}}/{\tau_{\text{s}}}$) for each feature are also indicated.}
\label{tab:deviation-PS} 
\tablewidth{\textwidth} 
\tablehead{
\colhead{} & \multicolumn{2}{c}{$-$OH (3.0--$\mu$m)}  & \multicolumn{2}{c}{$-$CH (3.4--$\mu$m)}  & \multicolumn{2}{c}{$-$SiO (10.0--$\mu$m)}  \\
\cline{2-7}
\colhead{} & \colhead{$\Delta$$\tau_{ps}$} & \colhead{$\Delta$$\tau_{ps}/\tau_{\text{s}}$} & \colhead{$\Delta$$\tau_{ps}$} & \colhead{$\Delta$$\tau_{ps}/\tau_{\text{s}}$} & \colhead{$\Delta$$\tau_{ps}$} & \colhead{$\Delta$$\tau_{ps}/\tau_{\text{s}}$} 
}
\startdata
Sgr A$^{*}$	&	0.06	&	0.07	&	0.05	&	0.08	&	0.42	&	0.22	\\
AFGL 2136	&	0.68	&	0.21	&	0.17	&	0.13	&	0.44	&	0.18	\\
NGC 7538 IRS 9	&	0.52	&	0.16	&	0.24	&	0.15	&	0.24	&	0.16	\\
Elias 29	&	0.34	&	0.22	&	0.06	&	0.13	&	0.19	&	0.17	\\
Orion BN	&	0.11	&	0.10	&	0.08	&	0.21	&	0.29	&	0.18	\\
RCrA IRS 1	&	0.44	&	0.31	&	0.11	&	0.23	&	0.05	&	0.10	\\
AFGL 989	&	0.31	&	0.25	&	0.04	&	0.10	&	0.09	&	0.20	\\
Mon R2 IRS 3	&	0.02	&	0.01	&	0.04	&	0.06	&	0.37	&	0.20	\\
S140	&	0.12	&	0.10	&	0.05	&	0.09	&	0.28	&	0.18	\\
AFGL 2591	&	0.11	&	0.09	&	0.09	&	0.14	&	0.33	&	0.18	\\
AFGL 490	&	0.01	&	0.04	&	0.00	&	0.06	&	0.20	&	0.29	\\
W3 IRS 5	&	0.24	&	0.06	&	0.25	&	0.13	&	1.33	&	0.26	\\
R CrA IRS 2	&	0.43	&	0.37	&	0.01	&	0.02	&	0.02	&	0.05	\\
Elias 16	&	0.33	&	0.27	&	0.02	&	0.11	&	-	&	-	\\
W33A	&	0.12	&	0.07	&	0.23	&	0.12	&	0.72	&	0.21	\\
Orion IRc2	&	0.36	&	0.34	&	0.24	&	0.49	&	0.22	&	0.13	\\
AFGL7009s	&	0.02	&	0.02	&	0.15	&	0.10	&	0.37	&	0.17	\\
\hline
Average	&	0.25	&	0.16	&	0.11	&	0.14	&	0.35	&	0.18	\\
\enddata
\end{deluxetable*}

\begin{deluxetable*}{c|cccc|cccc|cccc}
\tabletypesize{\small} 
\setlength{\tabcolsep}{6pt} 
\tablecaption{The absolute differences between photometric, spectroscopic and reported ([G04]) optical depths that arise from limited resolution ($\Delta \tau_{sr} = |\tau_s - \tau_r|$) and methodological differences ($\Delta \tau_{pr} = |\tau_p - \tau_r|$) are presented for the water ice $-$OH (3.0--$\mu$m) feature, the aliphatic hydrocarbon $-$CH (3.4--$\mu$m) feature, and the silicate $-$SiO (10.0--$\mu$m) feature. The relative (normalized) absolute differences (${\Delta \tau}/{\tau_{\text{s}}}$) for each feature are also indicated.}
\label{tab:deviation-PSR}  
\tablewidth{\textwidth} 
\tablehead{
\colhead{} & \multicolumn{4}{c}{$-$OH (3.0--$\mu$m)} & \multicolumn{4}{c}{$-$CH (3.4--$\mu$m)} & \multicolumn{4}{c}{$-$SiO (10.0--$\mu$m)} \\
\cline{2-13}
\colhead{} & \colhead{$\Delta$$\tau_{sr}$} & \colhead{$\Delta$$\tau_{pr}$} &  \colhead{$\Delta$$\tau_{sr}/\tau_{\text{s}}$} & \colhead{$\Delta$$\tau_{pr}/\tau_{\text{r}}$} & \colhead{$\Delta$$\tau_{sr}$} & \colhead{$\Delta$$\tau_{pr}$} & \colhead{$\Delta$$\tau_{sr}/\tau_{\text{s}}$} & \colhead{$\Delta$$\tau_{pr}/\tau_{\text{r}}$}  & \colhead{$\Delta$$\tau_{sr}$} & \colhead{$\Delta$$\tau_{pr}$} & \colhead{$\Delta$$\tau_{sr}/\tau_{\text{s}}$} & \colhead{$\Delta$$\tau_{pr}/\tau_{\text{r}}$}  
}
\startdata
Sgr A$^{*}$	&	0.34	&	0.28	&	0.68	&	0.57	&	0.33	&	0.29	&	1.58	&	1.37	&	0.41	&	0.83	&	0.18	&	0.36	\\
AFGL 2136	&	0.04	&	0.64	&	0.01	&	0.20	&	1.16	&	0.99	&	8.49	&	7.26	&	0.30	&	0.14	&	0.14	&	0.07	\\
NGC 7538 IRS 9	&	0.15	&	0.37	&	0.05	&	0.12	&	1.45	&	1.21	&	11.18	&	9.32	&	0.72	&	0.96	&	0.32	&	0.43	\\
Elias 29	&	0.27	&	0.61	&	0.14	&	0.33	&	0.41	&	0.47	&	0.46	&	0.53	&	0.19	&	0.39	&	0.15	&	0.29	\\
Orion BN	&	0.36	&	0.25	&	0.25	&	0.18	&	0.37	&	0.45	&	10.89	&	13.36	&	0.02	&	0.27	&	0.01	&	0.17	\\
RCrA IRS 1	&	0.03	&	0.46	&	0.02	&	0.32	&	0.45	&	0.34	&	10.93	&	8.21	&	0.14	&	0.18	&	0.23	&	0.30	\\
AFGL 989	&	0.16	&	0.47	&	0.11	&	0.33	&	0.31	&	0.28	&	6.26	&	5.51	&	0.26	&	0.35	&	0.36	&	0.49	\\
Mon R2 IRS 3	&	0.15	&	0.14	&	0.14	&	0.12	&	0.72	&	0.68	&	20.02	&	18.77	&	0.66	&	1.03	&	0.26	&	0.40	\\
S140	&	0.10	&	0.02	&	0.09	&	0.02	&	0.52	&	0.47	&	10.30	&	9.32	&	0.11	&	0.17	&	0.07	&	0.11	\\
AFGL 2591	&	0.43	&	0.32	&	0.58	&	0.43	&	0.58	&	0.49	&	12.84	&	10.91	&	0.40	&	0.73	&	0.18	&	0.32	\\
AFGL 490	&	0.05	&	0.04	&	0.15	&	0.11	&	0.04	&	0.05	&	1.45	&	1.59	&	0.05	&	0.15	&	0.07	&	0.24	\\
W3 IRS 5	&	1.22	&	0.98	&	0.44	&	0.35	&	1.76	&	1.51	&	13.23	&	11.35	&	0.79	&	2.12	&	0.14	&	0.37	\\
R CrA IRS 2	&	0.13	&	0.57	&	0.12	&	0.42	&	0.15	&	0.15	&	3.78	&	4.65	&	-	&	-	&	-	&	-	\\
Elias 16	&	0.04	&	0.29	&	0.03	&	0.25	&	0.25	&	0.27	&	7.69	&	6.49	&	-	&	-	&	-	&	-	\\
W33A	&	3.86	&	3.97	&	0.70	&	0.72	&	1.61	&	1.38	&	5.56	&	4.77	&	4.47	&	5.19	&	0.57	&	0.66	\\
Orion IRc2	&	0.41	&	0.05	&	0.28	&	0.03	&	-	&	-	&	-	&	-	&	0.10	&	0.32	&	0.05	&	0.17	\\
AFGL7009s	&	-	&	-	&	-	&	-	&	-	&	-	&	-	&	-	&	1.87	&	2.24	&	0.47	&	0.56	\\
\hline
Average	&	0.48	&	0.59	&	0.24	&	0.28	&	0.67	&	0.60	&	8.31	&	7.56	&	0.70	&	1.00	&	0.21	&	0.33	\\
\enddata
\end{deluxetable*}

\begin{figure*}[htbp]
  \begin{center}
    \begin{tabular}{ccc}
      {\includegraphics[angle=0,scale=0.20]{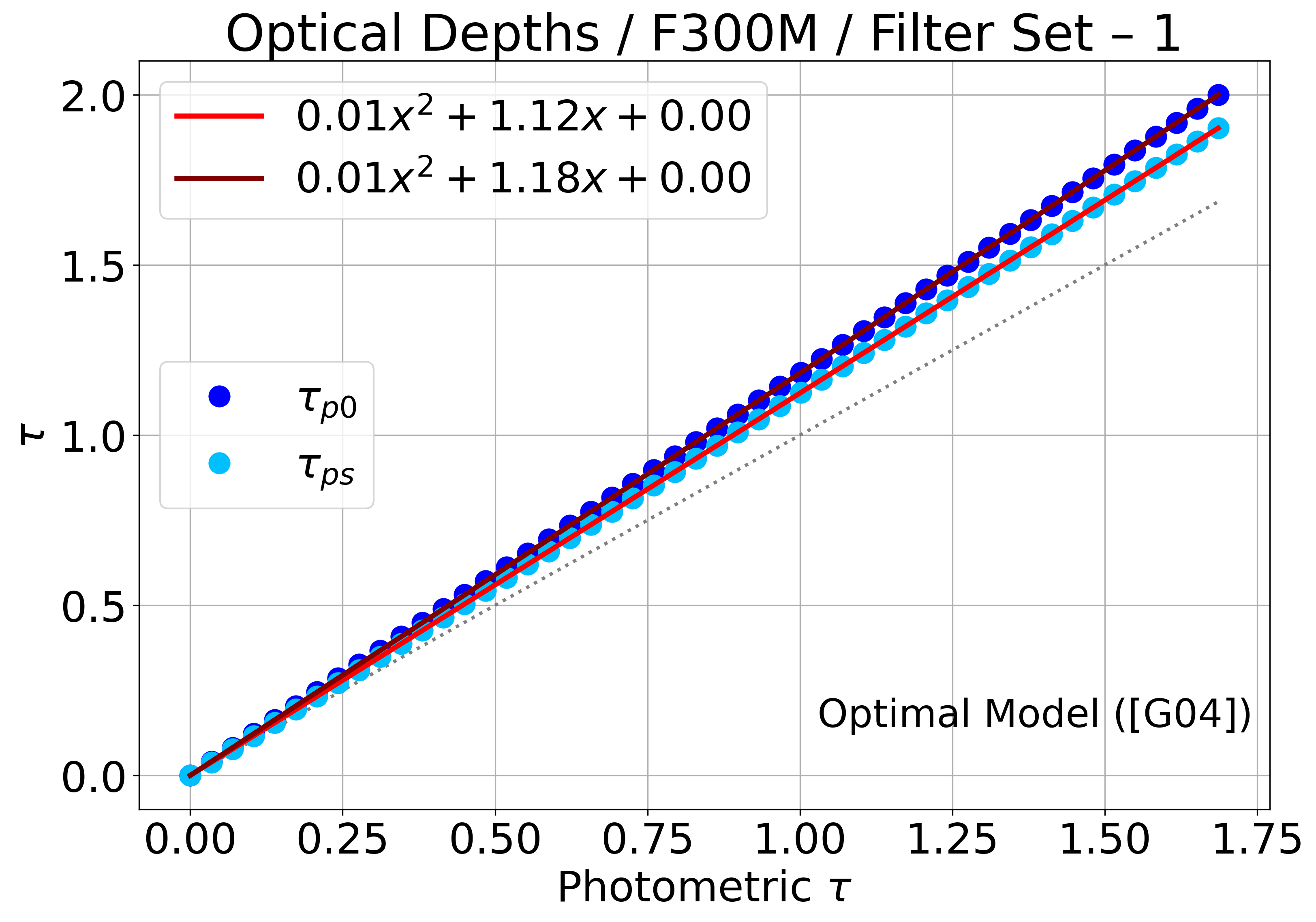}} &
      {\includegraphics[angle=0,scale=0.20]{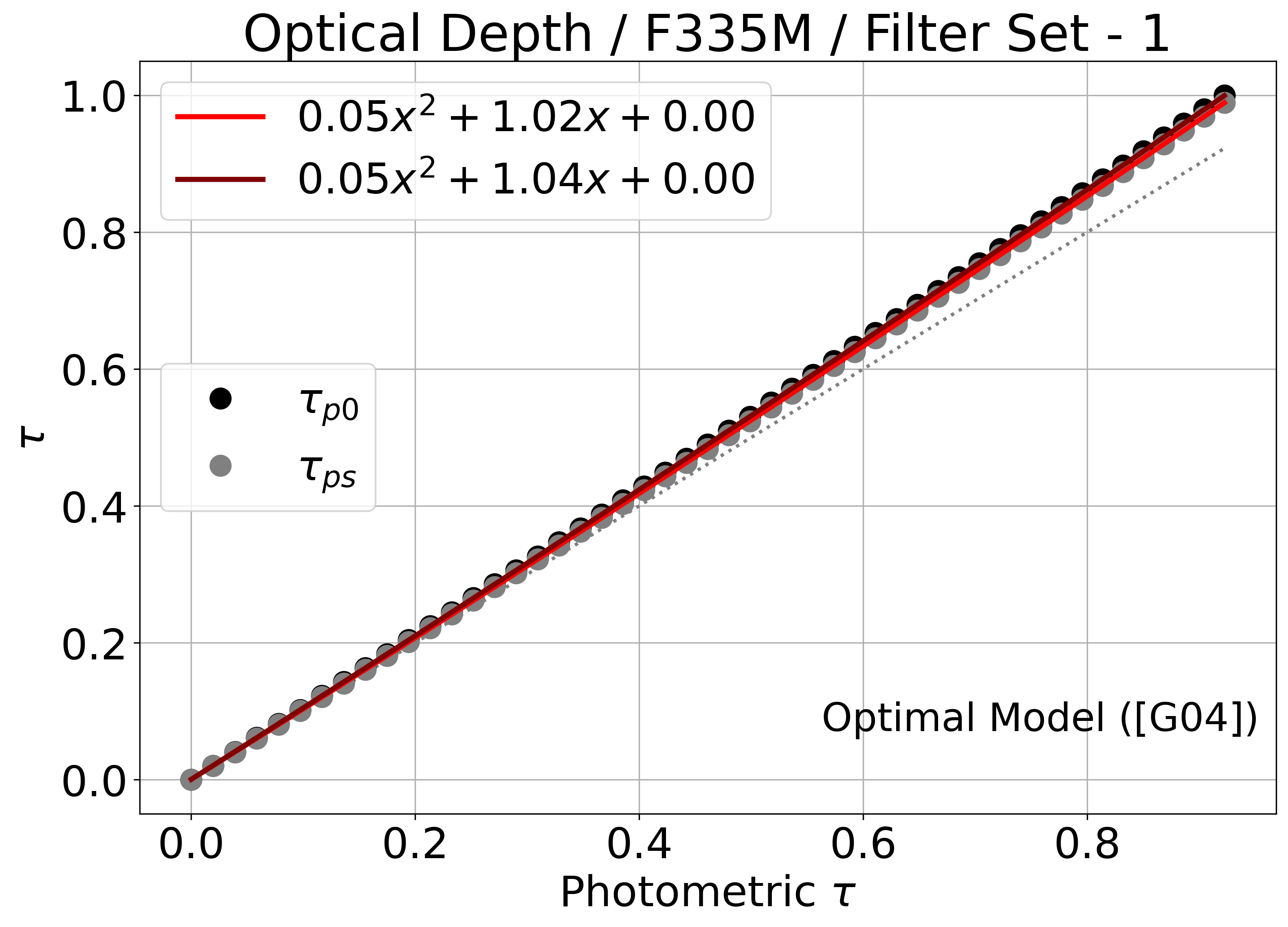}} &
      {\includegraphics[angle=0,scale=0.20]{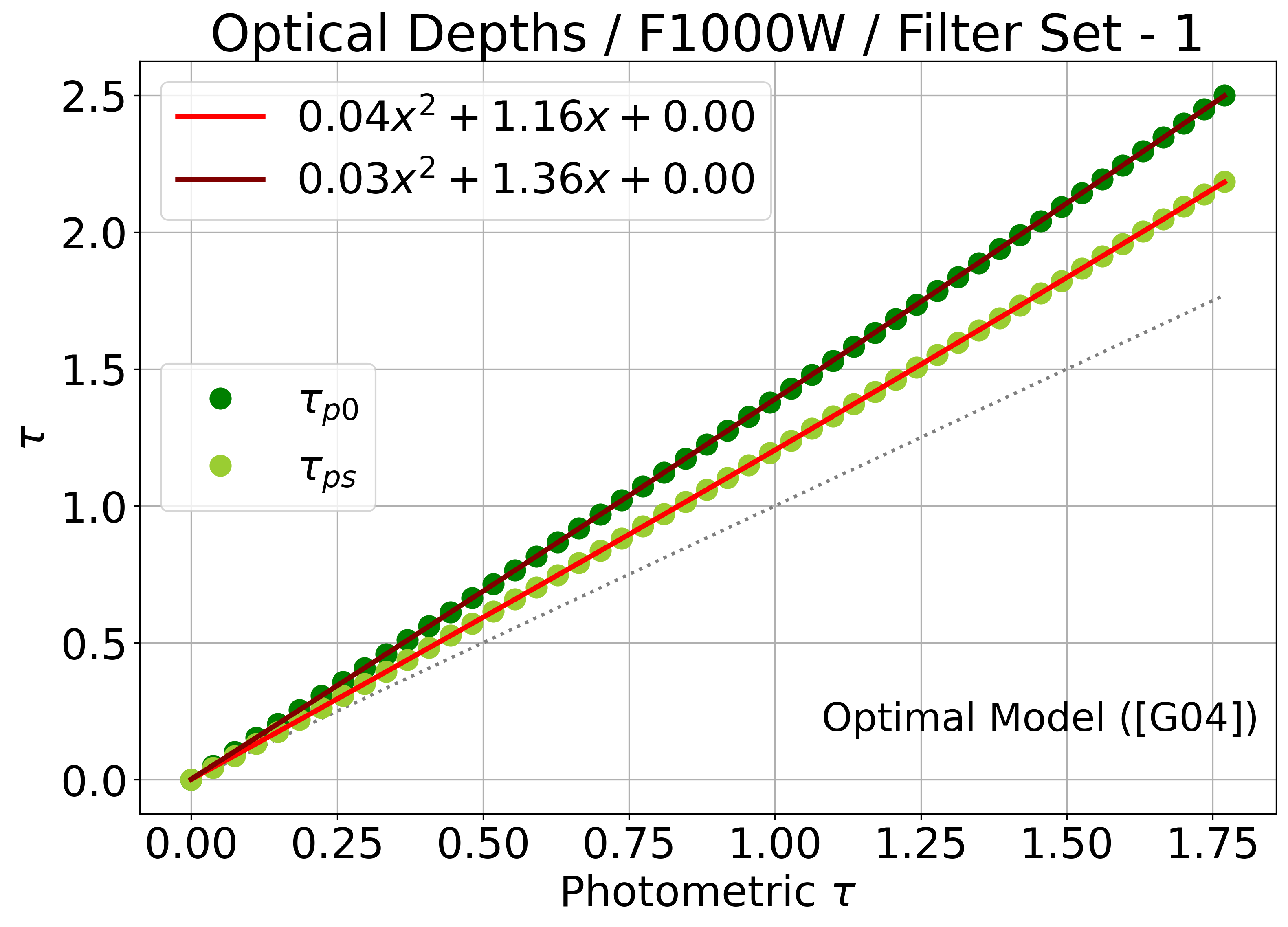}} \\
        \end{tabular}
    \caption{The photometric optical depths, spectroscopic and reference optical depths at 3.0--$\mu$m, 3.35–$\mu$m, and 10.0--$\mu$m are measured using \textit{optimal spectral models} and compared to investigate methodological biases. The polynomial equations (Table~\ref{tab:coeffs}) for calibrating methodological biases due to filter use ($P_{ps}(x)$, the correlation between $\tau_{p}$ and $\tau_{s}$) and total biases ($P_{p0}(x)$, correlation between $\tau_{p}$ and $\tau_{0}$, including those arising from limited resolution) are obtained based on this comparisons. The gray dotted line represents equal $\tau_{p}$ and $\tau_{s}$ values.} 
    \label{fig:Calib-Poly}
      \end{center}
\end{figure*}

\begin{deluxetable*}{cccc}[htbp]
\setlength{\tabcolsep}{22pt} 
\tablecaption{Polynomial equations from \textit{optimal models} used for calibrating methodological biases due to filter use ($P_{ps}(x)$, obtained from the correlation between $\tau_{p}$ and $\tau_{s}$) and total biases ($P_{p0}(x)$, obtained from the correlation between $\tau_{p}$ and $\tau_{0}$), including those arising from limited resolution (see figure*~\ref{fig:Calib-Poly}). \label{tab:coeffs}}
\tablehead{
\colhead{} & \colhead{$-$OH (3.0--$\mu$m)} & \colhead{$-$CH (3.4--$\mu$m)} & \colhead{$-$SiO (10.0--$\mu$m)}
}
\startdata
$P_{ps}(x)$
& \(0.01x^2 + 1.12x + 0.00\) 
& \(0.05x^2 + 1.02x + 0.00\) 
& \(0.04x^2 + 1.16x + 0.00\) \\
\hline
$P_{p0}(x)$
& \(0.01x^2 + 1.18x + 0.00\) 
& \(0.05x^2 + 1.04x + 0.00\) 
& \(0.03x^2 + 1.36x + 0.00\) \\
\enddata
\end{deluxetable*}

\begin{deluxetable*}{c|ccc|ccc|ccc}
\tabletypesize{\small} 
\setlength{\tabcolsep}{12 pt} 
\tablecaption{The photometric optical depths that are calibrated for methodological biases ($\tau_{c}$) due to filter use for the water ice $-$OH (3.0--$\mu$m) feature, the aliphatic hydrocarbon $-$CH (3.4--$\mu$m) feature, and the silicate $-$SiO (10.0--$\mu$m) feature are presented. The photometric optical depths corrected for total biases ($\tau_{c0}$) and adjusted based on blackbody continuum estimation ($\tau_{BB}$) are also presented for comparison.}
\label{tab:calibrated-improved}
\tablewidth{\textwidth}
\tablehead{
\colhead{} & \multicolumn{3}{c}{$-$OH (3.0--$\mu$m)} & \multicolumn{3}{c}{$-$CH (3.4--$\mu$m)} & \multicolumn{3}{c}{$-$SiO (10.0--$\mu$m)} \\
\cline{2-10}
\colhead{Source} & \colhead{$\tau_{c}$} & \colhead{$\tau_{c0}$} & \colhead{$\tau_{BB}$} & \colhead{$\tau_{c}$} & \colhead{$\tau_{c0}$} & \colhead{$\tau_{BB}$} & \colhead{$\tau_{c}$} & \colhead{$\tau_{c0}$} & \colhead{$\tau_{BB}$}
}
\startdata
Sgr A$^{*}$	&	0.88	&	0.93	&	0.83	&	0.52	&	0.53	&	0.49	&	1.80	&	2.08	&	2.17	\\
AFGL 2136	&	2.98	&	3.13	&	2.58	&	1.22	&	1.24	&	0.87	&	2.43	&	2.78	&	2.89	\\
NGC 7538 IRS 9	&	3.13	&	3.30	&	2.84	&	1.46	&	1.49	&	1.19	&	1.53	&	1.77	&	1.89	\\
Elias 29	&	1.41	&	1.48	&	1.14	&	0.44	&	0.44	&	0.23	&	1.12	&	1.29	&	1.37	\\
Orion BN	&	1.34	&	1.42	&	1.44	&	0.51	&	0.52	&	0.58	&	1.60	&	1.85	&	1.89	\\
RCrA IRS 1	&	1.10	&	1.16	&	1.19	&	0.39	&	0.40	&	0.46	&	0.50	&	0.58	&	0.70	\\
AFGL 989	&	1.08	&	1.14	&	1.17	&	0.34	&	0.34	&	0.41	&	0.43	&	0.51	&	0.63	\\
Mon R2 IRS 3	&	1.42	&	1.50	&	1.28	&	0.75	&	0.77	&	0.64	&	1.86	&	2.14	&	2.26	\\
S140	&	1.25	&	1.31	&	1.17	&	0.54	&	0.55	&	0.48	&	1.62	&	1.87	&	1.99	\\
AFGL 2591	&	1.20	&	1.26	&	0.86	&	0.56	&	0.57	&	0.31	&	1.86	&	2.14	&	2.26	\\
AFGL 490	&	0.32	&	0.34	&	0.39	&	0.08	&	0.08	&	0.16	&	0.58	&	0.67	&	0.79	\\
W3 IRS 5	&	4.35	&	4.58	&	3.09	&	1.81	&	1.84	&	0.89	&	4.81	&	5.41	&	5.52	\\
R CrA IRS 2	&	0.96	&	1.01	&	1.03	&	0.32	&	0.32	&	0.38	&	0.47	&	0.55	&	0.67	\\
Elias 16	&	0.89	&	0.94	&	0.94	&	0.19	&	0.19	&	0.19	&	-	&	-	&	-	\\
W33A	&	1.73	&	1.82	&	1.78	&	1.85	&	1.88	&	1.88	&	3.35	&	3.81	&	3.93	\\
OrionIRc2	&	1.62	&	1.71	&	1.75	&	0.76	&	0.77	&	0.84	&	1.89	&	2.18	&	2.16	\\
AFGL7009s	&	1.25	&	1.32	&	1.27	&	1.41	&	1.43	&	1.43	&	2.17	&	2.49	&	2.61	\\
\hline
Average	&	1.58	&	1.67	&	1.46	&	0.77	&	0.79	&	0.67	&	1.75	&	2.01	&	2.11	\\
\enddata
\end{deluxetable*}

\begin{deluxetable*}{c|cccc|cccc|cccc}
\tabletypesize{\small} 
\setlength{\tabcolsep}{6pt} 
\tablecaption{The absolute differences between calibrated, spectroscopic and reported ([G04]) optical depths ($\Delta \tau_{cs} = |\tau_c - \tau_s|$) and methodological differences ($\Delta \tau_{cr} = |\tau_c - \tau_r|$) are presented for the water ice $-$OH (3.0--$\mu$m) feature, the aliphatic hydrocarbon $-$CH (3.4--$\mu$m) feature, and the silicate $-$SiO (10.0--$\mu$m) feature. The relative (normalized) absolute differences (${\Delta \tau}/{\tau_{\text{s}}}$) for each feature are also indicated.}
\label{tab:deviation-CSR}  
\tablewidth{\textwidth} 
\tablehead{
\colhead{} & \multicolumn{4}{c}{$-$OH (3.0--$\mu$m)} & \multicolumn{4}{c}{$-$CH (3.4--$\mu$m)} & \multicolumn{4}{c}{$-$SiO (10.0--$\mu$m)} \\
\cline{2-13}
\colhead{} & \colhead{$\Delta$$\tau_{cs}$} & \colhead{$\Delta$$\tau_{cr}$} &  \colhead{$\Delta$$\tau_{cs}/\tau_{\text{s}}$} & \colhead{$\Delta$$\tau_{cr}/\tau_{\text{r}}$} & \colhead{$\Delta$$\tau_{cs}$} & \colhead{$\Delta$$\tau_{cr}$} &  \colhead{$\Delta$$\tau_{cs}/\tau_{\text{s}}$} & \colhead{$\Delta$$\tau_{cr}/\tau_{\text{r}}$} & \colhead{$\Delta$$\tau_{cs}$} & \colhead{$\Delta$$\tau_{cr}$} &  \colhead{$\Delta$$\tau_{cs}/\tau_{\text{s}}$} & \colhead{$\Delta$$\tau_{cr}/\tau_{\text{r}}$} 
}
\startdata
Sgr A$^{*}$	&	0.04	&	0.38	&	0.05	&	0.77	&	0.02	&	0.31	&	0.04	&	1.47	&	0.10	&	0.51	&	0.05	&	0.22	\\
AFGL 2136	&	0.31	&	0.26	&	0.09	&	0.08	&	0.08	&	1.08	&	0.06	&	7.89	&	0.03	&	0.33	&	0.01	&	0.16	\\
NGC 7538 IRS 9	&	0.12	&	0.03	&	0.04	&	0.01	&	0.12	&	1.33	&	0.08	&	10.22	&	0.02	&	0.70	&	0.01	&	0.31	\\
Elias 29	&	0.18	&	0.44	&	0.11	&	0.24	&	0.05	&	0.45	&	0.09	&	0.51	&	0.01	&	0.20	&	0.01	&	0.15	\\
Orion BN	&	0.26	&	0.10	&	0.24	&	0.07	&	0.11	&	0.48	&	0.26	&	14.00	&	0.01	&	0.01	&	0.01	&	0.01	\\
RCrA IRS 1	&	0.31	&	0.34	&	0.22	&	0.23	&	0.10	&	0.35	&	0.20	&	8.57	&	0.03	&	0.11	&	0.06	&	0.18	\\
AFGL 989	&	0.19	&	0.35	&	0.15	&	0.25	&	0.03	&	0.29	&	0.07	&	5.75	&	0.03	&	0.29	&	0.06	&	0.40	\\
Mon R2 IRS 3	&	0.15	&	0.30	&	0.12	&	0.27	&	0.01	&	0.72	&	0.01	&	19.87	&	0.04	&	0.69	&	0.02	&	0.27	\\
S140	&	0.03	&	0.13	&	0.02	&	0.11	&	0.03	&	0.49	&	0.05	&	9.79	&	0.00	&	0.11	&	0.00	&	0.07	\\
AFGL 2591	&	0.03	&	0.46	&	0.03	&	0.62	&	0.06	&	0.52	&	0.10	&	11.46	&	0.01	&	0.39	&	0.00	&	0.17	\\
AFGL 490	&	0.05	&	0.00	&	0.17	&	0.01	&	0.01	&	0.05	&	0.08	&	1.65	&	0.11	&	0.06	&	0.16	&	0.10	\\
W3 IRS 5	&	0.36	&	1.57	&	0.09	&	0.57	&	0.08	&	1.68	&	0.04	&	12.61	&	0.20	&	0.99	&	0.04	&	0.17	\\
R CrA IRS 2	&	0.33	&	0.47	&	0.28	&	0.34	&	0.00	&	0.15	&	0.00	&	4.81	&	0.09	&	-	&	0.24	&	-	\\
Elias 16	&	0.22	&	0.18	&	0.18	&	0.16	&	0.03	&	0.28	&	0.16	&	6.75	&	-	&	-	&	-	&	-	\\
W33A	&	0.09	&	3.77	&	0.05	&	0.69	&	0.05	&	1.56	&	0.03	&	5.37	&	0.01	&	4.49	&	0.00	&	0.57	\\
OrionIRc2	&	0.56	&	0.14	&	0.52	&	0.10	&	0.28	&	-	&	0.57	&	-	&	0.12	&	0.02	&	0.07	&	0.01	\\
AFGL7009s	&	0.17	&	-	&	0.16	&	-	&	0.04	&	-	&	0.03	&	-	&	0.03	&	1.83	&	0.02	&	0.46	\\
\hline
Average	&	0.20	&	0.56	&	0.15	&	0.28	&	0.06	&	0.65	&	0.11	&	8.05	&	0.05	&	0.72	&	0.05	&	0.22	\\
\enddata
\end{deluxetable*}

\newlength{\defaulttabcolsep}
\setlength{\defaulttabcolsep}{\tabcolsep}

\clearpage


\bibliography{A_List}{}
\bibliographystyle{aasjournal}



\end{document}